\documentclass[%
reprint,
superscriptaddress,
showpacs,preprintnumbers,
 amsmath,amssymb,
 aps,prl
]{revtex4-1}
\usepackage{mathtools,amssymb,lipsum}
\usepackage{multirow}
\usepackage{dcolumn}% Align table columns on decimal point
\usepackage{bm}% bold math
\usepackage{color}
\definecolor{mygreen}{rgb}{0.1, 0.6, 0.1}
\usepackage{amsmath}

\usepackage{graphicx}% added by Iimura
\usepackage{hyperref}% added by Iimura

\begin{document}

\title{A new study of the $N=32$ and $N=34$ shell gap for Ti and V by the first high-precision MRTOF mass measurements at BigRIPS-SLOWRI}% Force line breaks with \\

\author{S.~Iimura}
\email{%
shun.iimura@rikkyo.ac.jp
}%
\affiliation{%
 RIKEN Nishina Center for Accelerator-Based Science, Wako, Saitama 351-0198, Japan\\
}%
\affiliation{%
 Department of Physics, Graduate School of Science, Osaka University, 1-1 Machikaneyama, Toyonaka, Osaka 560-0043, Japan\\
 %this line break forced with \textbackslash\textbackslash
}%
\affiliation{%
 Wako Nuclear Science Center (WNSC), Institute of Particle and Nuclear Studies (IPNS), High Energy Accelerator Research Organization (KEK), Wako, Saitama 351-0198, Japan\\
 %this line break forced with \textbackslash\textbackslash
}%
\affiliation{%
 Department of Physics, College of Science, Rikkyo University, 3-34-1 Nishi-Ikebukuro, Tokyo 171-8501, Japan\\
 %this line break forced with \textbackslash\textbackslash
}%
\author{M.~Rosenbusch}
\email{%
rosmar@post.kek.jp
}%
\affiliation{%
 Wako Nuclear Science Center (WNSC), Institute of Particle and Nuclear Studies (IPNS), High Energy Accelerator Research Organization (KEK), Wako, Saitama 351-0198, Japan\\
 %this line break forced with \textbackslash\textbackslash
}%
\author{A.~Takamine}%
\affiliation{%
 RIKEN Nishina Center for Accelerator-Based Science, Wako, Saitama 351-0198, Japan\\
}%
\author{Y.~Tsunoda}%
\affiliation{%
 Center for Computational Sciences, University of Tsukuba, Tsukuba, 305-8577, Japan\\
}%
\author{M.~Wada}%
\affiliation{%
 Wako Nuclear Science Center (WNSC), Institute of Particle and Nuclear Studies (IPNS), High Energy Accelerator Research Organization (KEK), Wako, Saitama 351-0198, Japan\\
}
\author{S.~Chen}%
\affiliation{%
 Department of Physics, The University of Hong Kong, Pokufulam, China\\
}%
\author{D.~S.~Hou}%
\affiliation{%
Institute of Modern Physics, Chinese Academy of Sciences, Lanzhou 730000, China\\
}%
\affiliation{%
University of Chinese Academy of Sciences, Beijing 100049, China\\
}%
\affiliation{%
School of Nuclear Science and Technology, Lanzhou University,  Lanzhou 730000, China\\
}%
\author{W.~Xian}%
\affiliation{%
 Department of Physics, The University of Hong Kong, Pokufulam, China\\
}%
\author{H.~Ishiyama}
\affiliation{%
 RIKEN Nishina Center for Accelerator-Based Science, Wako, Saitama 351-0198, Japan\\
}%
\author{S.~Yan}%
\affiliation{%
 Institute of Mass Spectrometry and Atmospheric Environment, Jinan University, Guangzhou 510632, China\\
}%
\author{P.~Schury}%
\affiliation{%
 Wako Nuclear Science Center (WNSC), Institute of Particle and Nuclear Studies (IPNS), High Energy Accelerator Research Organization (KEK), Wako, Saitama 351-0198, Japan\\
}%
\author{H.~Crawford}%
\affiliation{%
 Nuclear Science Division, Lawrence Berkeley National Laboratory, Berkeley, CA, 94523, USA\\
}%
\author{P.~Doornenbal}
\affiliation{%
 RIKEN Nishina Center for Accelerator-Based Science, Wako, Saitama 351-0198, Japan\\
}%
\author{Y.~Hirayama}%
\affiliation{%
 Wako Nuclear Science Center (WNSC), Institute of Particle and Nuclear Studies (IPNS), High Energy Accelerator Research Organization (KEK), Wako, Saitama 351-0198, Japan\\
}%
\author{Y.~Ito}%
\affiliation{%
 Advanced Science Research Center, Japan Atomic Energy Agency, Ibaraki 319-1195, Japan\\
}%
\author{S.~Kimura}%
\affiliation{%
 RIKEN Nishina Center for Accelerator-Based Science, Wako, Saitama 351-0198, Japan\\
}%
\author{T.~Koiwai}%
\affiliation{%
 Department of Physics, The University of Tokyo, 7-3-1 Hongo, Bunkyo, Tokyo, 113-0033, Japan\\
 %this line break forced with \textbackslash\textbackslash
}%
\affiliation{%
 RIKEN Nishina Center for Accelerator-Based Science, Wako, Saitama 351-0198, Japan\\
}%
\author{T.~M.~Kojima}%
\affiliation{%
 RIKEN Nishina Center for Accelerator-Based Science, Wako, Saitama 351-0198, Japan\\
}%
\author{H.~Koura}%
\affiliation{%
 Advanced Science Research Center, Japan Atomic Energy Agency, Ibaraki 319-1195, Japan\\
}%
\author{J.~Lee}%
\affiliation{%
 Department of Physics, The University of Hong Kong, Pokufulam, China\\
}%
\author{J.~Liu}%
\affiliation{%
 Department of Physics, The University of Hong Kong, Pokufulam, China\\
}%
\affiliation{%
Institute of Modern Physics, Chinese Academy of Sciences, Lanzhou 730000, China\\
}%
\author{S.~Michimasa}%
\affiliation{%
 Center of Nuclear Study (CNS), The University of Tokyo, Bunkyo 113-0033, Japan\\
}%
\author{H.~Miyatake}%
\affiliation{%
 Wako Nuclear Science Center (WNSC), Institute of Particle and Nuclear Studies (IPNS), High Energy Accelerator Research Organization (KEK), Wako, Saitama 351-0198, Japan\\
}%
\author{J.~Y.~Moon}%
\affiliation{%
 Institute for Basic Science, 70, Yuseong-daero 1689-gil, Yusung-gu, Daejeon 305-811, Korea\\
}%
\author{S.~Nishimura}%
\affiliation{%
 RIKEN Nishina Center for Accelerator-Based Science, Wako, Saitama 351-0198, Japan\\
}%
\author{\\S.~Naimi}%
\affiliation{%
 RIKEN Nishina Center for Accelerator-Based Science, Wako, Saitama 351-0198, Japan\\
}%
\author{T.~Niwase}%
\affiliation{%
 RIKEN Nishina Center for Accelerator-Based Science, Wako, Saitama 351-0198, Japan\\
}%
\affiliation{%
 Kyushu University, Hakozaki, Higashi-ku, Fukuoka 812-8581, Japan\\
}%
\affiliation{%
 Wako Nuclear Science Center (WNSC), Institute of Particle and Nuclear Studies (IPNS), High Energy Accelerator Research Organization (KEK), Wako, Saitama 351-0198, Japan\\
}%
\author{A.~Odahara}
\affiliation{%
 Department of Physics, Graduate School of Science, Osaka University, 1-1 Machikaneyama, Toyonaka, Osaka 560-0043, Japan\\
 %this line break forced with \textbackslash\textbackslash
}%
\author{T.~Otsuka}
\affiliation{%
 Department of Physics, The University of Tokyo, 7-3-1 Hongo, Bunkyo, Tokyo, 113-0033, Japan\\
 %this line break forced with \textbackslash\textbackslash
}%
\affiliation{%
 RIKEN Nishina Center for Accelerator-Based Science, Wako, Saitama 351-0198, Japan\\
}%
\affiliation{%
 Advanced Science Research Center, Japan Atomic Energy Agency, Ibaraki 319-1195, Japan\\
}%
\author{S.~Paschalis}%
\affiliation{%
School of Physics, Engineering and Technology, University of York, York YO10 5DD, United Kingdom\\
}%
\author{M.~Petri}%
\affiliation{%
School of Physics, Engineering and Technology, University of York, York YO10 5DD, United Kingdom\\
}%
\author{N.~Shimizu}%
\affiliation{%
 Center for Computational Sciences, University of Tsukuba, Tsukuba, 305-8577, Japan\\
}%
\author{T.~Sonoda}%
\affiliation{%
 RIKEN Nishina Center for Accelerator-Based Science, Wako, Saitama 351-0198, Japan\\
% %This line break forced with \textbackslash\textbackslash
}%
\author{D.~Suzuki}
\affiliation{%
 RIKEN Nishina Center for Accelerator-Based Science, Wako, Saitama 351-0198, Japan\\
}%
\author{Y.~X.~Watanabe}%
\affiliation{%
 Wako Nuclear Science Center (WNSC), Institute of Particle and Nuclear Studies (IPNS), High Energy Accelerator Research Organization (KEK), Wako, Saitama 351-0198, Japan\\
}%
\author{K.~Wimmer}%
\affiliation{%
 Department of Physics, The University of Tokyo, 7-3-1 Hongo, Bunkyo, Tokyo, 113-0033, Japan\\
 %this line break forced with \textbackslash\textbackslash
}%
\affiliation{%
 GSI Helmholtzzentrum f\"{u}r Schwerionenforschung, 64291 Darmstadt, Germany\\
}%
\affiliation{%
 RIKEN Nishina Center for Accelerator-Based Science, Wako, Saitama 351-0198, Japan\\
}%
\author{H.~Wollnik}%
\affiliation{%
 New Mexico State University, Las Cruces, NM 88001, USA\\
}%

\date{\today}% It is always \today, today,

\begin{abstract}
The atomic masses of $^{55}$Sc, $^{56,58}$Ti, and $^{56-59}$V have been determined using the high-precision multi-reflection time-of-flight technique. The radioisotopes have been produced at RIKEN's RIBF facility and delivered to the novel designed gas cell and multi-reflection system (ZD MRTOF), which has been recently commissioned downstream of the ZeroDegree spectrometer following the BigRIPS separator. For $^{56,58}$Ti and $^{56-59}$V the mass uncertainties have been reduced down to the order of $10\,\mathrm{keV}$, shedding new light on the $N=34$ shell effect in Ti and V isotopes by the first high-precision mass measurements of the critical species $^{58}$Ti and $^{59}$V. With the new precision achieved, we reveal the non-existence of the $N=34$ empirical two-neutron shell gaps for Ti and V, and the enhanced energy gap above the occupied $\nu p_{3/2}$ orbit is identified as a feature unique to Ca. We perform new Monte Carlo shell model calculations including the $\nu d_{5/2}$ and $\nu g_{9/2}$ orbits and compare the results with conventional shell model calculations, which exclude the $\nu g_{9/2}$ and the $\nu d_{5/2}$ orbits. The comparison indicates that the shell gap reduction in Ti is related to a partial occupation of the higher orbitals for the outer two valence neutrons at $N=34$.
\end{abstract}

% PACS, Nuclear forces, Binding energies and masses, Shell model, Time of flight mass spectrometry
%Use showkeys class option if keyword
\maketitle

%\tableofcontents
Masses of neutron-rich isotopes with $N \geq 32$ between Ca and Ni have recently been studied intensely as valuable probes for the complex nuclear structure emerging from nucleon-nucleon interactions \cite{Reiter2018,Leistenschneider2018,Leistenschneider2021,Mougeot2018,Porter2022,Otsuka2020} and, furthermore, triggered major interest for nuclear astrophysics \cite{Schatz2014,Deibel2016}. About twenty years ago the major driving force for the strongly changing level structure was identified as the spin-isospin dependence of the tensor force between nucleons \cite{Otsuka2001}, which lowers the $\nu f_{5/2}$ orbit with increasing occupation of the $\pi f_{7/2}$ orbital for $Z>20$. An additional ingredient to the nuclear structure is a general decrease of the spin-orbit splitting for the neutron levels by a more diffuse surface of neutron-rich nuclei \cite{Sorlin2003,Gaudefroy2005}. The interconnection of the nuclear forces can cause the different orbits to be very close in energy, leading to the onset of collective behavior \cite{Sorlin2002, Sorlin2003, Aoi2009, Suzuki2013, Crawford2013}. For Ca isotopes, a pronounced reduction of tensor-interaction effects due to the decrease of proton valence particles occupying the $\pi f_{7/2}$ orbits has been confirmed by the discovery of two new magic neutron numbers \cite{Otsuka2005}, \textit{i.e.} $N=32$ in $^{52}$Ca by nuclear spectroscopy \cite{Huck1985} and atomic mass measurements \cite{Wienholtz2013}, and $N=34$ in $^{54}$Ca by in-beam $\gamma$-ray spectroscopy \cite{Steppenbeck2013} recently confirmed by the first mass spectroscopy \cite{Michimasa2018} of $^{55-57}$Ca. For systems with additional protons the level structure becomes more dense up to the pronounced collectivity in Cr isotopes, which has been investigated by in-beam $\gamma$-ray studies \cite{Gade2010, Crawford2013, Gade2021}, as well as new mass evaluations \cite{Mougeot2018} highlighting the necessity to include the full $pf$ shell, $g_{9/2}$, and $d_{5/2}$ orbits in modern theoretical calculations \cite{Lenzi2010}. In the Ti isotope chain a collective behavior by level intrusion of $\nu g_{9/2}$ has been found by decay spectroscopy for isomeric states of $^{61}$Ti \cite{Wimmer2019} and in-beam $\gamma$-ray measurements of $^{62}$Ti \cite{CORTES2020}. The first comprehensive mass studies of V and Ti above $N=34$ performed with the $B\rho$-TOF method are very recent \cite{Meisel2020, Michimasa2020}. A prominent onset of deformation was confirmed by an increase of binding energy toward $^{62}$Ti and $^{64}$V \cite{Michimasa2020}.\par
Multi-reflection time-of-flight (MRTOF) technology became state-of-the-art for nuclear mass measurements after 2010 and is being developed in several facilities worldwide (see references in \cite{Rosenbusch2022}). Precisely measured binding energies of nuclear ground states and metastable states are an essential benchmark for theoretical calculations of the nuclear level structure, and pave the way for accurate extrapolations to presently inaccessible nuclei. In this letter, we present the on-line debut of a new part of the SLOWRI project \cite{Wada2011}, the ZD MRTOF mass spectrograph \cite{Rosenbusch2022}, which has been put into operation and coupled to a cryogenic gas cell located downstream of the ZeroDegree spectrometer (ZDS) beamline at the RIBF facility. This unique configuration has been used in both stand-alone and symbiotic operations since an initial commissioning campaign performed together with an in-beam $\gamma$-ray spectroscopy campaign (HiCARI project \cite{Wimmer2021}). We report greatly improved mass precision for neutron-rich Ti and V isotopes up to $N=36$, and thus for their $N=34$ empirical two-neutron shell gaps for the first time by the key ingredients $^{58}$Ti and $^{59}$V. \par
\begin{figure}%[h]
\includegraphics[width=0.48\textwidth]{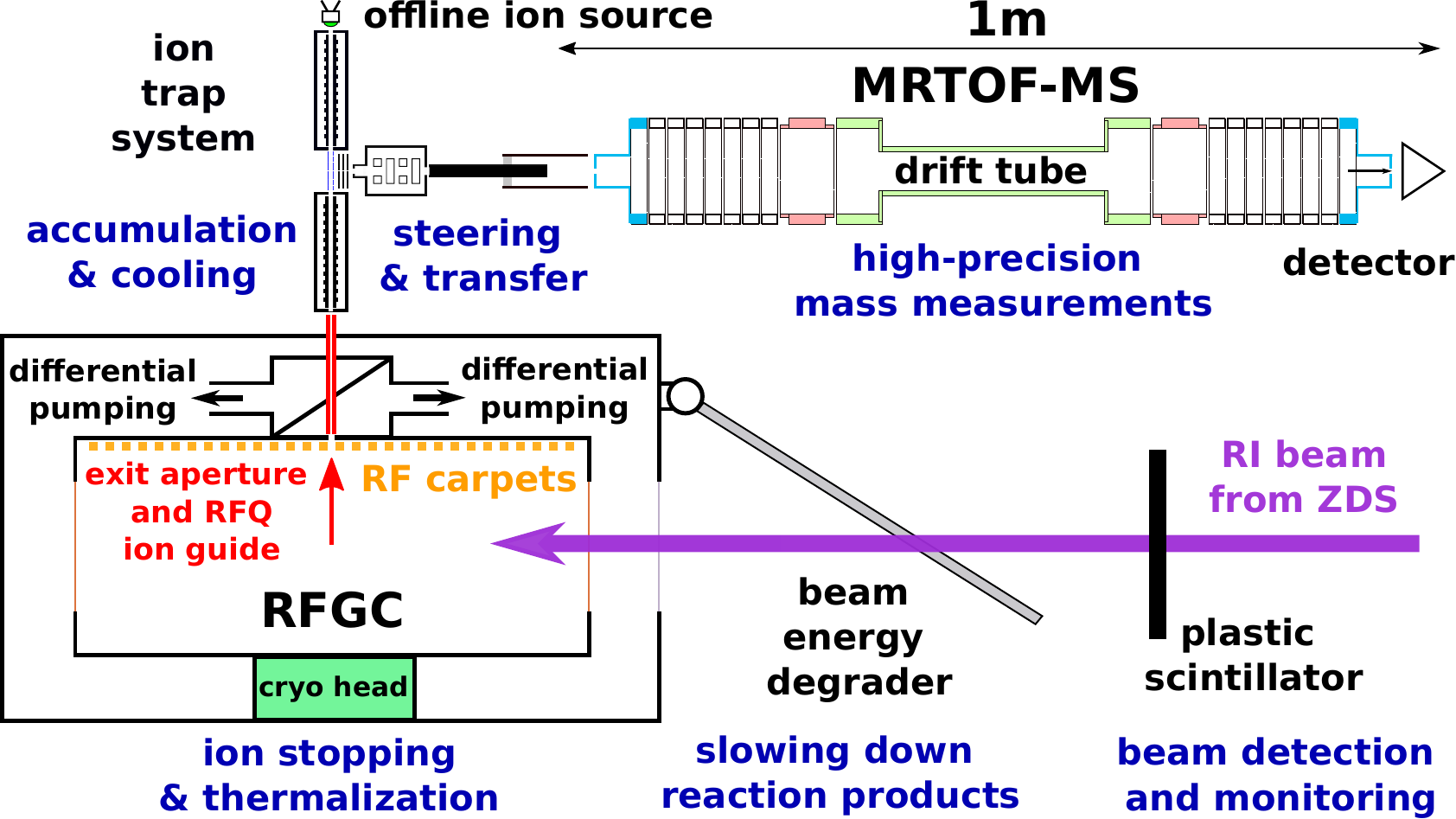}%
\caption{\label{fig:ZD-MRTOF_Setup} Sketch of the ZD MRTOF combined setup. Components from upstream to downstream: A plastic scintillator for beam monitoring, a rotatable energy degrader, the cryogenic RFGC, a radio-frequency quadrupole (RFQ)-based ion trap suite, and the MRTOF mass spectrograph.}
\end{figure}

The radioisotopes (RIs) were produced by projectile fragmentation of a $345\,\mathrm{MeV/nucleon}$ zinc beam from the RIKEN superconducting ring cyclotron (SRC) accelerator using a Be primary target of $2.03\,\mathrm{g/cm^2}$ thickness. The reaction products were selected by the BigRIPS separator \cite[]{Sakurai2010,Okuno2012} for upstream experiments, and the residues passed through the ZDS and reached the ZD MRTOF setup (shown in Fig.~\ref{fig:ZD-MRTOF_Setup}). The RI were slowed down using rotatable beam-energy degraders ($1-3\,\mathrm{mm}$ stainless steel \cite{Chen2021}), and subsequently stopped in a newly assembled cryogenic radio-frequency carpet type helium gas cell (RFGC) providing a stopping length of $50\,\mathrm{cm}$ in a $266\,\mathrm{mbar}$ pressure helium gas (room temperature equivalent) at $180\,\mathrm{K}$.\par 
The reaction products stopped in the He gas were extracted through the exit aperture of the gas cell using radio-frequency ion carpets \cite{WADA2003570,Takamine2005,ARAI201456,BOLLEN2011131}, and guided to a well-established ion-trap suite \cite{SCHURY201419,Ito2018} for accumulation, cooling, and preparation for the injection into the mass spectrograph. The ions have been reflected back and forth between the ion mirrors for $\approx 13\,\mathrm{ms}$ corresponding to about $600\,\mathrm{laps}$ with a maximum kinetic energy of $2.5\,\mathrm{keV}$ (in the central drift tube). During their multiple passes, the ion ensemble was purified from non-isobaric contaminant ions, which were extracted from the gas cell with orders of magnitude larger quantities than the ions of interest. To this end, a cleaning scheme using electrically pulsed mirror electrodes has been employed as described in \cite{Rosenbusch2022}. Ultimately, the ions were time focused onto a detector (ETP MagneToF) producing an impact signal whose time, relative to moment of ejection from the ion trap, was digitized using a multi-hit time-to-digital converter (MCS6A, Fast ComTech).\par
From the measured TOF, masses were calculated using the single reference method:
\begin{equation}
m_x=q_x\frac{m_r-q_r m_e}{q_r}\rho_t^2+q_x m_e\,,\quad \rho_t = \left(\frac{t_x-t_0}{t_r-t_0}\right)\,,
\end{equation}
where $m_x$ ($q_x$) and $m_r$ ($q_r$) are the masses (charges) of the ions of interest and the reference ions, respectively, and $m_e$ the electron mass. The TOF ratio $\rho_t$ linking the ion masses to each other is derived from the analyte and reference TOF $t_x$ and $t_r$, where $t_0$ is an offset time denoting the start time of the measurement. From test measuremnts in the same mass region the offset time was fixed to $t_0 = 150(10)\,\mathrm{ns}$, where uncertainties of $t_0$ introduce a systematic mass uncertainty (see \cite{Ito2013}). However, as isobars were used as references for all cases, the systematic mass uncertainties with $\delta m_{t0}^\mathrm{sys}/m < 10^{-9}$ become negligible and are not explicitly considered.\par
\begin{figure}[b]
\center
\includegraphics[width=1\linewidth]{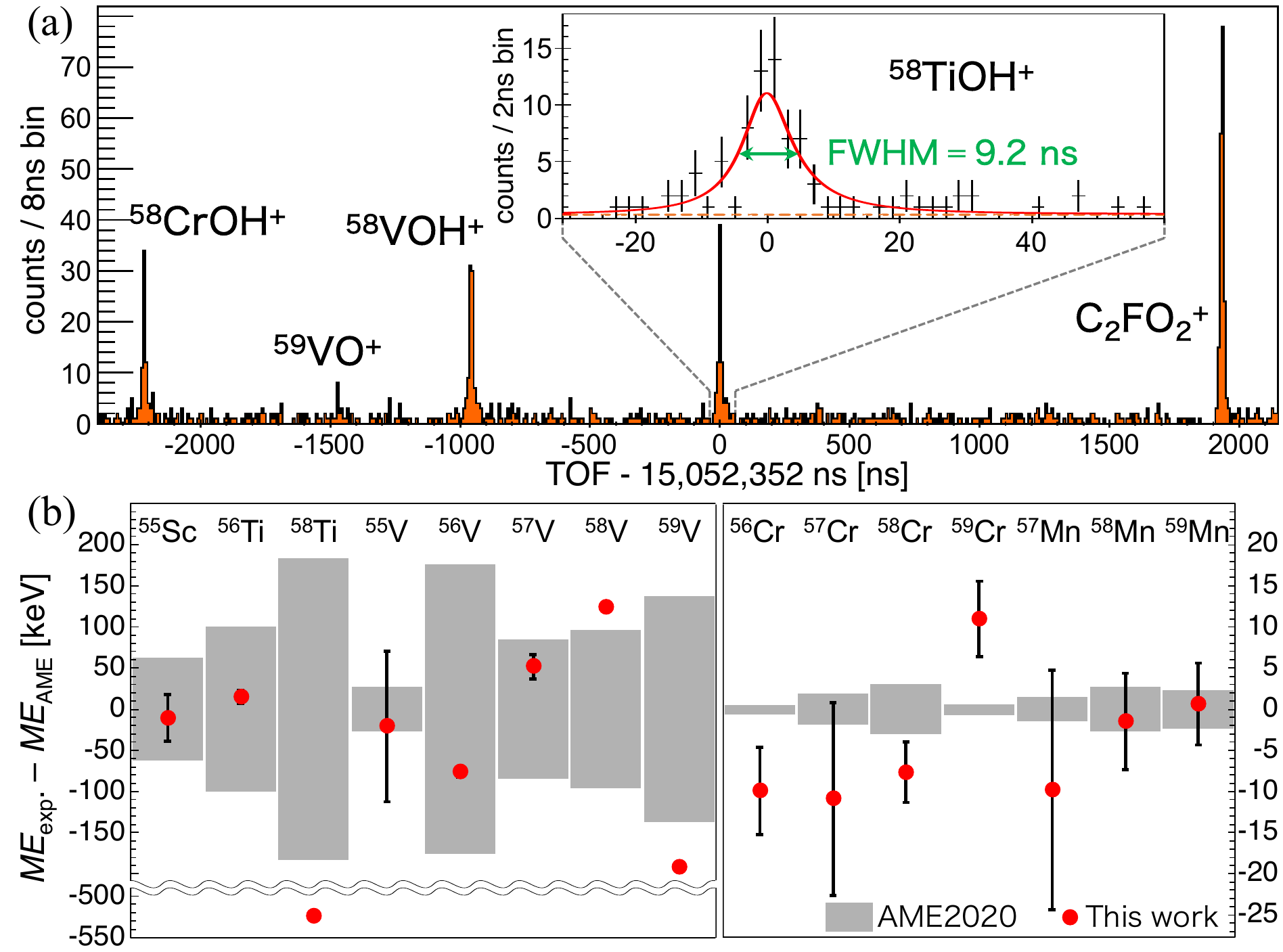}
\caption{(a) TOF spectrum including the most exotic species. The inset shows a magnified-view of the $^{58}$TiOH$^{+}$ peak together with the fitting function. The mass resolving power achieved for the ions of interest was $\approx\,820\,000$. (b) Mass-differences between our experimental values (red points) and the ones from AME2020 (grey bands).
}
\label{fig:spectrum}
\end{figure}

A software drift correction (see \textit{e.g.} \cite{Schury2017}) has been applied and the ion TOF signals were fitted using a Johnson's $S_U$ distribution \cite{Johnson1949a} as empirical fit function, which allows for additional shape parameters like skewness and kurtosis (see the inset of Fig.~\ref{fig:spectrum} (a)). An un-binned maximum log-likelihood method was used to perform TOF fits \cite{Verkerke2006,BRUN199781} employing simultaneous fitting for several analyte peaks and the mass reference.\par

The RIs were measured either as direct products from the incoming beam or as decay products with the beam components as precursors. In the case of $A=55$, $^{55}$Sc was the major beam component ($47\,\mathrm{\%}$, $850\,\mathrm{pps}$) while $^{55}$Ti, and $^{55}$V were produced from $\beta$-decay of $^{55}$Sc inside of the RFGC. The beam composition for $A=56$ and $A=57$ isotopes was $^{56}$Ti ($50\,\mathrm{\%}$, $4500\,\mathrm{pps}$) and $^{57}$V ($43\,\mathrm{\%}$, $3800\,\mathrm{cps}$), and for $A=58,59$ the major components were $^{58}$Ti ($22\,\mathrm{\%}$, $1900\,\mathrm{pps}$) and $^{59}$V ($72\,\mathrm{\%}$, $6200\,\mathrm{pps}$).\par
The ions of interest were extracted from the gas cell as atomic ions and as molecular compounds upon chemical reactions, depending on the conditions of the RFGC during the online commissioning tests. The presence of isobaric molecules in almost all spectra has been exploited for referencing, and for mass accuracy benchmarks if two or more well-known molecules were identified in the same spectrum. In the case of $^{55}$ScOH$^{+}$, no stable isobaric molecule was available, and $^{55}$TiOH$^{+}$ ions produced by $\beta$-decay have been used as a reference ($^{55}$Ti was recently measured at TITAN \cite{Leistenschneider2018}).\par

The experiment resulted in the measurement of 15 atomic masses, which are concluded in Table.~\ref{tab:result}. A TOF spectrum containing the key isotopes $^{58}$Ti and $^{59}$V is shown in Fig.~\ref{fig:spectrum} (a). Our experimentally determined mass values of $^{55}$V, $^{56-59}$Cr, and $^{57-59}$Mn are consistent with the previously adopted values in the AME2020 \cite{Wang_2021} (see Fig.~\ref{fig:spectrum} (b)). For the five isotopes $^{56,58}$Ti, and $^{56,58,59}$V we report an improvement of mass precision by more than an order of magnitude. The atomic mass of $^{55}$Sc published in \cite{Leistenschneider2021} was confirmed, and also $^{56,57}$V have been found to be in agreement to the previously known values. The new masses of $^{58,59}$V deviate from previous B$\rho$-TOF measurements \cite{Meisel2020,Michimasa2020} by more than one standard deviation $\sigma$, while particularly for $^{58}$Ti a sizable deviation of $2.5\,\mathrm{\sigma}$ from the recently reported values was found. Due to this deviation, the correct identification of $^{58}$TiOH$^{+}$ as an RI-molecule was additionally confirmed by time-of-flight spectra using a different beam with similar intensity but no content of $^{58}$Ti. Very recently, high-precision mass measurements of $^{56}$Ti and $^{56-58}$V were also published from TITAN \cite{Porter2022}, and are in agreement with our results. The high-precision mass measurements of $^{58}$Ti and $^{59}$V are the core achievement, and allow for the first complete study of the $N=34$ two-neutron shell gap for $Z=22,23$.\par

\begin{figure}%[h]
\includegraphics[width=0.48\textwidth]{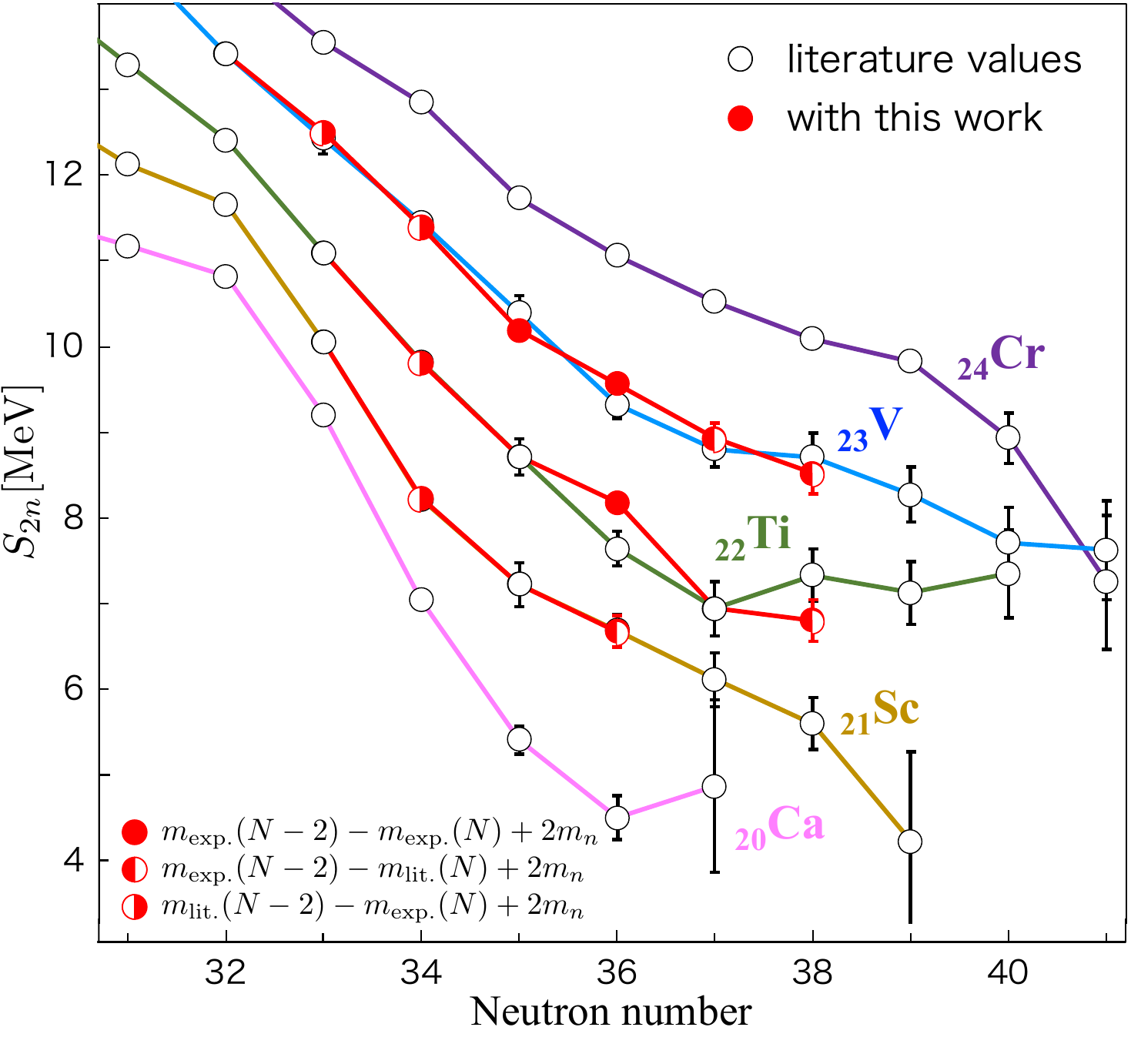}%
\caption{\label{fig.S2n} Two neutron separation energies of neutron-rich Ca, Sc, Ti, V, and Cr isotopes as a function of the neutron number. Black open circles are experimental values from AME2020 \cite{Wang_2021} including recent mass measurements \cite{Meisel2020,Michimasa2020}. The red filled circles are updated values from this work, where a split circle denotes the combination of the new data with values from AME2020 (see legend).}
\end{figure}

\begin{table}%[h]
  \caption{Results of the mass measurements: species of RI and the reference ions, number of measured events for the RI, TOF ratio, and the measured atomic mass excess of the RI. The chemical compounds refer to the most abundant stable isotope of each element.}
  %\caption{Results of the mass measurements: ion species of analyte and reference ions, TOF ratio for mass calibrations, measured atomic mass excess of each RIs, and the number of events measured for each isotope. The chemical components refer to the most abundant stable isotope of the element.}
  \label{tab:result}
  \resizebox{0.49\textwidth}{!}{%
    \begin{tabular}{ c c r @{} l l l }
      \hline \hline
      %Ion$_x$ & Ion$_r$ & $N_x$ /& $N_r$[cou&nts] & \multicolumn{1}{c}{$\rho_t$} & $M\!E_{\mathrm{exp.}}$[keV] \\
Ion$_x$ & Ion$_r$ & $N_x$[cou&nts] & \multicolumn{1}{c}{$\rho_t$} & $M\!E_{\mathrm{exp.}}$[keV] \\
      \hline
      $^{55}$ScOH$^{+}$ & $^{55}$TiOH$^{+}$	& 58&& 1.00008196(17)  & -30853(28)		\\
      \hline
      $^{56}$Ti$^{+}$  & N$_4^{+}$    	&  55&&0.999512461(71)	& -39408.2(7.4)  \\
      \hline
      $^{58}$TiOH$^{+}$ & C$_2$FO$_2^{+}$	   &  235&& 0.999871654(31)	& -31442.0(3.7) \\
      \hline
      $^{55}$VOH$^{+}$	& $^{55}$TiOH$^{+}$	&  4&& 0.99994550(68)	& -49146(92)	\\
      \hline
      $^{56}$V$^{+}$  & N$_4^{+}$	    &  342&& 0.999446769(59)	& -46259.6(6.2) \\
      \hline
      $^{57}$V$^{+}$  & ArOH$^{+}$	   &  95&& 0.99988791(14)	& -44383(15)	\\
      \hline
      $^{58}$V$^{+}$	& C$_2$H$_2$S$^{+}$	   & 105&& 0.999732737(82)	& \multirow{2}{*}{-40306.1(5.6)}  \\
      $^{58}$VOH$^{+}$	& C$_2$FO$_2^{+}$	   &  192&& 0.999808205(66)  &         \\
      \hline
      $^{59}$V$^{+}$	& $^{59}$Cr$^{+}$	&  270&& 1.00009378(10)	& \multirow{3}{*}{-37802.2(2.8)}  \\
      $^{59}$VO$^{+}$	& C$_2$FO$_2^{+}$	   &  109&& 0.999773864(47)  &  \\
      $^{59}$VOH$^{+}$	& CS$_2^{+}$	    &  244&& 1.000118625(24)  &        \\
      \hline
      $^{56}$Cr$^{+}$	& N$_4^{+}$	    &  222&& 0.999360129(51)	& -55295.0(5.3)  \\
      \hline
      $^{57}$Cr$^{+}$	& ArOH$^{+}$	   &  283&& 0.99981107(11)	& -52536(12)	 \\
      \hline
      $^{58}$Cr$^{+}$	& C$_2$H$_2$S$^{+}$	   &  192&& 0.999624418(45)	& \multirow{2}{*}{-51999.5(3.7)}  \\
      $^{58}$CrOH$^{+}$	& C$_2$FO$_2^{+}$	   & 131&& 0.999724521(55)  &  \\
      \hline
      $^{59}$CrOH$^{+}$	& CS$_2^{+}$	    &  99&& 1.000045786(33)	& -48105.0(4.6)  \\
      \hline
      $^{57}$Mn$^{+}$	& ArOH$^{+}$	   &  89&& 0.99976432(14)	& -57496(15)	\\
      \hline
      $^{58}$Mn$^{+}$	& C$_2$H$_2$S$^{+}$	   &  28&& 0.999588996(69)	& -55829.0(5.9)  \\
      \hline
      $^{59}$Mn$^{+}$	& $^{59}$Cr$^{+}$	&  509&& 0.999932535(69)	& -55524.7(5.0)  \\
      \hline \hline
    \end{tabular}
  }
\end{table}
We discuss the new insights into the structure of Ti, and V at $N=32,34$ using two-neutron separation energies $S_{2n}(N,Z) = m(N-2,Z) - m(N,Z) + 2m_n$, where $m(N,Z)$ is the atomic mass of a nucleus with $Z$ protons and $N$ neutrons, and $m_n$ the mass of a neutron. Figure \ref{fig.S2n} shows the $S_{2n}$ of the isotopic chains with $N = 30 - 41$ and $Z = 20 - 24$ including our new results. A pronounced steep decrease at neutron number 32 is visible for Ca, and Sc, but becomes weaker for Ti isotopes as previously reported. For Ti, a larger binding energy has been measured at $N=36$ and suggests an earlier onset of the deformation recently discovered toward $N=40$ \cite{Michimasa2020}. Comparing the previously reported data with the new results, the negative slope beyond $N=34$ decreases, which is similar to the recent findings in the Sc chain, but now confirmed for the even-proton number $Z=22$. It seems that the steep drop of $S_{2n}$ for $N>34$ is restricted to the Ca chain and possibly to systems with less protons (as observed for K isotopes at $N=32$ \cite{Rosenbusch2015}), and weakens for isotopes with $Z>20$. \par

\begin{figure}%[b]
\includegraphics[width=0.48\textwidth]{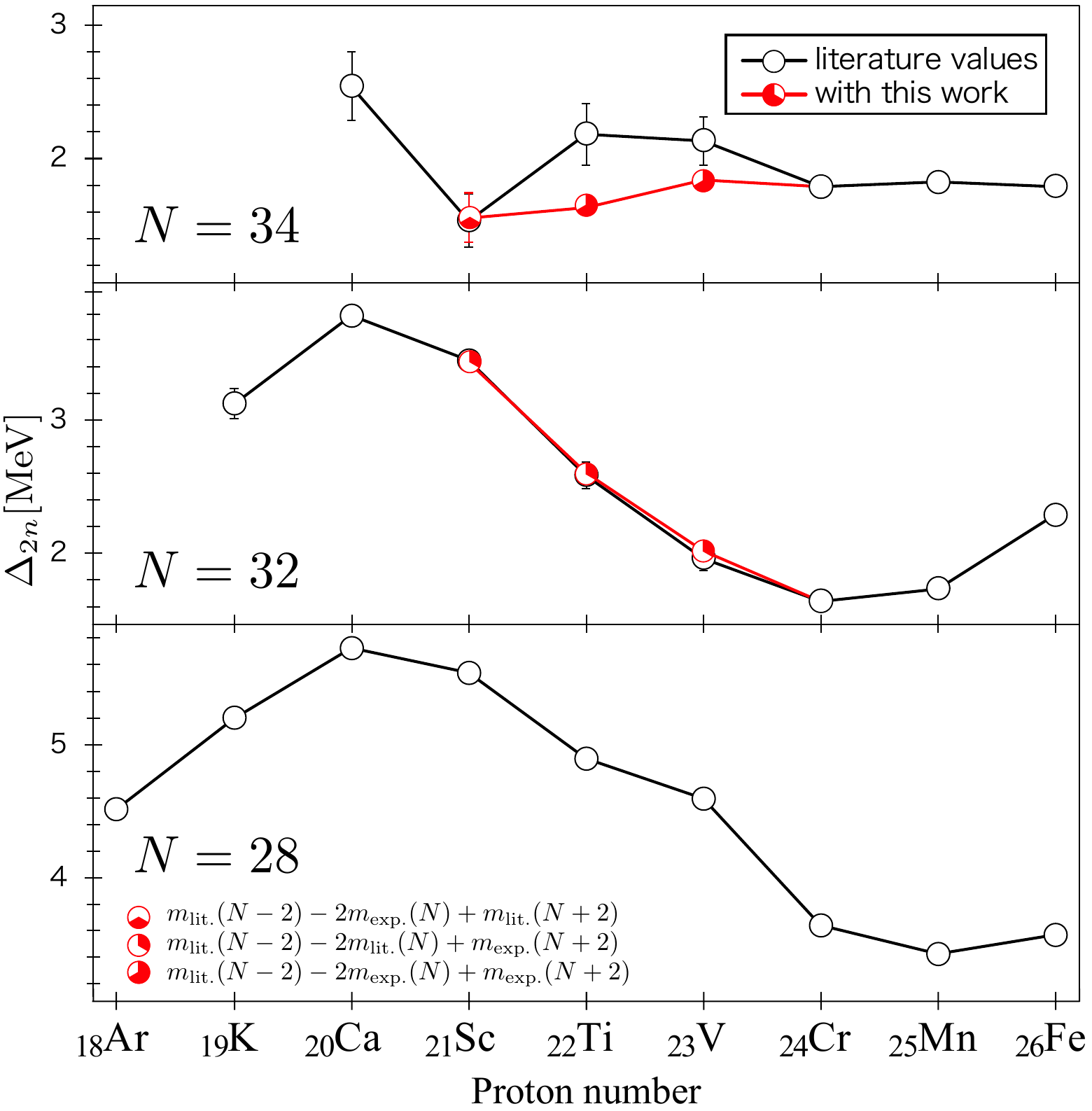}%
\caption{\label{fig.d2n_Z} Empirical shell gaps ($\Delta_{2n}$) for isotones with the canonical magic number of $N$ = 28, and the new magic numbers of $N$ = 32 and 34. Data are from the AME2020 with recent measurements as in Fig. \ref{fig.S2n} (black open circles), and our experimental values (red partly filled circles).}
\end{figure}
\begin{figure*}%[h]
\includegraphics[width=\textwidth]{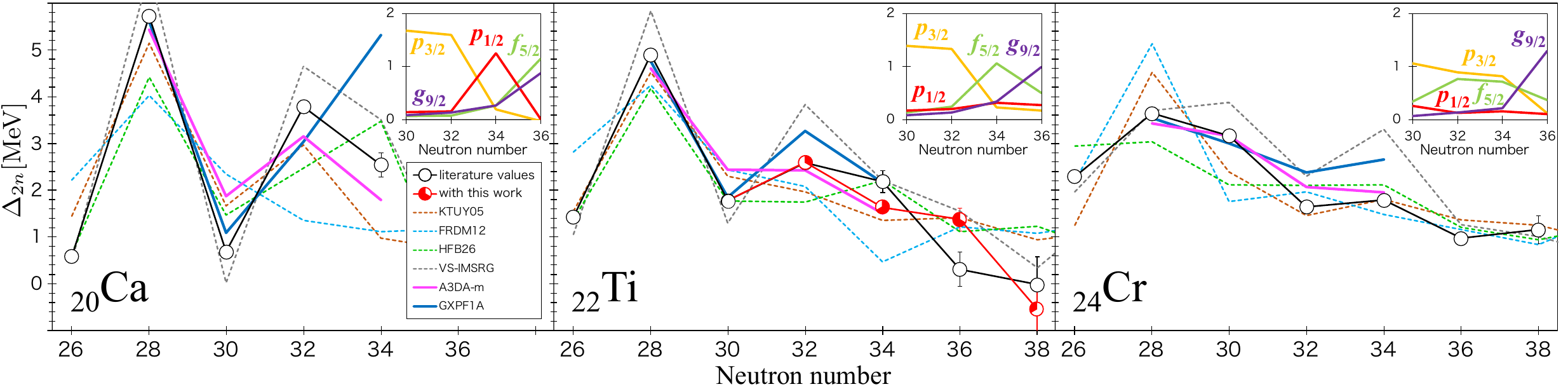}%
\caption{\label{fig.d2n_N} Empirical shell gaps ($\Delta_{2n}$) for Ca, Ti, and Cr isotopes as a function of neutron number. Data are from the AME2020 with recent measurements as in Fig. \ref{fig.S2n} (black open circles), and our experimental values as in Fig. \ref{fig.d2n_Z} (red partly filled circles). The lines colored with pink and blue show the theoretical values with A3DA-m and GXPF1A, respectively. The insets show the occupation number on each orbit for the last two neutrons, calculated with MCSM. Other chosen theoretical models are shown with dashed lines and given in the Legend. }
\end{figure*}
We further investigate the empirical neutron shell gaps defined by the differences of two-neutron separation energies $\Delta_{2n}(N,Z)=S_{2n}(Z, N)-S_{2n}(Z, N+2)$ as shown in Fig. \ref{fig.d2n_Z} as a function of proton number. For Ca, the height of the experimental $\Delta_{2n}$ peak at $N = 28$ measures close to $6\,\mathrm{MeV}$ while about $3.8\,\mathrm{MeV}$ are also observed at $N = 32$ \cite{Wienholtz2013} and still $2.6\,\mathrm{MeV}$ at $N = 34$ \cite{Michimasa2018}, which is due to the pronounced splitting between $\nu p _{3/2}$, $\nu p _{1/2}$, and $\nu f _{5/2}$ \cite{Otsuka2010,Otsuka2016}. For Sc isotopes, experimental $\Delta_{2n}$ are confirmed for $N = 32$, while for $N>32$ the present results have been combined with the existing data. For $N=32$, the hill of enhanced shell gaps (being maximum in Ca) gradually decreases when adding protons to the $\pi f _{7/2}$ shell, similar to---but weaker than---the effect in the $N=28$ isotones.\par
In turn, for the $N = 34$ isotones a new picture is obtained including our results. An increase of the shell gap from Sc to Ti was seen in the $\Delta_{2n}$ from AME2020 (black line), whereas a vanishing of this trend for Ti and V is observed when the new high-precision data is included. In contrast to $N=28$ and $32$, the pronounced $N=34$ gap turns out to be a unique feature for the Ca isotopes and shows no significant effect in the isotopes above. For V isotopes, the new studies reveal a similarly low shell gap at both $N=32$ and $N=34$.\par
We have performed advanced Monte Carlo shell model (MCSM) calculations \cite{Shimizu2012} for the even-$Z$ isotopic chains Ca, Ti, and Cr. During the last two decades, shell model approaches have made tremendous progress through the development of microscopic effective interactions employing the G-matrix equation \cite{Hjorth95}. In this work, the model space used in the calculations considers the full $pf$ shell, $g_{9/2}$ and $d_{5/2}$ for both protons and neutrons. Binding energies are calculated with respect to the doubly-magic $^{40}$Ca core. The nucleon-nucleon (NN) effective interactions in the model space are based on the A3DA Hamiltonian, which was developed using the GXPF1A \cite{Honma05}, JUN45 \cite{Honma09}, and G-matrix effective interactions. A further modified Hamiltonian including corrections for the $\nu g_{9/2}$ orbit adjusted to the Ni isotopic chain \cite{Tsunoda2014}, \textit{i.e.} A3DA-m, has been used in the present calculations. For comparison, we perform conventional shell model (SM) calculations employing the KSHELL code \cite{Shimizu2019} using the GXPF1A interaction, which excludes the $dg$ orbitals.\par
The new theoretical data of the empirical shell gaps from both MCSM (A3DA-m) and SM calculations (GXPF1A) are shown together with the experimental data in Fig.~\ref{fig.d2n_N} for the three calculated isotopes. The $\Delta_{2n}$ values measured by the present experiment are rather well reproduced by the MCSM calculations, where the limitation of orbits in GXPF1A leads to an overestimation of $\Delta_{2n}$ at $N=32$ for Ti and Cr, and at $N=34$ for all three isotopes. The insets in the figure show the calculated occupation numbers of the orbits in which the last two neutrons are located, resulting from the MCSM framework. The inclusion of orbits above the $pf$ shell leads to an increasing occupation of the $g_{9/2}$ orbit from Ca to Cr at $N=36$, which for the latter two isotopes dominates that of the $f_{5/2}$ orbit. This behavior produces an enhanced binding energy and lowers the calculated shell gap at $N=34$ for $Z>20$. For comparison with the new MSCM calculations in Fig.~\ref{fig.d2n_N}, we have selected mass models employing other theoretical techniques: the macroscopic-microscopic model FRDM12 \cite{Moeller2016}, the self-consistent mean-field model HFB26 \cite{Goriely2016}, the phenomenological mass model KTUY05 \cite{Koura2005}, and the recent ab-initio model VS-IMSRG \cite{Stroberg2021}.
\par
In conclusion, the masses of 15 neutron-rich nuclei have been measured with high precision and accuracy by the multi-reflection time-of-flight technique using the new ZD MRTOF-MS. Among the results, the mass precisions of $^{55}$Sc, $^{56}$Ti, $^{58}$Ti, $^{56}$V, $^{57}$V, $^{58}$V, and $^{59}$V have been significantly improved to the order of $10\,\mathrm{keV}$, or below. For $^{55}$Sc the recently measured value from TITAN (TRIUMF) \cite{Leistenschneider2021} has been confirmed. The newly determined masses of $^{58}$Ti and $^{59}$V were found to have significant deviations from previously measured values, where especially for $^{58}$Ti an increased binding energy has been measured. The newly determined two-neutron separation energies suggest that the $N=34$ shell effect in Ca isotopes is an exclusive feature of this chain, and does not reappear at or beyond Sc, as suggested by the previous data, and also as expected by the theoretical picture \cite{Otsuka2016}. New Monte Carlo shell model calculations using the A3DA-m Hamiltonian have been performed and reproduce the experimental findings, while conventional GXPF1A calculations produce an over-estimation of the shell gap. This result emphasizes the importance to include $dg$ orbits into the nuclear model space to explain the experimental findings.\par
We express our gratitude to the RIKEN Nishina Center for Accelerator-based Science, the Center for Nuclear Study at the University of Tokyo, and the HiCARI collaboration for their support of the online measurements. This work was supported by the Japan Society for the Promotion of Science KAKENHI (Grants No. 2200823, No. 24224008, No. 24740142, No. 15H02096, No. 15K05116, No. 17H01081, No. 17H06090, No. 18K13573, No. 18H05462, No. 19H00679, No. 19K14750, No. 20H05648, No. 21K13951, No. 22H01257, and No. 22H04946), RIKEN Junior Research Associate Program, the RIKEN programme for Evolution of Matter in the Universe (r-EMU), the UK STFC award ST/P003885/1, and the Royal Society. The MCSM and conventional shell-model calculations were performed on the supercomputer Fugaku at RIKEN AICS (hp210165, hp220174). This work was supported in part by MEXT as “Program for Promoting Researches on the Supercomputer Fugaku” (Simulation for basic science: from fundamental laws of particles to creation of nuclei) and by JICFuS.

\bibliographystyle{apsrev4-1} % Tell bibtex which bibliography style to use
%\bibliography{Ti_paper}% Produces the bibliography via BibTeX.

\begin{thebibliography}{58}%
\makeatletter
\providecommand \@ifxundefined [1]{%
 \@ifx{#1\undefined}
}%
\providecommand \@ifnum [1]{%
 \ifnum #1\expandafter \@firstoftwo
 \else \expandafter \@secondoftwo
 \fi
}%
\providecommand \@ifx [1]{%
 \ifx #1\expandafter \@firstoftwo
 \else \expandafter \@secondoftwo
 \fi
}%
\providecommand \natexlab [1]{#1}%
\providecommand \enquote  [1]{``#1''}%
\providecommand \bibnamefont  [1]{#1}%
\providecommand \bibfnamefont [1]{#1}%
\providecommand \citenamefont [1]{#1}%
\providecommand \href@noop [0]{\@secondoftwo}%
\providecommand \href [0]{\begingroup \@sanitize@url \@href}%
\providecommand \@href[1]{\@@startlink{#1}\@@href}%
\providecommand \@@href[1]{\endgroup#1\@@endlink}%
\providecommand \@sanitize@url [0]{\catcode `\\12\catcode `\$12\catcode
  `\&12\catcode `\#12\catcode `\^12\catcode `\_12\catcode `\%12\relax}%
\providecommand \@@startlink[1]{}%
\providecommand \@@endlink[0]{}%
\providecommand \url  [0]{\begingroup\@sanitize@url \@url }%
\providecommand \@url [1]{\endgroup\@href {#1}{\urlprefix }}%
\providecommand \urlprefix  [0]{URL }%
\providecommand \Eprint [0]{\href }%
\providecommand \doibase [0]{http://dx.doi.org/}%
\providecommand \selectlanguage [0]{\@gobble}%
\providecommand \bibinfo  [0]{\@secondoftwo}%
\providecommand \bibfield  [0]{\@secondoftwo}%
\providecommand \translation [1]{[#1]}%
\providecommand \BibitemOpen [0]{}%
\providecommand \bibitemStop [0]{}%
\providecommand \bibitemNoStop [0]{.\EOS\space}%
\providecommand \EOS [0]{\spacefactor3000\relax}%
\providecommand \BibitemShut  [1]{\csname bibitem#1\endcsname}%
\let\auto@bib@innerbib\@empty
%</preamble>
\bibitem [{\citenamefont {Reiter}\ \emph {et~al.}(2018)\citenamefont {Reiter},
  \citenamefont {{Ayet San Andr{\'{e}}s}}, \citenamefont {Dunling},
  \citenamefont {Kootte}, \citenamefont {Leistenschneider}, \citenamefont
  {Andreoiu}, \citenamefont {Babcock}, \citenamefont {Barquest}, \citenamefont
  {Bollig}, \citenamefont {Brunner}, \citenamefont {Dillmann}, \citenamefont
  {Finlay}, \citenamefont {Gwinner}, \citenamefont {Graham}, \citenamefont
  {Holt}, \citenamefont {Hornung}, \citenamefont {Jesch}, \citenamefont
  {Klawitter}, \citenamefont {Lan}, \citenamefont {Lascar}, \citenamefont
  {McKay}, \citenamefont {Paul}, \citenamefont {Steinbr{\"{u}}gge},
  \citenamefont {Thompson}, \citenamefont {Tracy}, \citenamefont {Wieser},
  \citenamefont {Will}, \citenamefont {Dickel}, \citenamefont {Pla{\ss}},
  \citenamefont {Scheidenberger}, \citenamefont {Kwiatkowski},\ and\
  \citenamefont {Dilling}}]{Reiter2018}%
  \BibitemOpen
  \bibfield  {author} {\bibinfo {author} {\bibfnamefont {M.~P.}\ \bibnamefont
  {Reiter}}, \bibinfo {author} {\bibfnamefont {S.}~\bibnamefont {{Ayet San
  Andr{\'{e}}s}}}, \bibinfo {author} {\bibfnamefont {E.}~\bibnamefont
  {Dunling}}, \bibinfo {author} {\bibfnamefont {B.}~\bibnamefont {Kootte}},
  \bibinfo {author} {\bibfnamefont {E.}~\bibnamefont {Leistenschneider}},
  \bibinfo {author} {\bibfnamefont {C.}~\bibnamefont {Andreoiu}}, \bibinfo
  {author} {\bibfnamefont {C.}~\bibnamefont {Babcock}}, \bibinfo {author}
  {\bibfnamefont {B.~R.}\ \bibnamefont {Barquest}}, \bibinfo {author}
  {\bibfnamefont {J.}~\bibnamefont {Bollig}}, \bibinfo {author} {\bibfnamefont
  {T.}~\bibnamefont {Brunner}}, \bibinfo {author} {\bibfnamefont
  {I.}~\bibnamefont {Dillmann}}, \bibinfo {author} {\bibfnamefont
  {A.}~\bibnamefont {Finlay}}, \bibinfo {author} {\bibfnamefont
  {G.}~\bibnamefont {Gwinner}}, \bibinfo {author} {\bibfnamefont
  {L.}~\bibnamefont {Graham}}, \bibinfo {author} {\bibfnamefont {J.~D.}\
  \bibnamefont {Holt}}, \bibinfo {author} {\bibfnamefont {C.}~\bibnamefont
  {Hornung}}, \bibinfo {author} {\bibfnamefont {C.}~\bibnamefont {Jesch}},
  \bibinfo {author} {\bibfnamefont {R.}~\bibnamefont {Klawitter}}, \bibinfo
  {author} {\bibfnamefont {Y.}~\bibnamefont {Lan}}, \bibinfo {author}
  {\bibfnamefont {D.}~\bibnamefont {Lascar}}, \bibinfo {author} {\bibfnamefont
  {J.~E.}\ \bibnamefont {McKay}}, \bibinfo {author} {\bibfnamefont {S.~F.}\
  \bibnamefont {Paul}}, \bibinfo {author} {\bibfnamefont {R.}~\bibnamefont
  {Steinbr{\"{u}}gge}}, \bibinfo {author} {\bibfnamefont {R.}~\bibnamefont
  {Thompson}}, \bibinfo {author} {\bibfnamefont {J.~L.}\ \bibnamefont {Tracy}},
  \bibinfo {author} {\bibfnamefont {M.~E.}\ \bibnamefont {Wieser}}, \bibinfo
  {author} {\bibfnamefont {C.}~\bibnamefont {Will}}, \bibinfo {author}
  {\bibfnamefont {T.}~\bibnamefont {Dickel}}, \bibinfo {author} {\bibfnamefont
  {W.~R.}\ \bibnamefont {Pla{\ss}}}, \bibinfo {author} {\bibfnamefont
  {C.}~\bibnamefont {Scheidenberger}}, \bibinfo {author} {\bibfnamefont
  {A.~A.}\ \bibnamefont {Kwiatkowski}}, \ and\ \bibinfo {author} {\bibfnamefont
  {J.}~\bibnamefont {Dilling}},\ }\href {\doibase 10.1103/PhysRevC.98.024310}
  {\bibfield  {journal} {\bibinfo  {journal} {Phys. Rev. C}\ }\textbf {\bibinfo
  {volume} {98}},\ \bibinfo {pages} {024310} (\bibinfo {year}
  {2018})}\BibitemShut {NoStop}%
\bibitem [{\citenamefont {Leistenschneider}\ \emph {et~al.}(2018)\citenamefont
  {Leistenschneider}, \citenamefont {Reiter}, \citenamefont {Ayet
  San~Andr\'es}, \citenamefont {Kootte}, \citenamefont {Holt}, \citenamefont
  {Navr\'atil}, \citenamefont {Babcock}, \citenamefont {Barbieri},
  \citenamefont {Barquest}, \citenamefont {Bergmann}, \citenamefont {Bollig},
  \citenamefont {Brunner}, \citenamefont {Dunling}, \citenamefont {Finlay},
  \citenamefont {Geissel}, \citenamefont {Graham}, \citenamefont {Greiner},
  \citenamefont {Hergert}, \citenamefont {Hornung}, \citenamefont {Jesch},
  \citenamefont {Klawitter}, \citenamefont {Lan}, \citenamefont {Lascar},
  \citenamefont {Leach}, \citenamefont {Lippert}, \citenamefont {McKay},
  \citenamefont {Paul}, \citenamefont {Schwenk}, \citenamefont {Short},
  \citenamefont {Simonis}, \citenamefont {Som\`a}, \citenamefont
  {Steinbr\"ugge}, \citenamefont {Stroberg}, \citenamefont {Thompson},
  \citenamefont {Wieser}, \citenamefont {Will}, \citenamefont {Yavor},
  \citenamefont {Andreoiu}, \citenamefont {Dickel}, \citenamefont {Dillmann},
  \citenamefont {Gwinner}, \citenamefont {Pla\ss{}}, \citenamefont
  {Scheidenberger}, \citenamefont {Kwiatkowski},\ and\ \citenamefont
  {Dilling}}]{Leistenschneider2018}%
  \BibitemOpen
  \bibfield  {author} {\bibinfo {author} {\bibfnamefont {E.}~\bibnamefont
  {Leistenschneider}}, \bibinfo {author} {\bibfnamefont {M.~P.}\ \bibnamefont
  {Reiter}}, \bibinfo {author} {\bibfnamefont {S.}~\bibnamefont {Ayet
  San~Andr\'es}}, \bibinfo {author} {\bibfnamefont {B.}~\bibnamefont {Kootte}},
  \bibinfo {author} {\bibfnamefont {J.~D.}\ \bibnamefont {Holt}}, \bibinfo
  {author} {\bibfnamefont {P.}~\bibnamefont {Navr\'atil}}, \bibinfo {author}
  {\bibfnamefont {C.}~\bibnamefont {Babcock}}, \bibinfo {author} {\bibfnamefont
  {C.}~\bibnamefont {Barbieri}}, \bibinfo {author} {\bibfnamefont {B.~R.}\
  \bibnamefont {Barquest}}, \bibinfo {author} {\bibfnamefont {J.}~\bibnamefont
  {Bergmann}}, \bibinfo {author} {\bibfnamefont {J.}~\bibnamefont {Bollig}},
  \bibinfo {author} {\bibfnamefont {T.}~\bibnamefont {Brunner}}, \bibinfo
  {author} {\bibfnamefont {E.}~\bibnamefont {Dunling}}, \bibinfo {author}
  {\bibfnamefont {A.}~\bibnamefont {Finlay}}, \bibinfo {author} {\bibfnamefont
  {H.}~\bibnamefont {Geissel}}, \bibinfo {author} {\bibfnamefont
  {L.}~\bibnamefont {Graham}}, \bibinfo {author} {\bibfnamefont
  {F.}~\bibnamefont {Greiner}}, \bibinfo {author} {\bibfnamefont
  {H.}~\bibnamefont {Hergert}}, \bibinfo {author} {\bibfnamefont
  {C.}~\bibnamefont {Hornung}}, \bibinfo {author} {\bibfnamefont
  {C.}~\bibnamefont {Jesch}}, \bibinfo {author} {\bibfnamefont
  {R.}~\bibnamefont {Klawitter}}, \bibinfo {author} {\bibfnamefont
  {Y.}~\bibnamefont {Lan}}, \bibinfo {author} {\bibfnamefont {D.}~\bibnamefont
  {Lascar}}, \bibinfo {author} {\bibfnamefont {K.~G.}\ \bibnamefont {Leach}},
  \bibinfo {author} {\bibfnamefont {W.}~\bibnamefont {Lippert}}, \bibinfo
  {author} {\bibfnamefont {J.~E.}\ \bibnamefont {McKay}}, \bibinfo {author}
  {\bibfnamefont {S.~F.}\ \bibnamefont {Paul}}, \bibinfo {author}
  {\bibfnamefont {A.}~\bibnamefont {Schwenk}}, \bibinfo {author} {\bibfnamefont
  {D.}~\bibnamefont {Short}}, \bibinfo {author} {\bibfnamefont
  {J.}~\bibnamefont {Simonis}}, \bibinfo {author} {\bibfnamefont
  {V.}~\bibnamefont {Som\`a}}, \bibinfo {author} {\bibfnamefont
  {R.}~\bibnamefont {Steinbr\"ugge}}, \bibinfo {author} {\bibfnamefont {S.~R.}\
  \bibnamefont {Stroberg}}, \bibinfo {author} {\bibfnamefont {R.}~\bibnamefont
  {Thompson}}, \bibinfo {author} {\bibfnamefont {M.~E.}\ \bibnamefont
  {Wieser}}, \bibinfo {author} {\bibfnamefont {C.}~\bibnamefont {Will}},
  \bibinfo {author} {\bibfnamefont {M.}~\bibnamefont {Yavor}}, \bibinfo
  {author} {\bibfnamefont {C.}~\bibnamefont {Andreoiu}}, \bibinfo {author}
  {\bibfnamefont {T.}~\bibnamefont {Dickel}}, \bibinfo {author} {\bibfnamefont
  {I.}~\bibnamefont {Dillmann}}, \bibinfo {author} {\bibfnamefont
  {G.}~\bibnamefont {Gwinner}}, \bibinfo {author} {\bibfnamefont {W.~R.}\
  \bibnamefont {Pla\ss{}}}, \bibinfo {author} {\bibfnamefont {C.}~\bibnamefont
  {Scheidenberger}}, \bibinfo {author} {\bibfnamefont {A.~A.}\ \bibnamefont
  {Kwiatkowski}}, \ and\ \bibinfo {author} {\bibfnamefont {J.}~\bibnamefont
  {Dilling}},\ }\href {\doibase 10.1103/PhysRevLett.120.062503} {\bibfield
  {journal} {\bibinfo  {journal} {Phys. Rev. Lett.}\ }\textbf {\bibinfo
  {volume} {120}},\ \bibinfo {pages} {062503} (\bibinfo {year}
  {2018})}\BibitemShut {NoStop}%
\bibitem [{\citenamefont {Leistenschneider}\ \emph {et~al.}(2021)\citenamefont
  {Leistenschneider}, \citenamefont {Dunling}, \citenamefont {Bollen},
  \citenamefont {Brown}, \citenamefont {Dilling}, \citenamefont {Hamaker},
  \citenamefont {Holt}, \citenamefont {Jacobs}, \citenamefont {Kwiatkowski},
  \citenamefont {Miyagi}, \citenamefont {Porter}, \citenamefont {Puentes},
  \citenamefont {Redshaw}, \citenamefont {Reiter}, \citenamefont {Ringle},
  \citenamefont {Sandler}, \citenamefont {Sumithrarachchi}, \citenamefont
  {Valverde},\ and\ \citenamefont {Yandow}}]{Leistenschneider2021}%
  \BibitemOpen
  \bibfield  {author} {\bibinfo {author} {\bibfnamefont {E.}~\bibnamefont
  {Leistenschneider}}, \bibinfo {author} {\bibfnamefont {E.}~\bibnamefont
  {Dunling}}, \bibinfo {author} {\bibfnamefont {G.}~\bibnamefont {Bollen}},
  \bibinfo {author} {\bibfnamefont {B.~A.}\ \bibnamefont {Brown}}, \bibinfo
  {author} {\bibfnamefont {J.}~\bibnamefont {Dilling}}, \bibinfo {author}
  {\bibfnamefont {A.}~\bibnamefont {Hamaker}}, \bibinfo {author} {\bibfnamefont
  {J.~D.}\ \bibnamefont {Holt}}, \bibinfo {author} {\bibfnamefont
  {A.}~\bibnamefont {Jacobs}}, \bibinfo {author} {\bibfnamefont {A.~A.}\
  \bibnamefont {Kwiatkowski}}, \bibinfo {author} {\bibfnamefont
  {T.}~\bibnamefont {Miyagi}}, \bibinfo {author} {\bibfnamefont {W.~S.}\
  \bibnamefont {Porter}}, \bibinfo {author} {\bibfnamefont {D.}~\bibnamefont
  {Puentes}}, \bibinfo {author} {\bibfnamefont {M.}~\bibnamefont {Redshaw}},
  \bibinfo {author} {\bibfnamefont {M.~P.}\ \bibnamefont {Reiter}}, \bibinfo
  {author} {\bibfnamefont {R.}~\bibnamefont {Ringle}}, \bibinfo {author}
  {\bibfnamefont {R.}~\bibnamefont {Sandler}}, \bibinfo {author} {\bibfnamefont
  {C.~S.}\ \bibnamefont {Sumithrarachchi}}, \bibinfo {author} {\bibfnamefont
  {A.~A.}\ \bibnamefont {Valverde}}, \ and\ \bibinfo {author} {\bibfnamefont
  {I.~T.}\ \bibnamefont {Yandow}},\ }\href {\doibase
  10.1103/PhysRevLett.126.042501} {\bibfield  {journal} {\bibinfo  {journal}
  {Phys. Rev. Lett.}\ }\textbf {\bibinfo {volume} {126}},\ \bibinfo {pages}
  {042501} (\bibinfo {year} {2021})}\BibitemShut {NoStop}%
\bibitem [{\citenamefont {Mougeot}\ \emph {et~al.}(2018)\citenamefont
  {Mougeot}, \citenamefont {Atanasov}, \citenamefont {Blaum}, \citenamefont
  {Chrysalidis}, \citenamefont {Goodacre}, \citenamefont {Fedorov},
  \citenamefont {Fedosseev}, \citenamefont {George}, \citenamefont {Herfurth},
  \citenamefont {Holt}, \citenamefont {Lunney}, \citenamefont {Manea},
  \citenamefont {Marsh}, \citenamefont {Neidherr}, \citenamefont {Rosenbusch},
  \citenamefont {Rothe}, \citenamefont {Schweikhard}, \citenamefont {Schwenk},
  \citenamefont {Seiffert}, \citenamefont {Simonis}, \citenamefont {Stroberg},
  \citenamefont {Welker}, \citenamefont {Wienholtz}, \citenamefont {Wolf},\
  and\ \citenamefont {Zuber}}]{Mougeot2018}%
  \BibitemOpen
  \bibfield  {author} {\bibinfo {author} {\bibfnamefont {M.}~\bibnamefont
  {Mougeot}}, \bibinfo {author} {\bibfnamefont {D.}~\bibnamefont {Atanasov}},
  \bibinfo {author} {\bibfnamefont {K.}~\bibnamefont {Blaum}}, \bibinfo
  {author} {\bibfnamefont {K.}~\bibnamefont {Chrysalidis}}, \bibinfo {author}
  {\bibfnamefont {T.~D.}\ \bibnamefont {Goodacre}}, \bibinfo {author}
  {\bibfnamefont {D.}~\bibnamefont {Fedorov}}, \bibinfo {author} {\bibfnamefont
  {V.}~\bibnamefont {Fedosseev}}, \bibinfo {author} {\bibfnamefont
  {S.}~\bibnamefont {George}}, \bibinfo {author} {\bibfnamefont
  {F.}~\bibnamefont {Herfurth}}, \bibinfo {author} {\bibfnamefont {J.~D.}\
  \bibnamefont {Holt}}, \bibinfo {author} {\bibfnamefont {D.}~\bibnamefont
  {Lunney}}, \bibinfo {author} {\bibfnamefont {V.}~\bibnamefont {Manea}},
  \bibinfo {author} {\bibfnamefont {B.}~\bibnamefont {Marsh}}, \bibinfo
  {author} {\bibfnamefont {D.}~\bibnamefont {Neidherr}}, \bibinfo {author}
  {\bibfnamefont {M.}~\bibnamefont {Rosenbusch}}, \bibinfo {author}
  {\bibfnamefont {S.}~\bibnamefont {Rothe}}, \bibinfo {author} {\bibfnamefont
  {L.}~\bibnamefont {Schweikhard}}, \bibinfo {author} {\bibfnamefont
  {A.}~\bibnamefont {Schwenk}}, \bibinfo {author} {\bibfnamefont
  {C.}~\bibnamefont {Seiffert}}, \bibinfo {author} {\bibfnamefont
  {J.}~\bibnamefont {Simonis}}, \bibinfo {author} {\bibfnamefont {S.~R.}\
  \bibnamefont {Stroberg}}, \bibinfo {author} {\bibfnamefont {A.}~\bibnamefont
  {Welker}}, \bibinfo {author} {\bibfnamefont {F.}~\bibnamefont {Wienholtz}},
  \bibinfo {author} {\bibfnamefont {R.~N.}\ \bibnamefont {Wolf}}, \ and\
  \bibinfo {author} {\bibfnamefont {K.}~\bibnamefont {Zuber}},\ }\href
  {\doibase 10.1103/PhysRevLett.120.232501} {\bibfield  {journal} {\bibinfo
  {journal} {Phys. Rev. Lett.}\ }\textbf {\bibinfo {volume} {120}},\ \bibinfo
  {pages} {232501} (\bibinfo {year} {2018})}\BibitemShut {NoStop}%
\bibitem [{\citenamefont {Porter}\ \emph {et~al.}(2022)\citenamefont {Porter},
  \citenamefont {Dunling}, \citenamefont {Leistenschneider}, \citenamefont
  {Bergmann}, \citenamefont {Bollen}, \citenamefont {Dickel}, \citenamefont
  {Dietrich}, \citenamefont {Hamaker}, \citenamefont {Hockenbery},
  \citenamefont {Izzo}, \citenamefont {Jacobs}, \citenamefont {Javaji},
  \citenamefont {Kootte}, \citenamefont {Lan}, \citenamefont {Miskun},
  \citenamefont {Mukul}, \citenamefont {Murb\"ock}, \citenamefont {Paul},
  \citenamefont {Pla\ss{}}, \citenamefont {Puentes}, \citenamefont {Redshaw},
  \citenamefont {Reiter}, \citenamefont {Ringle}, \citenamefont {Ringuette},
  \citenamefont {Sandler}, \citenamefont {Scheidenberger}, \citenamefont
  {Silwal}, \citenamefont {Simpson}, \citenamefont {Sumithrarachchi},
  \citenamefont {Teigelh\"ofer}, \citenamefont {Valverde}, \citenamefont
  {Weil}, \citenamefont {Yandow}, \citenamefont {Dilling},\ and\ \citenamefont
  {Kwiatkowski}}]{Porter2022}%
  \BibitemOpen
  \bibfield  {author} {\bibinfo {author} {\bibfnamefont {W.~S.}\ \bibnamefont
  {Porter}}, \bibinfo {author} {\bibfnamefont {E.}~\bibnamefont {Dunling}},
  \bibinfo {author} {\bibfnamefont {E.}~\bibnamefont {Leistenschneider}},
  \bibinfo {author} {\bibfnamefont {J.}~\bibnamefont {Bergmann}}, \bibinfo
  {author} {\bibfnamefont {G.}~\bibnamefont {Bollen}}, \bibinfo {author}
  {\bibfnamefont {T.}~\bibnamefont {Dickel}}, \bibinfo {author} {\bibfnamefont
  {K.~A.}\ \bibnamefont {Dietrich}}, \bibinfo {author} {\bibfnamefont
  {A.}~\bibnamefont {Hamaker}}, \bibinfo {author} {\bibfnamefont
  {Z.}~\bibnamefont {Hockenbery}}, \bibinfo {author} {\bibfnamefont
  {C.}~\bibnamefont {Izzo}}, \bibinfo {author} {\bibfnamefont {A.}~\bibnamefont
  {Jacobs}}, \bibinfo {author} {\bibfnamefont {A.}~\bibnamefont {Javaji}},
  \bibinfo {author} {\bibfnamefont {B.}~\bibnamefont {Kootte}}, \bibinfo
  {author} {\bibfnamefont {Y.}~\bibnamefont {Lan}}, \bibinfo {author}
  {\bibfnamefont {I.}~\bibnamefont {Miskun}}, \bibinfo {author} {\bibfnamefont
  {I.}~\bibnamefont {Mukul}}, \bibinfo {author} {\bibfnamefont
  {T.}~\bibnamefont {Murb\"ock}}, \bibinfo {author} {\bibfnamefont {S.~F.}\
  \bibnamefont {Paul}}, \bibinfo {author} {\bibfnamefont {W.~R.}\ \bibnamefont
  {Pla\ss{}}}, \bibinfo {author} {\bibfnamefont {D.}~\bibnamefont {Puentes}},
  \bibinfo {author} {\bibfnamefont {M.}~\bibnamefont {Redshaw}}, \bibinfo
  {author} {\bibfnamefont {M.~P.}\ \bibnamefont {Reiter}}, \bibinfo {author}
  {\bibfnamefont {R.}~\bibnamefont {Ringle}}, \bibinfo {author} {\bibfnamefont
  {J.}~\bibnamefont {Ringuette}}, \bibinfo {author} {\bibfnamefont
  {R.}~\bibnamefont {Sandler}}, \bibinfo {author} {\bibfnamefont
  {C.}~\bibnamefont {Scheidenberger}}, \bibinfo {author} {\bibfnamefont
  {R.}~\bibnamefont {Silwal}}, \bibinfo {author} {\bibfnamefont
  {R.}~\bibnamefont {Simpson}}, \bibinfo {author} {\bibfnamefont {C.~S.}\
  \bibnamefont {Sumithrarachchi}}, \bibinfo {author} {\bibfnamefont
  {A.}~\bibnamefont {Teigelh\"ofer}}, \bibinfo {author} {\bibfnamefont {A.~A.}\
  \bibnamefont {Valverde}}, \bibinfo {author} {\bibfnamefont {R.}~\bibnamefont
  {Weil}}, \bibinfo {author} {\bibfnamefont {I.~T.}\ \bibnamefont {Yandow}},
  \bibinfo {author} {\bibfnamefont {J.}~\bibnamefont {Dilling}}, \ and\
  \bibinfo {author} {\bibfnamefont {A.~A.}\ \bibnamefont {Kwiatkowski}},\
  }\href {\doibase 10.1103/PhysRevC.106.024312} {\bibfield  {journal} {\bibinfo
   {journal} {Phys. Rev. C}\ }\textbf {\bibinfo {volume} {106}},\ \bibinfo
  {pages} {024312} (\bibinfo {year} {2022})}\BibitemShut {NoStop}%
\bibitem [{\citenamefont {Otsuka}\ \emph {et~al.}(2020)\citenamefont {Otsuka},
  \citenamefont {Gade}, \citenamefont {Sorlin}, \citenamefont {Suzuki},\ and\
  \citenamefont {Utsuno}}]{Otsuka2020}%
  \BibitemOpen
  \bibfield  {author} {\bibinfo {author} {\bibfnamefont {T.}~\bibnamefont
  {Otsuka}}, \bibinfo {author} {\bibfnamefont {A.}~\bibnamefont {Gade}},
  \bibinfo {author} {\bibfnamefont {O.}~\bibnamefont {Sorlin}}, \bibinfo
  {author} {\bibfnamefont {T.}~\bibnamefont {Suzuki}}, \ and\ \bibinfo {author}
  {\bibfnamefont {Y.}~\bibnamefont {Utsuno}},\ }\href {\doibase
  10.1103/RevModPhys.92.015002} {\bibfield  {journal} {\bibinfo  {journal}
  {Rev. Mod. Phys.}\ }\textbf {\bibinfo {volume} {92}},\ \bibinfo {pages}
  {015002} (\bibinfo {year} {2020})}\BibitemShut {NoStop}%
\bibitem [{\citenamefont {Schatz}\ \emph {et~al.}(2014)\citenamefont {Schatz},
  \citenamefont {Gupta}, \citenamefont {M{\"o}ller}, \citenamefont {Beard},
  \citenamefont {Brown}, \citenamefont {Deibel}, \citenamefont {Gasques},
  \citenamefont {Hix}, \citenamefont {Keek}, \citenamefont {Lau}, \citenamefont
  {Steiner},\ and\ \citenamefont {Wiescher}}]{Schatz2014}%
  \BibitemOpen
  \bibfield  {author} {\bibinfo {author} {\bibfnamefont {H.}~\bibnamefont
  {Schatz}}, \bibinfo {author} {\bibfnamefont {S.}~\bibnamefont {Gupta}},
  \bibinfo {author} {\bibfnamefont {P.}~\bibnamefont {M{\"o}ller}}, \bibinfo
  {author} {\bibfnamefont {M.}~\bibnamefont {Beard}}, \bibinfo {author}
  {\bibfnamefont {E.~F.}\ \bibnamefont {Brown}}, \bibinfo {author}
  {\bibfnamefont {A.~T.}\ \bibnamefont {Deibel}}, \bibinfo {author}
  {\bibfnamefont {L.~R.}\ \bibnamefont {Gasques}}, \bibinfo {author}
  {\bibfnamefont {W.~R.}\ \bibnamefont {Hix}}, \bibinfo {author} {\bibfnamefont
  {L.}~\bibnamefont {Keek}}, \bibinfo {author} {\bibfnamefont {R.}~\bibnamefont
  {Lau}}, \bibinfo {author} {\bibfnamefont {A.~W.}\ \bibnamefont {Steiner}}, \
  and\ \bibinfo {author} {\bibfnamefont {M.}~\bibnamefont {Wiescher}},\ }\href
  {\doibase 10.1038/nature12757} {\bibfield  {journal} {\bibinfo  {journal}
  {Nature}\ }\textbf {\bibinfo {volume} {505}},\ \bibinfo {pages} {62}
  (\bibinfo {year} {2014})}\BibitemShut {NoStop}%
\bibitem [{\citenamefont {Deibel}\ \emph {et~al.}(2016)\citenamefont {Deibel},
  \citenamefont {Meisel}, \citenamefont {Schatz}, \citenamefont {Brown},\ and\
  \citenamefont {Cumming}}]{Deibel2016}%
  \BibitemOpen
  \bibfield  {author} {\bibinfo {author} {\bibfnamefont {A.}~\bibnamefont
  {Deibel}}, \bibinfo {author} {\bibfnamefont {Z.}~\bibnamefont {Meisel}},
  \bibinfo {author} {\bibfnamefont {H.}~\bibnamefont {Schatz}}, \bibinfo
  {author} {\bibfnamefont {E.~F.}\ \bibnamefont {Brown}}, \ and\ \bibinfo
  {author} {\bibfnamefont {A.}~\bibnamefont {Cumming}},\ }\href {\doibase
  10.3847/0004-637x/831/1/13} {\bibfield  {journal} {\bibinfo  {journal}
  {Astrophys. J.}\ }\textbf {\bibinfo {volume} {831}},\ \bibinfo {pages} {13}
  (\bibinfo {year} {2016})}\BibitemShut {NoStop}%
\bibitem [{\citenamefont {Otsuka}\ \emph {et~al.}(2001)\citenamefont {Otsuka},
  \citenamefont {Fujimoto}, \citenamefont {Utsuno}, \citenamefont {Brown},
  \citenamefont {Honma},\ and\ \citenamefont {Mizusaki}}]{Otsuka2001}%
  \BibitemOpen
  \bibfield  {author} {\bibinfo {author} {\bibfnamefont {T.}~\bibnamefont
  {Otsuka}}, \bibinfo {author} {\bibfnamefont {R.}~\bibnamefont {Fujimoto}},
  \bibinfo {author} {\bibfnamefont {Y.}~\bibnamefont {Utsuno}}, \bibinfo
  {author} {\bibfnamefont {B.~A.}\ \bibnamefont {Brown}}, \bibinfo {author}
  {\bibfnamefont {M.}~\bibnamefont {Honma}}, \ and\ \bibinfo {author}
  {\bibfnamefont {T.}~\bibnamefont {Mizusaki}},\ }\href {\doibase
  10.1103/PhysRevLett.87.082502} {\bibfield  {journal} {\bibinfo  {journal}
  {Phys. Rev. Lett.}\ }\textbf {\bibinfo {volume} {87}},\ \bibinfo {pages}
  {082502} (\bibinfo {year} {2001})}\BibitemShut {NoStop}%
\bibitem [{\citenamefont {Sorlin}\ \emph {et~al.}(2003)\citenamefont {Sorlin},
  \citenamefont {Donzaud}, \citenamefont {Nowacki}, \citenamefont
  {Ang{\'{e}}lique}, \citenamefont {Azaiez}, \citenamefont {Bourgeois},
  \citenamefont {Chiste}, \citenamefont {Dlouhy}, \citenamefont {Gr{\'{e}}vy},
  \citenamefont {Guillemaud-Mueller}, \citenamefont {Ibrahim}, \citenamefont
  {Kratz}, \citenamefont {Lewitowicz}, \citenamefont {Lukyanov}, \citenamefont
  {Mrasek}, \citenamefont {Penionzhkevich}, \citenamefont {{De Oliveira
  Santos}}, \citenamefont {Pfeiffer}, \citenamefont {Pougheon}, \citenamefont
  {Poves}, \citenamefont {Saint-Laurent},\ and\ \citenamefont
  {Stanoiu}}]{Sorlin2003}%
  \BibitemOpen
  \bibfield  {author} {\bibinfo {author} {\bibfnamefont {O.}~\bibnamefont
  {Sorlin}}, \bibinfo {author} {\bibfnamefont {C.}~\bibnamefont {Donzaud}},
  \bibinfo {author} {\bibfnamefont {F.}~\bibnamefont {Nowacki}}, \bibinfo
  {author} {\bibfnamefont {J.~C.}\ \bibnamefont {Ang{\'{e}}lique}}, \bibinfo
  {author} {\bibfnamefont {F.}~\bibnamefont {Azaiez}}, \bibinfo {author}
  {\bibfnamefont {C.}~\bibnamefont {Bourgeois}}, \bibinfo {author}
  {\bibfnamefont {V.}~\bibnamefont {Chiste}}, \bibinfo {author} {\bibfnamefont
  {Z.}~\bibnamefont {Dlouhy}}, \bibinfo {author} {\bibfnamefont
  {S.}~\bibnamefont {Gr{\'{e}}vy}}, \bibinfo {author} {\bibfnamefont
  {D.}~\bibnamefont {Guillemaud-Mueller}}, \bibinfo {author} {\bibfnamefont
  {F.}~\bibnamefont {Ibrahim}}, \bibinfo {author} {\bibfnamefont {K.~L.}\
  \bibnamefont {Kratz}}, \bibinfo {author} {\bibfnamefont {M.}~\bibnamefont
  {Lewitowicz}}, \bibinfo {author} {\bibfnamefont {S.~M.}\ \bibnamefont
  {Lukyanov}}, \bibinfo {author} {\bibfnamefont {J.}~\bibnamefont {Mrasek}},
  \bibinfo {author} {\bibfnamefont {Y.~E.}\ \bibnamefont {Penionzhkevich}},
  \bibinfo {author} {\bibfnamefont {F.}~\bibnamefont {{De Oliveira Santos}}},
  \bibinfo {author} {\bibfnamefont {B.}~\bibnamefont {Pfeiffer}}, \bibinfo
  {author} {\bibfnamefont {F.}~\bibnamefont {Pougheon}}, \bibinfo {author}
  {\bibfnamefont {A.}~\bibnamefont {Poves}}, \bibinfo {author} {\bibfnamefont
  {M.~G.}\ \bibnamefont {Saint-Laurent}}, \ and\ \bibinfo {author}
  {\bibfnamefont {M.}~\bibnamefont {Stanoiu}},\ }\href {\doibase
  10.1140/epja/i2002-10069-9} {\bibfield  {journal} {\bibinfo  {journal} {Eur.
  Phys. J. A}\ }\textbf {\bibinfo {volume} {16}},\ \bibinfo {pages} {55}
  (\bibinfo {year} {2003})}\BibitemShut {NoStop}%
\bibitem [{\citenamefont {Gaudefroy}\ \emph {et~al.}(2005)\citenamefont
  {Gaudefroy}, \citenamefont {Sorlin}, \citenamefont {Donzaud}, \citenamefont
  {Ang{\'e}lique}, \citenamefont {Azaiez}, \citenamefont {Bourgeois},
  \citenamefont {Chiste}, \citenamefont {Dlouhy}, \citenamefont {Gr{\'e}vy},
  \citenamefont {Guillemaud-Mueller}, \citenamefont {Ibrahim}, \citenamefont
  {Kratz}, \citenamefont {Lewitowicz}, \citenamefont {Lukyanov}, \citenamefont
  {Matea}, \citenamefont {Mrasek}, \citenamefont {Nowacki}, \citenamefont
  {de~Oliveira~Santos}, \citenamefont {Penionzhkevich}, \citenamefont
  {Pfeiffer}, \citenamefont {Pougheon}, \citenamefont {Saint-Laurent},\ and\
  \citenamefont {Stanoiu}}]{Gaudefroy2005}%
  \BibitemOpen
  \bibfield  {author} {\bibinfo {author} {\bibfnamefont {L.}~\bibnamefont
  {Gaudefroy}}, \bibinfo {author} {\bibfnamefont {O.}~\bibnamefont {Sorlin}},
  \bibinfo {author} {\bibfnamefont {C.}~\bibnamefont {Donzaud}}, \bibinfo
  {author} {\bibfnamefont {J.~C.}\ \bibnamefont {Ang{\'e}lique}}, \bibinfo
  {author} {\bibfnamefont {F.}~\bibnamefont {Azaiez}}, \bibinfo {author}
  {\bibfnamefont {C.}~\bibnamefont {Bourgeois}}, \bibinfo {author}
  {\bibfnamefont {V.}~\bibnamefont {Chiste}}, \bibinfo {author} {\bibfnamefont
  {Z.}~\bibnamefont {Dlouhy}}, \bibinfo {author} {\bibfnamefont
  {S.}~\bibnamefont {Gr{\'e}vy}}, \bibinfo {author} {\bibfnamefont
  {D.}~\bibnamefont {Guillemaud-Mueller}}, \bibinfo {author} {\bibfnamefont
  {F.}~\bibnamefont {Ibrahim}}, \bibinfo {author} {\bibfnamefont {K.~L.}\
  \bibnamefont {Kratz}}, \bibinfo {author} {\bibfnamefont {M.}~\bibnamefont
  {Lewitowicz}}, \bibinfo {author} {\bibfnamefont {S.~M.}\ \bibnamefont
  {Lukyanov}}, \bibinfo {author} {\bibfnamefont {I.}~\bibnamefont {Matea}},
  \bibinfo {author} {\bibfnamefont {J.}~\bibnamefont {Mrasek}}, \bibinfo
  {author} {\bibfnamefont {F.}~\bibnamefont {Nowacki}}, \bibinfo {author}
  {\bibfnamefont {F.}~\bibnamefont {de~Oliveira~Santos}}, \bibinfo {author}
  {\bibfnamefont {Y.~E.}\ \bibnamefont {Penionzhkevich}}, \bibinfo {author}
  {\bibfnamefont {B.}~\bibnamefont {Pfeiffer}}, \bibinfo {author}
  {\bibfnamefont {F.}~\bibnamefont {Pougheon}}, \bibinfo {author}
  {\bibfnamefont {M.~G.}\ \bibnamefont {Saint-Laurent}}, \ and\ \bibinfo
  {author} {\bibfnamefont {M.}~\bibnamefont {Stanoiu}},\ }\href {\doibase
  10.1140/epja/i2004-10068-x} {\bibfield  {journal} {\bibinfo  {journal} {Eur.
  Phys. J. A}\ }\textbf {\bibinfo {volume} {23}},\ \bibinfo {pages} {41}
  (\bibinfo {year} {2005})}\BibitemShut {NoStop}%
\bibitem [{\citenamefont {Sorlin}\ \emph {et~al.}(2002)\citenamefont {Sorlin},
  \citenamefont {Leenhardt}, \citenamefont {Donzaud}, \citenamefont {Duprat},
  \citenamefont {Azaiez}, \citenamefont {Nowacki}, \citenamefont {Grawe},
  \citenamefont {Dombr{\'{a}}di}, \citenamefont {Amorini}, \citenamefont
  {Astier}, \citenamefont {Baiborodin}, \citenamefont {Belleguic},
  \citenamefont {Borcea}, \citenamefont {Bourgeois}, \citenamefont {Cullen},
  \citenamefont {Dlouhy}, \citenamefont {Dragulescu}, \citenamefont
  {G{\'{o}}rska}, \citenamefont {Gr{\'{e}}vy}, \citenamefont
  {Guillemaud-Mueller}, \citenamefont {Hagemann}, \citenamefont {Herskind},
  \citenamefont {Kiener}, \citenamefont {Lemmon}, \citenamefont {Lewitowicz},
  \citenamefont {Lukyanov}, \citenamefont {Mayet}, \citenamefont {{de Oliveira
  Santos}}, \citenamefont {Pantalica}, \citenamefont {Penionzhkevich},
  \citenamefont {Pougheon}, \citenamefont {Poves}, \citenamefont {Redon},
  \citenamefont {Saint-Laurent}, \citenamefont {Scarpaci}, \citenamefont
  {Sletten}, \citenamefont {Stanoiu}, \citenamefont {Tarasov},\ and\
  \citenamefont {Theisen}}]{Sorlin2002}%
  \BibitemOpen
  \bibfield  {author} {\bibinfo {author} {\bibfnamefont {O.}~\bibnamefont
  {Sorlin}}, \bibinfo {author} {\bibfnamefont {S.}~\bibnamefont {Leenhardt}},
  \bibinfo {author} {\bibfnamefont {C.}~\bibnamefont {Donzaud}}, \bibinfo
  {author} {\bibfnamefont {J.}~\bibnamefont {Duprat}}, \bibinfo {author}
  {\bibfnamefont {F.}~\bibnamefont {Azaiez}}, \bibinfo {author} {\bibfnamefont
  {F.}~\bibnamefont {Nowacki}}, \bibinfo {author} {\bibfnamefont
  {H.}~\bibnamefont {Grawe}}, \bibinfo {author} {\bibfnamefont
  {Z.}~\bibnamefont {Dombr{\'{a}}di}}, \bibinfo {author} {\bibfnamefont
  {F.}~\bibnamefont {Amorini}}, \bibinfo {author} {\bibfnamefont
  {A.}~\bibnamefont {Astier}}, \bibinfo {author} {\bibfnamefont
  {D.}~\bibnamefont {Baiborodin}}, \bibinfo {author} {\bibfnamefont
  {M.}~\bibnamefont {Belleguic}}, \bibinfo {author} {\bibfnamefont
  {C.}~\bibnamefont {Borcea}}, \bibinfo {author} {\bibfnamefont
  {C.}~\bibnamefont {Bourgeois}}, \bibinfo {author} {\bibfnamefont {D.~M.}\
  \bibnamefont {Cullen}}, \bibinfo {author} {\bibfnamefont {Z.}~\bibnamefont
  {Dlouhy}}, \bibinfo {author} {\bibfnamefont {E.}~\bibnamefont {Dragulescu}},
  \bibinfo {author} {\bibfnamefont {M.}~\bibnamefont {G{\'{o}}rska}}, \bibinfo
  {author} {\bibfnamefont {S.}~\bibnamefont {Gr{\'{e}}vy}}, \bibinfo {author}
  {\bibfnamefont {D.}~\bibnamefont {Guillemaud-Mueller}}, \bibinfo {author}
  {\bibfnamefont {G.}~\bibnamefont {Hagemann}}, \bibinfo {author}
  {\bibfnamefont {B.}~\bibnamefont {Herskind}}, \bibinfo {author}
  {\bibfnamefont {J.}~\bibnamefont {Kiener}}, \bibinfo {author} {\bibfnamefont
  {R.}~\bibnamefont {Lemmon}}, \bibinfo {author} {\bibfnamefont
  {M.}~\bibnamefont {Lewitowicz}}, \bibinfo {author} {\bibfnamefont {S.~M.}\
  \bibnamefont {Lukyanov}}, \bibinfo {author} {\bibfnamefont {P.}~\bibnamefont
  {Mayet}}, \bibinfo {author} {\bibfnamefont {F.}~\bibnamefont {{de Oliveira
  Santos}}}, \bibinfo {author} {\bibfnamefont {D.}~\bibnamefont {Pantalica}},
  \bibinfo {author} {\bibfnamefont {Y.~E.}\ \bibnamefont {Penionzhkevich}},
  \bibinfo {author} {\bibfnamefont {F.}~\bibnamefont {Pougheon}}, \bibinfo
  {author} {\bibfnamefont {A.}~\bibnamefont {Poves}}, \bibinfo {author}
  {\bibfnamefont {N.}~\bibnamefont {Redon}}, \bibinfo {author} {\bibfnamefont
  {M.~G.}\ \bibnamefont {Saint-Laurent}}, \bibinfo {author} {\bibfnamefont
  {J.~A.}\ \bibnamefont {Scarpaci}}, \bibinfo {author} {\bibfnamefont
  {G.}~\bibnamefont {Sletten}}, \bibinfo {author} {\bibfnamefont
  {M.}~\bibnamefont {Stanoiu}}, \bibinfo {author} {\bibfnamefont
  {O.}~\bibnamefont {Tarasov}}, \ and\ \bibinfo {author} {\bibfnamefont
  {C.}~\bibnamefont {Theisen}},\ }\href {\doibase
  10.1103/PhysRevLett.88.092501} {\bibfield  {journal} {\bibinfo  {journal}
  {Phys. Rev. Lett.}\ }\textbf {\bibinfo {volume} {88}},\ \bibinfo {pages}
  {092501} (\bibinfo {year} {2002})}\BibitemShut {NoStop}%
\bibitem [{\citenamefont {Aoi}\ \emph {et~al.}(2009)\citenamefont {Aoi},
  \citenamefont {Takeshita}, \citenamefont {Suzuki}, \citenamefont {Takeuchi},
  \citenamefont {Ota}, \citenamefont {Baba}, \citenamefont {Bishop},
  \citenamefont {Fukui}, \citenamefont {Hashimoto}, \citenamefont {Ong},
  \citenamefont {Ideguchi}, \citenamefont {Ieki}, \citenamefont {Imai},
  \citenamefont {Ishihara}, \citenamefont {Iwasaki}, \citenamefont {Kanno},
  \citenamefont {Kondo}, \citenamefont {Kubo}, \citenamefont {Kurita},
  \citenamefont {Kusaka}, \citenamefont {Minemura}, \citenamefont
  {Motobayashi}, \citenamefont {Nakabayashi}, \citenamefont {Nakamura},
  \citenamefont {Nakao}, \citenamefont {Niikura}, \citenamefont {Okumura},
  \citenamefont {Ohnishi}, \citenamefont {Sakurai}, \citenamefont {Shimoura},
  \citenamefont {Sugo}, \citenamefont {Suzuki}, \citenamefont {Suzuki},
  \citenamefont {Tamaki}, \citenamefont {Tanaka}, \citenamefont {Togano},\ and\
  \citenamefont {Yamada}}]{Aoi2009}%
  \BibitemOpen
  \bibfield  {author} {\bibinfo {author} {\bibfnamefont {N.}~\bibnamefont
  {Aoi}}, \bibinfo {author} {\bibfnamefont {E.}~\bibnamefont {Takeshita}},
  \bibinfo {author} {\bibfnamefont {H.}~\bibnamefont {Suzuki}}, \bibinfo
  {author} {\bibfnamefont {S.}~\bibnamefont {Takeuchi}}, \bibinfo {author}
  {\bibfnamefont {S.}~\bibnamefont {Ota}}, \bibinfo {author} {\bibfnamefont
  {H.}~\bibnamefont {Baba}}, \bibinfo {author} {\bibfnamefont {S.}~\bibnamefont
  {Bishop}}, \bibinfo {author} {\bibfnamefont {T.}~\bibnamefont {Fukui}},
  \bibinfo {author} {\bibfnamefont {Y.}~\bibnamefont {Hashimoto}}, \bibinfo
  {author} {\bibfnamefont {H.~J.}\ \bibnamefont {Ong}}, \bibinfo {author}
  {\bibfnamefont {E.}~\bibnamefont {Ideguchi}}, \bibinfo {author}
  {\bibfnamefont {K.}~\bibnamefont {Ieki}}, \bibinfo {author} {\bibfnamefont
  {N.}~\bibnamefont {Imai}}, \bibinfo {author} {\bibfnamefont {M.}~\bibnamefont
  {Ishihara}}, \bibinfo {author} {\bibfnamefont {H.}~\bibnamefont {Iwasaki}},
  \bibinfo {author} {\bibfnamefont {S.}~\bibnamefont {Kanno}}, \bibinfo
  {author} {\bibfnamefont {Y.}~\bibnamefont {Kondo}}, \bibinfo {author}
  {\bibfnamefont {T.}~\bibnamefont {Kubo}}, \bibinfo {author} {\bibfnamefont
  {K.}~\bibnamefont {Kurita}}, \bibinfo {author} {\bibfnamefont
  {K.}~\bibnamefont {Kusaka}}, \bibinfo {author} {\bibfnamefont
  {T.}~\bibnamefont {Minemura}}, \bibinfo {author} {\bibfnamefont
  {T.}~\bibnamefont {Motobayashi}}, \bibinfo {author} {\bibfnamefont
  {T.}~\bibnamefont {Nakabayashi}}, \bibinfo {author} {\bibfnamefont
  {T.}~\bibnamefont {Nakamura}}, \bibinfo {author} {\bibfnamefont
  {T.}~\bibnamefont {Nakao}}, \bibinfo {author} {\bibfnamefont
  {M.}~\bibnamefont {Niikura}}, \bibinfo {author} {\bibfnamefont
  {T.}~\bibnamefont {Okumura}}, \bibinfo {author} {\bibfnamefont {T.~K.}\
  \bibnamefont {Ohnishi}}, \bibinfo {author} {\bibfnamefont {H.}~\bibnamefont
  {Sakurai}}, \bibinfo {author} {\bibfnamefont {S.}~\bibnamefont {Shimoura}},
  \bibinfo {author} {\bibfnamefont {R.}~\bibnamefont {Sugo}}, \bibinfo {author}
  {\bibfnamefont {D.}~\bibnamefont {Suzuki}}, \bibinfo {author} {\bibfnamefont
  {M.~K.}\ \bibnamefont {Suzuki}}, \bibinfo {author} {\bibfnamefont
  {M.}~\bibnamefont {Tamaki}}, \bibinfo {author} {\bibfnamefont
  {K.}~\bibnamefont {Tanaka}}, \bibinfo {author} {\bibfnamefont
  {Y.}~\bibnamefont {Togano}}, \ and\ \bibinfo {author} {\bibfnamefont
  {K.}~\bibnamefont {Yamada}},\ }\href {\doibase
  10.1103/PhysRevLett.102.012502} {\bibfield  {journal} {\bibinfo  {journal}
  {Phys. Rev. Lett.}\ }\textbf {\bibinfo {volume} {102}},\ \bibinfo {pages}
  {012502} (\bibinfo {year} {2009})}\BibitemShut {NoStop}%
\bibitem [{\citenamefont {Suzuki}\ \emph {et~al.}(2013)\citenamefont {Suzuki},
  \citenamefont {Aoi}, \citenamefont {Takeshita}, \citenamefont {Takeuchi},
  \citenamefont {Ota}, \citenamefont {Baba}, \citenamefont {Bishop},
  \citenamefont {Fukui}, \citenamefont {Hashimoto}, \citenamefont {Ideguchi},
  \citenamefont {Ieki}, \citenamefont {Imai}, \citenamefont {Ishihara},
  \citenamefont {Iwasaki}, \citenamefont {Kanno}, \citenamefont {Kondo},
  \citenamefont {Kubo}, \citenamefont {Kurita}, \citenamefont {Kusaka},
  \citenamefont {Minemura}, \citenamefont {Motobayashi}, \citenamefont
  {Nakabayashi}, \citenamefont {Nakamura}, \citenamefont {Nakao}, \citenamefont
  {Niikura}, \citenamefont {Okumura}, \citenamefont {Ohnishi}, \citenamefont
  {Ong}, \citenamefont {Sakurai}, \citenamefont {Shimoura}, \citenamefont
  {Sugo}, \citenamefont {Suzuki}, \citenamefont {Suzuki}, \citenamefont
  {Tamaki}, \citenamefont {Tanaka}, \citenamefont {Togano},\ and\ \citenamefont
  {Yamada}}]{Suzuki2013}%
  \BibitemOpen
  \bibfield  {author} {\bibinfo {author} {\bibfnamefont {H.}~\bibnamefont
  {Suzuki}}, \bibinfo {author} {\bibfnamefont {N.}~\bibnamefont {Aoi}},
  \bibinfo {author} {\bibfnamefont {E.}~\bibnamefont {Takeshita}}, \bibinfo
  {author} {\bibfnamefont {S.}~\bibnamefont {Takeuchi}}, \bibinfo {author}
  {\bibfnamefont {S.}~\bibnamefont {Ota}}, \bibinfo {author} {\bibfnamefont
  {H.}~\bibnamefont {Baba}}, \bibinfo {author} {\bibfnamefont {S.}~\bibnamefont
  {Bishop}}, \bibinfo {author} {\bibfnamefont {T.}~\bibnamefont {Fukui}},
  \bibinfo {author} {\bibfnamefont {Y.}~\bibnamefont {Hashimoto}}, \bibinfo
  {author} {\bibfnamefont {E.}~\bibnamefont {Ideguchi}}, \bibinfo {author}
  {\bibfnamefont {K.}~\bibnamefont {Ieki}}, \bibinfo {author} {\bibfnamefont
  {N.}~\bibnamefont {Imai}}, \bibinfo {author} {\bibfnamefont {M.}~\bibnamefont
  {Ishihara}}, \bibinfo {author} {\bibfnamefont {H.}~\bibnamefont {Iwasaki}},
  \bibinfo {author} {\bibfnamefont {S.}~\bibnamefont {Kanno}}, \bibinfo
  {author} {\bibfnamefont {Y.}~\bibnamefont {Kondo}}, \bibinfo {author}
  {\bibfnamefont {T.}~\bibnamefont {Kubo}}, \bibinfo {author} {\bibfnamefont
  {K.}~\bibnamefont {Kurita}}, \bibinfo {author} {\bibfnamefont
  {K.}~\bibnamefont {Kusaka}}, \bibinfo {author} {\bibfnamefont
  {T.}~\bibnamefont {Minemura}}, \bibinfo {author} {\bibfnamefont
  {T.}~\bibnamefont {Motobayashi}}, \bibinfo {author} {\bibfnamefont
  {T.}~\bibnamefont {Nakabayashi}}, \bibinfo {author} {\bibfnamefont
  {T.}~\bibnamefont {Nakamura}}, \bibinfo {author} {\bibfnamefont
  {T.}~\bibnamefont {Nakao}}, \bibinfo {author} {\bibfnamefont
  {M.}~\bibnamefont {Niikura}}, \bibinfo {author} {\bibfnamefont
  {T.}~\bibnamefont {Okumura}}, \bibinfo {author} {\bibfnamefont {T.~K.}\
  \bibnamefont {Ohnishi}}, \bibinfo {author} {\bibfnamefont {H.~J.}\
  \bibnamefont {Ong}}, \bibinfo {author} {\bibfnamefont {H.}~\bibnamefont
  {Sakurai}}, \bibinfo {author} {\bibfnamefont {S.}~\bibnamefont {Shimoura}},
  \bibinfo {author} {\bibfnamefont {R.}~\bibnamefont {Sugo}}, \bibinfo {author}
  {\bibfnamefont {D.}~\bibnamefont {Suzuki}}, \bibinfo {author} {\bibfnamefont
  {M.~K.}\ \bibnamefont {Suzuki}}, \bibinfo {author} {\bibfnamefont
  {M.}~\bibnamefont {Tamaki}}, \bibinfo {author} {\bibfnamefont
  {K.}~\bibnamefont {Tanaka}}, \bibinfo {author} {\bibfnamefont
  {Y.}~\bibnamefont {Togano}}, \ and\ \bibinfo {author} {\bibfnamefont
  {K.}~\bibnamefont {Yamada}},\ }\href {\doibase 10.1103/PhysRevC.88.024326}
  {\bibfield  {journal} {\bibinfo  {journal} {Phys. Rev. C}\ }\textbf {\bibinfo
  {volume} {88}},\ \bibinfo {pages} {024326} (\bibinfo {year}
  {2013})}\BibitemShut {NoStop}%
\bibitem [{\citenamefont {Crawford}\ \emph {et~al.}(2013)\citenamefont
  {Crawford}, \citenamefont {Clark}, \citenamefont {Fallon}, \citenamefont
  {Macchiavelli}, \citenamefont {Baugher}, \citenamefont {Bazin}, \citenamefont
  {Beausang}, \citenamefont {Berryman}, \citenamefont {Bleuel}, \citenamefont
  {Campbell}, \citenamefont {Cromaz}, \citenamefont {de~Angelis}, \citenamefont
  {Gade}, \citenamefont {Hughes}, \citenamefont {Lee}, \citenamefont {Lenzi},
  \citenamefont {Nowacki}, \citenamefont {Paschalis}, \citenamefont {Petri},
  \citenamefont {Poves}, \citenamefont {Ratkiewicz}, \citenamefont {Ross},
  \citenamefont {Sahin}, \citenamefont {Weisshaar}, \citenamefont {Wimmer},\
  and\ \citenamefont {Winkler}}]{Crawford2013}%
  \BibitemOpen
  \bibfield  {author} {\bibinfo {author} {\bibfnamefont {H.~L.}\ \bibnamefont
  {Crawford}}, \bibinfo {author} {\bibfnamefont {R.~M.}\ \bibnamefont {Clark}},
  \bibinfo {author} {\bibfnamefont {P.}~\bibnamefont {Fallon}}, \bibinfo
  {author} {\bibfnamefont {A.~O.}\ \bibnamefont {Macchiavelli}}, \bibinfo
  {author} {\bibfnamefont {T.}~\bibnamefont {Baugher}}, \bibinfo {author}
  {\bibfnamefont {D.}~\bibnamefont {Bazin}}, \bibinfo {author} {\bibfnamefont
  {C.~W.}\ \bibnamefont {Beausang}}, \bibinfo {author} {\bibfnamefont {J.~S.}\
  \bibnamefont {Berryman}}, \bibinfo {author} {\bibfnamefont {D.~L.}\
  \bibnamefont {Bleuel}}, \bibinfo {author} {\bibfnamefont {C.~M.}\
  \bibnamefont {Campbell}}, \bibinfo {author} {\bibfnamefont {M.}~\bibnamefont
  {Cromaz}}, \bibinfo {author} {\bibfnamefont {G.}~\bibnamefont {de~Angelis}},
  \bibinfo {author} {\bibfnamefont {A.}~\bibnamefont {Gade}}, \bibinfo {author}
  {\bibfnamefont {R.~O.}\ \bibnamefont {Hughes}}, \bibinfo {author}
  {\bibfnamefont {I.~Y.}\ \bibnamefont {Lee}}, \bibinfo {author} {\bibfnamefont
  {S.~M.}\ \bibnamefont {Lenzi}}, \bibinfo {author} {\bibfnamefont
  {F.}~\bibnamefont {Nowacki}}, \bibinfo {author} {\bibfnamefont
  {S.}~\bibnamefont {Paschalis}}, \bibinfo {author} {\bibfnamefont
  {M.}~\bibnamefont {Petri}}, \bibinfo {author} {\bibfnamefont
  {A.}~\bibnamefont {Poves}}, \bibinfo {author} {\bibfnamefont
  {A.}~\bibnamefont {Ratkiewicz}}, \bibinfo {author} {\bibfnamefont {T.~J.}\
  \bibnamefont {Ross}}, \bibinfo {author} {\bibfnamefont {E.}~\bibnamefont
  {Sahin}}, \bibinfo {author} {\bibfnamefont {D.}~\bibnamefont {Weisshaar}},
  \bibinfo {author} {\bibfnamefont {K.}~\bibnamefont {Wimmer}}, \ and\ \bibinfo
  {author} {\bibfnamefont {R.}~\bibnamefont {Winkler}},\ }\href {\doibase
  10.1103/PhysRevLett.110.242701} {\bibfield  {journal} {\bibinfo  {journal}
  {Phys. Rev. Lett.}\ }\textbf {\bibinfo {volume} {110}},\ \bibinfo {pages}
  {242701} (\bibinfo {year} {2013})}\BibitemShut {NoStop}%
\bibitem [{\citenamefont {Otsuka}\ \emph {et~al.}(2005)\citenamefont {Otsuka},
  \citenamefont {Suzuki}, \citenamefont {Fujimoto}, \citenamefont {Grawe},\
  and\ \citenamefont {Akaishi}}]{Otsuka2005}%
  \BibitemOpen
  \bibfield  {author} {\bibinfo {author} {\bibfnamefont {T.}~\bibnamefont
  {Otsuka}}, \bibinfo {author} {\bibfnamefont {T.}~\bibnamefont {Suzuki}},
  \bibinfo {author} {\bibfnamefont {R.}~\bibnamefont {Fujimoto}}, \bibinfo
  {author} {\bibfnamefont {H.}~\bibnamefont {Grawe}}, \ and\ \bibinfo {author}
  {\bibfnamefont {Y.}~\bibnamefont {Akaishi}},\ }\href {\doibase
  10.1103/PhysRevLett.95.232502} {\bibfield  {journal} {\bibinfo  {journal}
  {Phys. Rev. Lett.}\ }\textbf {\bibinfo {volume} {95}},\ \bibinfo {pages}
  {232502} (\bibinfo {year} {2005})}\BibitemShut {NoStop}%
\bibitem [{\citenamefont {Huck}\ \emph {et~al.}(1985)\citenamefont {Huck},
  \citenamefont {Klotz}, \citenamefont {Knipper}, \citenamefont {Mieh\'e},
  \citenamefont {Richard-Serre}, \citenamefont {Walter}, \citenamefont {Poves},
  \citenamefont {Ravn},\ and\ \citenamefont {Marguier}}]{Huck1985}%
  \BibitemOpen
  \bibfield  {author} {\bibinfo {author} {\bibfnamefont {A.}~\bibnamefont
  {Huck}}, \bibinfo {author} {\bibfnamefont {G.}~\bibnamefont {Klotz}},
  \bibinfo {author} {\bibfnamefont {A.}~\bibnamefont {Knipper}}, \bibinfo
  {author} {\bibfnamefont {C.}~\bibnamefont {Mieh\'e}}, \bibinfo {author}
  {\bibfnamefont {C.}~\bibnamefont {Richard-Serre}}, \bibinfo {author}
  {\bibfnamefont {G.}~\bibnamefont {Walter}}, \bibinfo {author} {\bibfnamefont
  {A.}~\bibnamefont {Poves}}, \bibinfo {author} {\bibfnamefont {H.~L.}\
  \bibnamefont {Ravn}}, \ and\ \bibinfo {author} {\bibfnamefont
  {G.}~\bibnamefont {Marguier}},\ }\href {\doibase 10.1103/PhysRevC.31.2226}
  {\bibfield  {journal} {\bibinfo  {journal} {Phys. Rev. C}\ }\textbf {\bibinfo
  {volume} {31}},\ \bibinfo {pages} {2226} (\bibinfo {year}
  {1985})}\BibitemShut {NoStop}%
\bibitem [{\citenamefont {Wienholtz}\ \emph {et~al.}(2013)\citenamefont
  {Wienholtz}, \citenamefont {Beck}, \citenamefont {Blaum}, \citenamefont
  {Borgmann}, \citenamefont {Breitenfeldt}, \citenamefont {Cakirli},
  \citenamefont {George}, \citenamefont {Herfurth}, \citenamefont {Holt},
  \citenamefont {Kowalska}, \citenamefont {Kreim}, \citenamefont {Lunney},
  \citenamefont {Manea}, \citenamefont {Men\`endez}, \citenamefont {Neidherr},
  \citenamefont {Rosenbusch}, \citenamefont {Schweikhard}, \citenamefont
  {Schwenk}, \citenamefont {Simonis}, \citenamefont {Stanja}, \citenamefont
  {Wolf},\ and\ \citenamefont {Zuber}}]{Wienholtz2013}%
  \BibitemOpen
  \bibfield  {author} {\bibinfo {author} {\bibfnamefont {F.}~\bibnamefont
  {Wienholtz}}, \bibinfo {author} {\bibfnamefont {D.}~\bibnamefont {Beck}},
  \bibinfo {author} {\bibfnamefont {K.}~\bibnamefont {Blaum}}, \bibinfo
  {author} {\bibfnamefont {C.}~\bibnamefont {Borgmann}}, \bibinfo {author}
  {\bibfnamefont {M.}~\bibnamefont {Breitenfeldt}}, \bibinfo {author}
  {\bibfnamefont {R.~B.}\ \bibnamefont {Cakirli}}, \bibinfo {author}
  {\bibfnamefont {S.}~\bibnamefont {George}}, \bibinfo {author} {\bibfnamefont
  {F.}~\bibnamefont {Herfurth}}, \bibinfo {author} {\bibfnamefont {J.~D.}\
  \bibnamefont {Holt}}, \bibinfo {author} {\bibfnamefont {M.}~\bibnamefont
  {Kowalska}}, \bibinfo {author} {\bibfnamefont {S.}~\bibnamefont {Kreim}},
  \bibinfo {author} {\bibfnamefont {D.}~\bibnamefont {Lunney}}, \bibinfo
  {author} {\bibfnamefont {V.}~\bibnamefont {Manea}}, \bibinfo {author}
  {\bibfnamefont {J.}~\bibnamefont {Men\`endez}}, \bibinfo {author}
  {\bibfnamefont {D.}~\bibnamefont {Neidherr}}, \bibinfo {author}
  {\bibfnamefont {M.}~\bibnamefont {Rosenbusch}}, \bibinfo {author}
  {\bibfnamefont {L.}~\bibnamefont {Schweikhard}}, \bibinfo {author}
  {\bibfnamefont {A.}~\bibnamefont {Schwenk}}, \bibinfo {author} {\bibfnamefont
  {J.}~\bibnamefont {Simonis}}, \bibinfo {author} {\bibfnamefont
  {J.}~\bibnamefont {Stanja}}, \bibinfo {author} {\bibfnamefont {R.~N.}\
  \bibnamefont {Wolf}}, \ and\ \bibinfo {author} {\bibfnamefont
  {K.}~\bibnamefont {Zuber}},\ }\href {http://dx.doi.org/10.1038/nature12226}
  {\bibfield  {journal} {\bibinfo  {journal} {Nature}\ }\textbf {\bibinfo
  {volume} {498}},\ \bibinfo {pages} {346} (\bibinfo {year}
  {2013})}\BibitemShut {NoStop}%
\bibitem [{\citenamefont {Steppenbeck}\ \emph {et~al.}(2013)\citenamefont
  {Steppenbeck}, \citenamefont {Takeuchi}, \citenamefont {Aoi}, \citenamefont
  {Doornenbal}, \citenamefont {Matsushita}, \citenamefont {Wang}, \citenamefont
  {Baba}, \citenamefont {Fukuda}, \citenamefont {Go}, \citenamefont {Honma},
  \citenamefont {Lee}, \citenamefont {Matsui}, \citenamefont {Michimasa},
  \citenamefont {Motobayashi}, \citenamefont {Nishimura}, \citenamefont
  {Otsuka}, \citenamefont {Sakurai}, \citenamefont {Shiga}, \citenamefont
  {S{\"o}derstr{\"o}m}, \citenamefont {Sumikama}, \citenamefont {Suzuki},
  \citenamefont {Taniuchi}, \citenamefont {Utsuno}, \citenamefont
  {Valiente-Dob{\'o}n},\ and\ \citenamefont {Yoneda}}]{Steppenbeck2013}%
  \BibitemOpen
  \bibfield  {author} {\bibinfo {author} {\bibfnamefont {D.}~\bibnamefont
  {Steppenbeck}}, \bibinfo {author} {\bibfnamefont {S.}~\bibnamefont
  {Takeuchi}}, \bibinfo {author} {\bibfnamefont {N.}~\bibnamefont {Aoi}},
  \bibinfo {author} {\bibfnamefont {P.}~\bibnamefont {Doornenbal}}, \bibinfo
  {author} {\bibfnamefont {M.}~\bibnamefont {Matsushita}}, \bibinfo {author}
  {\bibfnamefont {H.}~\bibnamefont {Wang}}, \bibinfo {author} {\bibfnamefont
  {H.}~\bibnamefont {Baba}}, \bibinfo {author} {\bibfnamefont {N.}~\bibnamefont
  {Fukuda}}, \bibinfo {author} {\bibfnamefont {S.}~\bibnamefont {Go}}, \bibinfo
  {author} {\bibfnamefont {M.}~\bibnamefont {Honma}}, \bibinfo {author}
  {\bibfnamefont {J.}~\bibnamefont {Lee}}, \bibinfo {author} {\bibfnamefont
  {K.}~\bibnamefont {Matsui}}, \bibinfo {author} {\bibfnamefont
  {S.}~\bibnamefont {Michimasa}}, \bibinfo {author} {\bibfnamefont
  {T.}~\bibnamefont {Motobayashi}}, \bibinfo {author} {\bibfnamefont
  {D.}~\bibnamefont {Nishimura}}, \bibinfo {author} {\bibfnamefont
  {T.}~\bibnamefont {Otsuka}}, \bibinfo {author} {\bibfnamefont
  {H.}~\bibnamefont {Sakurai}}, \bibinfo {author} {\bibfnamefont
  {Y.}~\bibnamefont {Shiga}}, \bibinfo {author} {\bibfnamefont {P.-A.}\
  \bibnamefont {S{\"o}derstr{\"o}m}}, \bibinfo {author} {\bibfnamefont
  {T.}~\bibnamefont {Sumikama}}, \bibinfo {author} {\bibfnamefont
  {H.}~\bibnamefont {Suzuki}}, \bibinfo {author} {\bibfnamefont
  {R.}~\bibnamefont {Taniuchi}}, \bibinfo {author} {\bibfnamefont
  {Y.}~\bibnamefont {Utsuno}}, \bibinfo {author} {\bibfnamefont {J.~J.}\
  \bibnamefont {Valiente-Dob{\'o}n}}, \ and\ \bibinfo {author} {\bibfnamefont
  {K.}~\bibnamefont {Yoneda}},\ }\href {https://doi.org/10.1038/nature12522}
  {\bibfield  {journal} {\bibinfo  {journal} {Nature}\ }\textbf {\bibinfo
  {volume} {502}},\ \bibinfo {pages} {207} (\bibinfo {year}
  {2013})}\BibitemShut {NoStop}%
\bibitem [{\citenamefont {Michimasa}\ \emph {et~al.}(2018)\citenamefont
  {Michimasa}, \citenamefont {Kobayashi}, \citenamefont {Kiyokawa},
  \citenamefont {Ota}, \citenamefont {Ahn}, \citenamefont {Baba}, \citenamefont
  {Berg}, \citenamefont {Dozono}, \citenamefont {Fukuda}, \citenamefont
  {Furuno}, \citenamefont {Ideguchi}, \citenamefont {Inabe}, \citenamefont
  {Kawabata}, \citenamefont {Kawase}, \citenamefont {Kisamori}, \citenamefont
  {Kobayashi}, \citenamefont {Kubo}, \citenamefont {Kubota}, \citenamefont
  {Lee}, \citenamefont {Matsushita}, \citenamefont {Miya}, \citenamefont
  {Mizukami}, \citenamefont {Nagakura}, \citenamefont {Nishimura},
  \citenamefont {Oikawa}, \citenamefont {Sakai}, \citenamefont {Shimizu},
  \citenamefont {Stolz}, \citenamefont {Suzuki}, \citenamefont {Takaki},
  \citenamefont {Takeda}, \citenamefont {Takeuchi}, \citenamefont {Tokieda},
  \citenamefont {Uesaka}, \citenamefont {Yako}, \citenamefont {Yamaguchi},
  \citenamefont {Yanagisawa}, \citenamefont {Yokoyama}, \citenamefont
  {Yoshida},\ and\ \citenamefont {Shimoura}}]{Michimasa2018}%
  \BibitemOpen
  \bibfield  {author} {\bibinfo {author} {\bibfnamefont {S.}~\bibnamefont
  {Michimasa}}, \bibinfo {author} {\bibfnamefont {M.}~\bibnamefont
  {Kobayashi}}, \bibinfo {author} {\bibfnamefont {Y.}~\bibnamefont {Kiyokawa}},
  \bibinfo {author} {\bibfnamefont {S.}~\bibnamefont {Ota}}, \bibinfo {author}
  {\bibfnamefont {D.~S.}\ \bibnamefont {Ahn}}, \bibinfo {author} {\bibfnamefont
  {H.}~\bibnamefont {Baba}}, \bibinfo {author} {\bibfnamefont {G.~P.~A.}\
  \bibnamefont {Berg}}, \bibinfo {author} {\bibfnamefont {M.}~\bibnamefont
  {Dozono}}, \bibinfo {author} {\bibfnamefont {N.}~\bibnamefont {Fukuda}},
  \bibinfo {author} {\bibfnamefont {T.}~\bibnamefont {Furuno}}, \bibinfo
  {author} {\bibfnamefont {E.}~\bibnamefont {Ideguchi}}, \bibinfo {author}
  {\bibfnamefont {N.}~\bibnamefont {Inabe}}, \bibinfo {author} {\bibfnamefont
  {T.}~\bibnamefont {Kawabata}}, \bibinfo {author} {\bibfnamefont
  {S.}~\bibnamefont {Kawase}}, \bibinfo {author} {\bibfnamefont
  {K.}~\bibnamefont {Kisamori}}, \bibinfo {author} {\bibfnamefont
  {K.}~\bibnamefont {Kobayashi}}, \bibinfo {author} {\bibfnamefont
  {T.}~\bibnamefont {Kubo}}, \bibinfo {author} {\bibfnamefont {Y.}~\bibnamefont
  {Kubota}}, \bibinfo {author} {\bibfnamefont {C.~S.}\ \bibnamefont {Lee}},
  \bibinfo {author} {\bibfnamefont {M.}~\bibnamefont {Matsushita}}, \bibinfo
  {author} {\bibfnamefont {H.}~\bibnamefont {Miya}}, \bibinfo {author}
  {\bibfnamefont {A.}~\bibnamefont {Mizukami}}, \bibinfo {author}
  {\bibfnamefont {H.}~\bibnamefont {Nagakura}}, \bibinfo {author}
  {\bibfnamefont {D.}~\bibnamefont {Nishimura}}, \bibinfo {author}
  {\bibfnamefont {H.}~\bibnamefont {Oikawa}}, \bibinfo {author} {\bibfnamefont
  {H.}~\bibnamefont {Sakai}}, \bibinfo {author} {\bibfnamefont
  {Y.}~\bibnamefont {Shimizu}}, \bibinfo {author} {\bibfnamefont
  {A.}~\bibnamefont {Stolz}}, \bibinfo {author} {\bibfnamefont
  {H.}~\bibnamefont {Suzuki}}, \bibinfo {author} {\bibfnamefont
  {M.}~\bibnamefont {Takaki}}, \bibinfo {author} {\bibfnamefont
  {H.}~\bibnamefont {Takeda}}, \bibinfo {author} {\bibfnamefont
  {S.}~\bibnamefont {Takeuchi}}, \bibinfo {author} {\bibfnamefont
  {H.}~\bibnamefont {Tokieda}}, \bibinfo {author} {\bibfnamefont
  {T.}~\bibnamefont {Uesaka}}, \bibinfo {author} {\bibfnamefont
  {K.}~\bibnamefont {Yako}}, \bibinfo {author} {\bibfnamefont {Y.}~\bibnamefont
  {Yamaguchi}}, \bibinfo {author} {\bibfnamefont {Y.}~\bibnamefont
  {Yanagisawa}}, \bibinfo {author} {\bibfnamefont {R.}~\bibnamefont
  {Yokoyama}}, \bibinfo {author} {\bibfnamefont {K.}~\bibnamefont {Yoshida}}, \
  and\ \bibinfo {author} {\bibfnamefont {S.}~\bibnamefont {Shimoura}},\ }\href
  {\doibase 10.1103/PhysRevLett.121.022506} {\bibfield  {journal} {\bibinfo
  {journal} {Phys. Rev. Lett.}\ }\textbf {\bibinfo {volume} {121}},\ \bibinfo
  {pages} {022506} (\bibinfo {year} {2018})}\BibitemShut {NoStop}%
\bibitem [{\citenamefont {Gade}\ \emph {et~al.}(2010)\citenamefont {Gade},
  \citenamefont {Janssens}, \citenamefont {Baugher}, \citenamefont {Bazin},
  \citenamefont {Brown}, \citenamefont {Carpenter}, \citenamefont {Chiara},
  \citenamefont {Deacon}, \citenamefont {Freeman}, \citenamefont {Grinyer},
  \citenamefont {Hoffman}, \citenamefont {Kay}, \citenamefont {Kondev},
  \citenamefont {Lauritsen}, \citenamefont {McDaniel}, \citenamefont
  {Meierbachtol}, \citenamefont {Ratkiewicz}, \citenamefont {Stroberg},
  \citenamefont {Walsh}, \citenamefont {Weisshaar}, \citenamefont {Winkler},\
  and\ \citenamefont {Zhu}}]{Gade2010}%
  \BibitemOpen
  \bibfield  {author} {\bibinfo {author} {\bibfnamefont {A.}~\bibnamefont
  {Gade}}, \bibinfo {author} {\bibfnamefont {R.~V.~F.}\ \bibnamefont
  {Janssens}}, \bibinfo {author} {\bibfnamefont {T.}~\bibnamefont {Baugher}},
  \bibinfo {author} {\bibfnamefont {D.}~\bibnamefont {Bazin}}, \bibinfo
  {author} {\bibfnamefont {B.~A.}\ \bibnamefont {Brown}}, \bibinfo {author}
  {\bibfnamefont {M.~P.}\ \bibnamefont {Carpenter}}, \bibinfo {author}
  {\bibfnamefont {C.~J.}\ \bibnamefont {Chiara}}, \bibinfo {author}
  {\bibfnamefont {A.~N.}\ \bibnamefont {Deacon}}, \bibinfo {author}
  {\bibfnamefont {S.~J.}\ \bibnamefont {Freeman}}, \bibinfo {author}
  {\bibfnamefont {G.~F.}\ \bibnamefont {Grinyer}}, \bibinfo {author}
  {\bibfnamefont {C.~R.}\ \bibnamefont {Hoffman}}, \bibinfo {author}
  {\bibfnamefont {B.~P.}\ \bibnamefont {Kay}}, \bibinfo {author} {\bibfnamefont
  {F.~G.}\ \bibnamefont {Kondev}}, \bibinfo {author} {\bibfnamefont
  {T.}~\bibnamefont {Lauritsen}}, \bibinfo {author} {\bibfnamefont
  {S.}~\bibnamefont {McDaniel}}, \bibinfo {author} {\bibfnamefont
  {K.}~\bibnamefont {Meierbachtol}}, \bibinfo {author} {\bibfnamefont
  {A.}~\bibnamefont {Ratkiewicz}}, \bibinfo {author} {\bibfnamefont {S.~R.}\
  \bibnamefont {Stroberg}}, \bibinfo {author} {\bibfnamefont {K.~A.}\
  \bibnamefont {Walsh}}, \bibinfo {author} {\bibfnamefont {D.}~\bibnamefont
  {Weisshaar}}, \bibinfo {author} {\bibfnamefont {R.}~\bibnamefont {Winkler}},
  \ and\ \bibinfo {author} {\bibfnamefont {S.}~\bibnamefont {Zhu}},\ }\href
  {\doibase 10.1103/PhysRevC.81.051304} {\bibfield  {journal} {\bibinfo
  {journal} {Phys. Rev. C}\ }\textbf {\bibinfo {volume} {81}},\ \bibinfo
  {pages} {051304(R)} (\bibinfo {year} {2010})}\BibitemShut {NoStop}%
\bibitem [{\citenamefont {Gade}\ \emph {et~al.}(2021)\citenamefont {Gade},
  \citenamefont {Janssens}, \citenamefont {Bazin}, \citenamefont {Farris},
  \citenamefont {Hill}, \citenamefont {Lenzi}, \citenamefont {Li},
  \citenamefont {Little}, \citenamefont {Longfellow}, \citenamefont {Nowacki},
  \citenamefont {Poves}, \citenamefont {Rhodes}, \citenamefont {Tostevin},\
  and\ \citenamefont {Weisshaar}}]{Gade2021}%
  \BibitemOpen
  \bibfield  {author} {\bibinfo {author} {\bibfnamefont {A.}~\bibnamefont
  {Gade}}, \bibinfo {author} {\bibfnamefont {R.~V.~F.}\ \bibnamefont
  {Janssens}}, \bibinfo {author} {\bibfnamefont {D.}~\bibnamefont {Bazin}},
  \bibinfo {author} {\bibfnamefont {P.}~\bibnamefont {Farris}}, \bibinfo
  {author} {\bibfnamefont {A.~M.}\ \bibnamefont {Hill}}, \bibinfo {author}
  {\bibfnamefont {S.~M.}\ \bibnamefont {Lenzi}}, \bibinfo {author}
  {\bibfnamefont {J.}~\bibnamefont {Li}}, \bibinfo {author} {\bibfnamefont
  {D.}~\bibnamefont {Little}}, \bibinfo {author} {\bibfnamefont
  {B.}~\bibnamefont {Longfellow}}, \bibinfo {author} {\bibfnamefont
  {F.}~\bibnamefont {Nowacki}}, \bibinfo {author} {\bibfnamefont
  {A.}~\bibnamefont {Poves}}, \bibinfo {author} {\bibfnamefont
  {D.}~\bibnamefont {Rhodes}}, \bibinfo {author} {\bibfnamefont {J.~A.}\
  \bibnamefont {Tostevin}}, \ and\ \bibinfo {author} {\bibfnamefont
  {D.}~\bibnamefont {Weisshaar}},\ }\href {\doibase
  10.1103/PhysRevC.103.014314} {\bibfield  {journal} {\bibinfo  {journal}
  {Phys. Rev. C}\ }\textbf {\bibinfo {volume} {103}},\ \bibinfo {pages}
  {014314} (\bibinfo {year} {2021})}\BibitemShut {NoStop}%
\bibitem [{\citenamefont {Lenzi}\ \emph {et~al.}(2010)\citenamefont {Lenzi},
  \citenamefont {Nowacki}, \citenamefont {Poves},\ and\ \citenamefont
  {Sieja}}]{Lenzi2010}%
  \BibitemOpen
  \bibfield  {author} {\bibinfo {author} {\bibfnamefont {S.~M.}\ \bibnamefont
  {Lenzi}}, \bibinfo {author} {\bibfnamefont {F.}~\bibnamefont {Nowacki}},
  \bibinfo {author} {\bibfnamefont {A.}~\bibnamefont {Poves}}, \ and\ \bibinfo
  {author} {\bibfnamefont {K.}~\bibnamefont {Sieja}},\ }\href {\doibase
  10.1103/PhysRevC.82.054301} {\bibfield  {journal} {\bibinfo  {journal} {Phys.
  Rev. C}\ }\textbf {\bibinfo {volume} {82}},\ \bibinfo {pages} {054301}
  (\bibinfo {year} {2010})}\BibitemShut {NoStop}%
\bibitem [{\citenamefont {Wimmer}\ \emph {et~al.}(2019)\citenamefont {Wimmer},
  \citenamefont {Recchia}, \citenamefont {Lenzi}, \citenamefont {Riccetto},
  \citenamefont {Davinson}, \citenamefont {Estrade}, \citenamefont {Griffin},
  \citenamefont {Nishimura}, \citenamefont {Nowacki}, \citenamefont {Phong},
  \citenamefont {Poves}, \citenamefont {S{\"{o}}derstr{\"{o}}m}, \citenamefont
  {Aktas}, \citenamefont {Al-Aqeel}, \citenamefont {Ando}, \citenamefont
  {Baba}, \citenamefont {Bae}, \citenamefont {Choi}, \citenamefont
  {Doornenbal}, \citenamefont {Ha}, \citenamefont {Harkness-Brennan},
  \citenamefont {Isobe}, \citenamefont {John}, \citenamefont {Kahl},
  \citenamefont {Kiss}, \citenamefont {Kojouharov}, \citenamefont {Kurz},
  \citenamefont {Labiche}, \citenamefont {Matsui}, \citenamefont {Momiyama},
  \citenamefont {Napoli}, \citenamefont {Niikura}, \citenamefont {Nita},
  \citenamefont {Saito}, \citenamefont {Sakurai}, \citenamefont {Schaffner},
  \citenamefont {Schrock}, \citenamefont {Stahl}, \citenamefont {Sumikama},
  \citenamefont {Werner}, \citenamefont {Witt},\ and\ \citenamefont
  {Woods}}]{Wimmer2019}%
  \BibitemOpen
  \bibfield  {author} {\bibinfo {author} {\bibfnamefont {K.}~\bibnamefont
  {Wimmer}}, \bibinfo {author} {\bibfnamefont {F.}~\bibnamefont {Recchia}},
  \bibinfo {author} {\bibfnamefont {S.~M.}\ \bibnamefont {Lenzi}}, \bibinfo
  {author} {\bibfnamefont {S.}~\bibnamefont {Riccetto}}, \bibinfo {author}
  {\bibfnamefont {T.}~\bibnamefont {Davinson}}, \bibinfo {author}
  {\bibfnamefont {A.}~\bibnamefont {Estrade}}, \bibinfo {author} {\bibfnamefont
  {C.~J.}\ \bibnamefont {Griffin}}, \bibinfo {author} {\bibfnamefont
  {S.}~\bibnamefont {Nishimura}}, \bibinfo {author} {\bibfnamefont
  {F.}~\bibnamefont {Nowacki}}, \bibinfo {author} {\bibfnamefont
  {V.}~\bibnamefont {Phong}}, \bibinfo {author} {\bibfnamefont
  {A.}~\bibnamefont {Poves}}, \bibinfo {author} {\bibfnamefont {P.~A.}\
  \bibnamefont {S{\"{o}}derstr{\"{o}}m}}, \bibinfo {author} {\bibfnamefont
  {O.}~\bibnamefont {Aktas}}, \bibinfo {author} {\bibfnamefont
  {M.}~\bibnamefont {Al-Aqeel}}, \bibinfo {author} {\bibfnamefont
  {T.}~\bibnamefont {Ando}}, \bibinfo {author} {\bibfnamefont {H.}~\bibnamefont
  {Baba}}, \bibinfo {author} {\bibfnamefont {S.}~\bibnamefont {Bae}}, \bibinfo
  {author} {\bibfnamefont {S.}~\bibnamefont {Choi}}, \bibinfo {author}
  {\bibfnamefont {P.}~\bibnamefont {Doornenbal}}, \bibinfo {author}
  {\bibfnamefont {J.}~\bibnamefont {Ha}}, \bibinfo {author} {\bibfnamefont
  {L.}~\bibnamefont {Harkness-Brennan}}, \bibinfo {author} {\bibfnamefont
  {T.}~\bibnamefont {Isobe}}, \bibinfo {author} {\bibfnamefont {P.~R.}\
  \bibnamefont {John}}, \bibinfo {author} {\bibfnamefont {D.}~\bibnamefont
  {Kahl}}, \bibinfo {author} {\bibfnamefont {G.}~\bibnamefont {Kiss}}, \bibinfo
  {author} {\bibfnamefont {I.}~\bibnamefont {Kojouharov}}, \bibinfo {author}
  {\bibfnamefont {N.}~\bibnamefont {Kurz}}, \bibinfo {author} {\bibfnamefont
  {M.}~\bibnamefont {Labiche}}, \bibinfo {author} {\bibfnamefont
  {K.}~\bibnamefont {Matsui}}, \bibinfo {author} {\bibfnamefont
  {S.}~\bibnamefont {Momiyama}}, \bibinfo {author} {\bibfnamefont {D.~R.}\
  \bibnamefont {Napoli}}, \bibinfo {author} {\bibfnamefont {M.}~\bibnamefont
  {Niikura}}, \bibinfo {author} {\bibfnamefont {C.}~\bibnamefont {Nita}},
  \bibinfo {author} {\bibfnamefont {Y.}~\bibnamefont {Saito}}, \bibinfo
  {author} {\bibfnamefont {H.}~\bibnamefont {Sakurai}}, \bibinfo {author}
  {\bibfnamefont {H.}~\bibnamefont {Schaffner}}, \bibinfo {author}
  {\bibfnamefont {P.}~\bibnamefont {Schrock}}, \bibinfo {author} {\bibfnamefont
  {C.}~\bibnamefont {Stahl}}, \bibinfo {author} {\bibfnamefont
  {T.}~\bibnamefont {Sumikama}}, \bibinfo {author} {\bibfnamefont
  {V.}~\bibnamefont {Werner}}, \bibinfo {author} {\bibfnamefont
  {W.}~\bibnamefont {Witt}}, \ and\ \bibinfo {author} {\bibfnamefont {P.~J.}\
  \bibnamefont {Woods}},\ }\href {\doibase 10.1016/j.physletb.2019.03.018}
  {\bibfield  {journal} {\bibinfo  {journal} {Physics Letters, Section B:
  Nuclear, Elementary Particle and High-Energy Physics}\ }\textbf {\bibinfo
  {volume} {792}},\ \bibinfo {pages} {16} (\bibinfo {year} {2019})}\BibitemShut
  {NoStop}%
\bibitem [{\citenamefont {Cort{\'e}s}\ \emph {et~al.}(2020)\citenamefont
  {Cort{\'e}s}, \citenamefont {Rodriguez}, \citenamefont {Doornenbal},
  \citenamefont {Obertelli}, \citenamefont {Holt}, \citenamefont {Lenzi},
  \citenamefont {Men{\'e}ndez}, \citenamefont {Nowacki}, \citenamefont {Ogata},
  \citenamefont {Poves}, \citenamefont {Rodr{\'\i}guez}, \citenamefont
  {Schwenk}, \citenamefont {Simonis}, \citenamefont {Stroberg}, \citenamefont
  {Yoshida}, \citenamefont {Achouri}, \citenamefont {Baba}, \citenamefont
  {Browne}, \citenamefont {Calvet}, \citenamefont {Ch{\^a}teau}, \citenamefont
  {Chen}, \citenamefont {Chiga}, \citenamefont {Corsi}, \citenamefont
  {Delbart}, \citenamefont {Gheller}, \citenamefont {Giganon}, \citenamefont
  {Gillibert}, \citenamefont {Hilaire}, \citenamefont {Isobe}, \citenamefont
  {Kobayashi}, \citenamefont {Kubota}, \citenamefont {Lapoux}, \citenamefont
  {Liu}, \citenamefont {Motobayashi}, \citenamefont {Murray}, \citenamefont
  {Otsu}, \citenamefont {Panin}, \citenamefont {Paul}, \citenamefont {Sakurai},
  \citenamefont {Sasano}, \citenamefont {Steppenbeck}, \citenamefont {Stuhl},
  \citenamefont {Sun}, \citenamefont {Togano}, \citenamefont {Uesaka},
  \citenamefont {Wimmer}, \citenamefont {Yoneda}, \citenamefont {Aktas},
  \citenamefont {Aumann}, \citenamefont {Chung}, \citenamefont {Flavigny},
  \citenamefont {Franchoo}, \citenamefont {Ga{\v s}pari{\'c}}, \citenamefont
  {Gerst}, \citenamefont {Gibelin}, \citenamefont {Hahn}, \citenamefont {Kim},
  \citenamefont {Koiwai}, \citenamefont {Kondo}, \citenamefont {Koseoglou},
  \citenamefont {Lee}, \citenamefont {Lehr}, \citenamefont {Linh},
  \citenamefont {Lokotko}, \citenamefont {MacCormick}, \citenamefont
  {Moschner}, \citenamefont {Nakamura}, \citenamefont {Park}, \citenamefont
  {Rossi}, \citenamefont {Sahin}, \citenamefont {Sohler}, \citenamefont
  {S{\"o}derstr{\"o}m}, \citenamefont {Takeuchi}, \citenamefont {Toernqvist},
  \citenamefont {Vaquero}, \citenamefont {Wagner}, \citenamefont {Wang},
  \citenamefont {Werner}, \citenamefont {Xu}, \citenamefont {Yamada},
  \citenamefont {Yan}, \citenamefont {Yang}, \citenamefont {Yasuda},\ and\
  \citenamefont {Zanetti}}]{CORTES2020}%
  \BibitemOpen
  \bibfield  {author} {\bibinfo {author} {\bibfnamefont {M.}~\bibnamefont
  {Cort{\'e}s}}, \bibinfo {author} {\bibfnamefont {W.}~\bibnamefont
  {Rodriguez}}, \bibinfo {author} {\bibfnamefont {P.}~\bibnamefont
  {Doornenbal}}, \bibinfo {author} {\bibfnamefont {A.}~\bibnamefont
  {Obertelli}}, \bibinfo {author} {\bibfnamefont {J.}~\bibnamefont {Holt}},
  \bibinfo {author} {\bibfnamefont {S.}~\bibnamefont {Lenzi}}, \bibinfo
  {author} {\bibfnamefont {J.}~\bibnamefont {Men{\'e}ndez}}, \bibinfo {author}
  {\bibfnamefont {F.}~\bibnamefont {Nowacki}}, \bibinfo {author} {\bibfnamefont
  {K.}~\bibnamefont {Ogata}}, \bibinfo {author} {\bibfnamefont
  {A.}~\bibnamefont {Poves}}, \bibinfo {author} {\bibfnamefont
  {T.}~\bibnamefont {Rodr{\'\i}guez}}, \bibinfo {author} {\bibfnamefont
  {A.}~\bibnamefont {Schwenk}}, \bibinfo {author} {\bibfnamefont
  {J.}~\bibnamefont {Simonis}}, \bibinfo {author} {\bibfnamefont
  {S.}~\bibnamefont {Stroberg}}, \bibinfo {author} {\bibfnamefont
  {K.}~\bibnamefont {Yoshida}}, \bibinfo {author} {\bibfnamefont
  {L.}~\bibnamefont {Achouri}}, \bibinfo {author} {\bibfnamefont
  {H.}~\bibnamefont {Baba}}, \bibinfo {author} {\bibfnamefont {F.}~\bibnamefont
  {Browne}}, \bibinfo {author} {\bibfnamefont {D.}~\bibnamefont {Calvet}},
  \bibinfo {author} {\bibfnamefont {F.}~\bibnamefont {Ch{\^a}teau}}, \bibinfo
  {author} {\bibfnamefont {S.}~\bibnamefont {Chen}}, \bibinfo {author}
  {\bibfnamefont {N.}~\bibnamefont {Chiga}}, \bibinfo {author} {\bibfnamefont
  {A.}~\bibnamefont {Corsi}}, \bibinfo {author} {\bibfnamefont
  {A.}~\bibnamefont {Delbart}}, \bibinfo {author} {\bibfnamefont {J.-M.}\
  \bibnamefont {Gheller}}, \bibinfo {author} {\bibfnamefont {A.}~\bibnamefont
  {Giganon}}, \bibinfo {author} {\bibfnamefont {A.}~\bibnamefont {Gillibert}},
  \bibinfo {author} {\bibfnamefont {C.}~\bibnamefont {Hilaire}}, \bibinfo
  {author} {\bibfnamefont {T.}~\bibnamefont {Isobe}}, \bibinfo {author}
  {\bibfnamefont {T.}~\bibnamefont {Kobayashi}}, \bibinfo {author}
  {\bibfnamefont {Y.}~\bibnamefont {Kubota}}, \bibinfo {author} {\bibfnamefont
  {V.}~\bibnamefont {Lapoux}}, \bibinfo {author} {\bibfnamefont
  {H.}~\bibnamefont {Liu}}, \bibinfo {author} {\bibfnamefont {T.}~\bibnamefont
  {Motobayashi}}, \bibinfo {author} {\bibfnamefont {I.}~\bibnamefont {Murray}},
  \bibinfo {author} {\bibfnamefont {H.}~\bibnamefont {Otsu}}, \bibinfo {author}
  {\bibfnamefont {V.}~\bibnamefont {Panin}}, \bibinfo {author} {\bibfnamefont
  {N.}~\bibnamefont {Paul}}, \bibinfo {author} {\bibfnamefont {H.}~\bibnamefont
  {Sakurai}}, \bibinfo {author} {\bibfnamefont {M.}~\bibnamefont {Sasano}},
  \bibinfo {author} {\bibfnamefont {D.}~\bibnamefont {Steppenbeck}}, \bibinfo
  {author} {\bibfnamefont {L.}~\bibnamefont {Stuhl}}, \bibinfo {author}
  {\bibfnamefont {Y.}~\bibnamefont {Sun}}, \bibinfo {author} {\bibfnamefont
  {Y.}~\bibnamefont {Togano}}, \bibinfo {author} {\bibfnamefont
  {T.}~\bibnamefont {Uesaka}}, \bibinfo {author} {\bibfnamefont
  {K.}~\bibnamefont {Wimmer}}, \bibinfo {author} {\bibfnamefont
  {K.}~\bibnamefont {Yoneda}}, \bibinfo {author} {\bibfnamefont
  {O.}~\bibnamefont {Aktas}}, \bibinfo {author} {\bibfnamefont
  {T.}~\bibnamefont {Aumann}}, \bibinfo {author} {\bibfnamefont
  {L.}~\bibnamefont {Chung}}, \bibinfo {author} {\bibfnamefont
  {F.}~\bibnamefont {Flavigny}}, \bibinfo {author} {\bibfnamefont
  {S.}~\bibnamefont {Franchoo}}, \bibinfo {author} {\bibfnamefont
  {I.}~\bibnamefont {Ga{\v s}pari{\'c}}}, \bibinfo {author} {\bibfnamefont
  {R.-B.}\ \bibnamefont {Gerst}}, \bibinfo {author} {\bibfnamefont
  {J.}~\bibnamefont {Gibelin}}, \bibinfo {author} {\bibfnamefont
  {K.}~\bibnamefont {Hahn}}, \bibinfo {author} {\bibfnamefont {D.}~\bibnamefont
  {Kim}}, \bibinfo {author} {\bibfnamefont {T.}~\bibnamefont {Koiwai}},
  \bibinfo {author} {\bibfnamefont {Y.}~\bibnamefont {Kondo}}, \bibinfo
  {author} {\bibfnamefont {P.}~\bibnamefont {Koseoglou}}, \bibinfo {author}
  {\bibfnamefont {J.}~\bibnamefont {Lee}}, \bibinfo {author} {\bibfnamefont
  {C.}~\bibnamefont {Lehr}}, \bibinfo {author} {\bibfnamefont {B.}~\bibnamefont
  {Linh}}, \bibinfo {author} {\bibfnamefont {T.}~\bibnamefont {Lokotko}},
  \bibinfo {author} {\bibfnamefont {M.}~\bibnamefont {MacCormick}}, \bibinfo
  {author} {\bibfnamefont {K.}~\bibnamefont {Moschner}}, \bibinfo {author}
  {\bibfnamefont {T.}~\bibnamefont {Nakamura}}, \bibinfo {author}
  {\bibfnamefont {S.}~\bibnamefont {Park}}, \bibinfo {author} {\bibfnamefont
  {D.}~\bibnamefont {Rossi}}, \bibinfo {author} {\bibfnamefont
  {E.}~\bibnamefont {Sahin}}, \bibinfo {author} {\bibfnamefont
  {D.}~\bibnamefont {Sohler}}, \bibinfo {author} {\bibfnamefont {P.-A.}\
  \bibnamefont {S{\"o}derstr{\"o}m}}, \bibinfo {author} {\bibfnamefont
  {S.}~\bibnamefont {Takeuchi}}, \bibinfo {author} {\bibfnamefont
  {H.}~\bibnamefont {Toernqvist}}, \bibinfo {author} {\bibfnamefont
  {V.}~\bibnamefont {Vaquero}}, \bibinfo {author} {\bibfnamefont
  {V.}~\bibnamefont {Wagner}}, \bibinfo {author} {\bibfnamefont
  {S.}~\bibnamefont {Wang}}, \bibinfo {author} {\bibfnamefont {V.}~\bibnamefont
  {Werner}}, \bibinfo {author} {\bibfnamefont {X.}~\bibnamefont {Xu}}, \bibinfo
  {author} {\bibfnamefont {H.}~\bibnamefont {Yamada}}, \bibinfo {author}
  {\bibfnamefont {D.}~\bibnamefont {Yan}}, \bibinfo {author} {\bibfnamefont
  {Z.}~\bibnamefont {Yang}}, \bibinfo {author} {\bibfnamefont {M.}~\bibnamefont
  {Yasuda}}, \ and\ \bibinfo {author} {\bibfnamefont {L.}~\bibnamefont
  {Zanetti}},\ }\href {\doibase https://doi.org/10.1016/j.physletb.2019.135071}
  {\bibfield  {journal} {\bibinfo  {journal} {Phys. Lett. B}\ }\textbf
  {\bibinfo {volume} {800}},\ \bibinfo {pages} {135071} (\bibinfo {year}
  {2020})}\BibitemShut {NoStop}%
\bibitem [{\citenamefont {Meisel}\ \emph {et~al.}(2020)\citenamefont {Meisel},
  \citenamefont {George}, \citenamefont {Ahn}, \citenamefont {Bazin},
  \citenamefont {Brown}, \citenamefont {Browne}, \citenamefont {Carpino},
  \citenamefont {Chung}, \citenamefont {Cyburt}, \citenamefont {Estrad\'e},
  \citenamefont {Famiano}, \citenamefont {Gade}, \citenamefont {Langer},
  \citenamefont {Mato\ifmmode~\check{s}\else \v{s}\fi{}}, \citenamefont
  {Mittig}, \citenamefont {Montes}, \citenamefont {Morrissey}, \citenamefont
  {Pereira}, \citenamefont {Schatz}, \citenamefont {Schatz}, \citenamefont
  {Scott}, \citenamefont {Shapira}, \citenamefont {Smith}, \citenamefont
  {Stevens}, \citenamefont {Tan}, \citenamefont {Tarasov}, \citenamefont
  {Towers}, \citenamefont {Wimmer}, \citenamefont {Winkelbauer}, \citenamefont
  {Yurkon},\ and\ \citenamefont {Zegers}}]{Meisel2020}%
  \BibitemOpen
  \bibfield  {author} {\bibinfo {author} {\bibfnamefont {Z.}~\bibnamefont
  {Meisel}}, \bibinfo {author} {\bibfnamefont {S.}~\bibnamefont {George}},
  \bibinfo {author} {\bibfnamefont {S.}~\bibnamefont {Ahn}}, \bibinfo {author}
  {\bibfnamefont {D.}~\bibnamefont {Bazin}}, \bibinfo {author} {\bibfnamefont
  {B.~A.}\ \bibnamefont {Brown}}, \bibinfo {author} {\bibfnamefont
  {J.}~\bibnamefont {Browne}}, \bibinfo {author} {\bibfnamefont {J.~F.}\
  \bibnamefont {Carpino}}, \bibinfo {author} {\bibfnamefont {H.}~\bibnamefont
  {Chung}}, \bibinfo {author} {\bibfnamefont {R.~H.}\ \bibnamefont {Cyburt}},
  \bibinfo {author} {\bibfnamefont {A.}~\bibnamefont {Estrad\'e}}, \bibinfo
  {author} {\bibfnamefont {M.}~\bibnamefont {Famiano}}, \bibinfo {author}
  {\bibfnamefont {A.}~\bibnamefont {Gade}}, \bibinfo {author} {\bibfnamefont
  {C.}~\bibnamefont {Langer}}, \bibinfo {author} {\bibfnamefont
  {M.}~\bibnamefont {Mato\ifmmode~\check{s}\else \v{s}\fi{}}}, \bibinfo
  {author} {\bibfnamefont {W.}~\bibnamefont {Mittig}}, \bibinfo {author}
  {\bibfnamefont {F.}~\bibnamefont {Montes}}, \bibinfo {author} {\bibfnamefont
  {D.~J.}\ \bibnamefont {Morrissey}}, \bibinfo {author} {\bibfnamefont
  {J.}~\bibnamefont {Pereira}}, \bibinfo {author} {\bibfnamefont
  {H.}~\bibnamefont {Schatz}}, \bibinfo {author} {\bibfnamefont
  {J.}~\bibnamefont {Schatz}}, \bibinfo {author} {\bibfnamefont
  {M.}~\bibnamefont {Scott}}, \bibinfo {author} {\bibfnamefont
  {D.}~\bibnamefont {Shapira}}, \bibinfo {author} {\bibfnamefont
  {K.}~\bibnamefont {Smith}}, \bibinfo {author} {\bibfnamefont
  {J.}~\bibnamefont {Stevens}}, \bibinfo {author} {\bibfnamefont
  {W.}~\bibnamefont {Tan}}, \bibinfo {author} {\bibfnamefont {O.}~\bibnamefont
  {Tarasov}}, \bibinfo {author} {\bibfnamefont {S.}~\bibnamefont {Towers}},
  \bibinfo {author} {\bibfnamefont {K.}~\bibnamefont {Wimmer}}, \bibinfo
  {author} {\bibfnamefont {J.~R.}\ \bibnamefont {Winkelbauer}}, \bibinfo
  {author} {\bibfnamefont {J.}~\bibnamefont {Yurkon}}, \ and\ \bibinfo {author}
  {\bibfnamefont {R.~G.~T.}\ \bibnamefont {Zegers}},\ }\href {\doibase
  10.1103/PhysRevC.101.052801} {\bibfield  {journal} {\bibinfo  {journal}
  {Phys. Rev. C}\ }\textbf {\bibinfo {volume} {101}},\ \bibinfo {pages}
  {052801(R)} (\bibinfo {year} {2020})}\BibitemShut {NoStop}%
\bibitem [{\citenamefont {Michimasa}\ \emph {et~al.}(2020)\citenamefont
  {Michimasa}, \citenamefont {Kobayashi}, \citenamefont {Kiyokawa},
  \citenamefont {Ota}, \citenamefont {Yokoyama}, \citenamefont {Nishimura},
  \citenamefont {Ahn}, \citenamefont {Baba}, \citenamefont {Berg},
  \citenamefont {Dozono}, \citenamefont {Fukuda}, \citenamefont {Furuno},
  \citenamefont {Ideguchi}, \citenamefont {Inabe}, \citenamefont {Kawabata},
  \citenamefont {Kawase}, \citenamefont {Kisamori}, \citenamefont {Kobayashi},
  \citenamefont {Kubo}, \citenamefont {Kubota}, \citenamefont {Lee},
  \citenamefont {Matsushita}, \citenamefont {Miya}, \citenamefont {Mizukami},
  \citenamefont {Nagakura}, \citenamefont {Oikawa}, \citenamefont {Sakai},
  \citenamefont {Shimizu}, \citenamefont {Stolz}, \citenamefont {Suzuki},
  \citenamefont {Takaki}, \citenamefont {Takeda}, \citenamefont {Takeuchi},
  \citenamefont {Tokieda}, \citenamefont {Uesaka}, \citenamefont {Yako},
  \citenamefont {Yamaguchi}, \citenamefont {Yanagisawa}, \citenamefont
  {Yoshida},\ and\ \citenamefont {Shimoura}}]{Michimasa2020}%
  \BibitemOpen
  \bibfield  {author} {\bibinfo {author} {\bibfnamefont {S.}~\bibnamefont
  {Michimasa}}, \bibinfo {author} {\bibfnamefont {M.}~\bibnamefont
  {Kobayashi}}, \bibinfo {author} {\bibfnamefont {Y.}~\bibnamefont {Kiyokawa}},
  \bibinfo {author} {\bibfnamefont {S.}~\bibnamefont {Ota}}, \bibinfo {author}
  {\bibfnamefont {R.}~\bibnamefont {Yokoyama}}, \bibinfo {author}
  {\bibfnamefont {D.}~\bibnamefont {Nishimura}}, \bibinfo {author}
  {\bibfnamefont {D.~S.}\ \bibnamefont {Ahn}}, \bibinfo {author} {\bibfnamefont
  {H.}~\bibnamefont {Baba}}, \bibinfo {author} {\bibfnamefont {G.~P.~A.}\
  \bibnamefont {Berg}}, \bibinfo {author} {\bibfnamefont {M.}~\bibnamefont
  {Dozono}}, \bibinfo {author} {\bibfnamefont {N.}~\bibnamefont {Fukuda}},
  \bibinfo {author} {\bibfnamefont {T.}~\bibnamefont {Furuno}}, \bibinfo
  {author} {\bibfnamefont {E.}~\bibnamefont {Ideguchi}}, \bibinfo {author}
  {\bibfnamefont {N.}~\bibnamefont {Inabe}}, \bibinfo {author} {\bibfnamefont
  {T.}~\bibnamefont {Kawabata}}, \bibinfo {author} {\bibfnamefont
  {S.}~\bibnamefont {Kawase}}, \bibinfo {author} {\bibfnamefont
  {K.}~\bibnamefont {Kisamori}}, \bibinfo {author} {\bibfnamefont
  {K.}~\bibnamefont {Kobayashi}}, \bibinfo {author} {\bibfnamefont
  {T.}~\bibnamefont {Kubo}}, \bibinfo {author} {\bibfnamefont {Y.}~\bibnamefont
  {Kubota}}, \bibinfo {author} {\bibfnamefont {C.~S.}\ \bibnamefont {Lee}},
  \bibinfo {author} {\bibfnamefont {M.}~\bibnamefont {Matsushita}}, \bibinfo
  {author} {\bibfnamefont {H.}~\bibnamefont {Miya}}, \bibinfo {author}
  {\bibfnamefont {A.}~\bibnamefont {Mizukami}}, \bibinfo {author}
  {\bibfnamefont {H.}~\bibnamefont {Nagakura}}, \bibinfo {author}
  {\bibfnamefont {H.}~\bibnamefont {Oikawa}}, \bibinfo {author} {\bibfnamefont
  {H.}~\bibnamefont {Sakai}}, \bibinfo {author} {\bibfnamefont
  {Y.}~\bibnamefont {Shimizu}}, \bibinfo {author} {\bibfnamefont
  {A.}~\bibnamefont {Stolz}}, \bibinfo {author} {\bibfnamefont
  {H.}~\bibnamefont {Suzuki}}, \bibinfo {author} {\bibfnamefont
  {M.}~\bibnamefont {Takaki}}, \bibinfo {author} {\bibfnamefont
  {H.}~\bibnamefont {Takeda}}, \bibinfo {author} {\bibfnamefont
  {S.}~\bibnamefont {Takeuchi}}, \bibinfo {author} {\bibfnamefont
  {H.}~\bibnamefont {Tokieda}}, \bibinfo {author} {\bibfnamefont
  {T.}~\bibnamefont {Uesaka}}, \bibinfo {author} {\bibfnamefont
  {K.}~\bibnamefont {Yako}}, \bibinfo {author} {\bibfnamefont {Y.}~\bibnamefont
  {Yamaguchi}}, \bibinfo {author} {\bibfnamefont {Y.}~\bibnamefont
  {Yanagisawa}}, \bibinfo {author} {\bibfnamefont {K.}~\bibnamefont {Yoshida}},
  \ and\ \bibinfo {author} {\bibfnamefont {S.}~\bibnamefont {Shimoura}},\
  }\href {\doibase 10.1103/PhysRevLett.125.122501} {\bibfield  {journal}
  {\bibinfo  {journal} {Phys. Rev. Lett.}\ }\textbf {\bibinfo {volume} {125}},\
  \bibinfo {pages} {122501} (\bibinfo {year} {2020})}\BibitemShut {NoStop}%
\bibitem [{\citenamefont {Rosenbusch}\ \emph {et~al.}(2022)\citenamefont
  {Rosenbusch}, \citenamefont {Wada}, \citenamefont {Chen}, \citenamefont
  {Takamine}, \citenamefont {Iimura}, \citenamefont {Hou}, \citenamefont
  {Xian}, \citenamefont {Yan}, \citenamefont {Schury}, \citenamefont {Ito},
  \citenamefont {Ishiyama}, \citenamefont {Kimura}, \citenamefont {Lee},
  \citenamefont {Liu}, \citenamefont {Michimasa}, \citenamefont {Miyatake},
  \citenamefont {Moon}, \citenamefont {Nishimura}, \citenamefont {Naimi},
  \citenamefont {Niwase},\ and\ \citenamefont {Wollnik}}]{Rosenbusch2022}%
  \BibitemOpen
  \bibfield  {author} {\bibinfo {author} {\bibfnamefont {M.}~\bibnamefont
  {Rosenbusch}}, \bibinfo {author} {\bibfnamefont {M.}~\bibnamefont {Wada}},
  \bibinfo {author} {\bibfnamefont {S.}~\bibnamefont {Chen}}, \bibinfo {author}
  {\bibfnamefont {A.}~\bibnamefont {Takamine}}, \bibinfo {author}
  {\bibfnamefont {S.}~\bibnamefont {Iimura}}, \bibinfo {author} {\bibfnamefont
  {D.}~\bibnamefont {Hou}}, \bibinfo {author} {\bibfnamefont {W.}~\bibnamefont
  {Xian}}, \bibinfo {author} {\bibfnamefont {S.}~\bibnamefont {Yan}}, \bibinfo
  {author} {\bibfnamefont {P.}~\bibnamefont {Schury}}, \bibinfo {author}
  {\bibfnamefont {Y.}~\bibnamefont {Ito}}, \bibinfo {author} {\bibfnamefont
  {H.}~\bibnamefont {Ishiyama}}, \bibinfo {author} {\bibfnamefont
  {S.}~\bibnamefont {Kimura}}, \bibinfo {author} {\bibfnamefont
  {J.}~\bibnamefont {Lee}}, \bibinfo {author} {\bibfnamefont {J.}~\bibnamefont
  {Liu}}, \bibinfo {author} {\bibfnamefont {S.}~\bibnamefont {Michimasa}},
  \bibinfo {author} {\bibfnamefont {H.}~\bibnamefont {Miyatake}}, \bibinfo
  {author} {\bibfnamefont {Y.~J.}\ \bibnamefont {Moon}}, \bibinfo {author}
  {\bibfnamefont {S.}~\bibnamefont {Nishimura}}, \bibinfo {author}
  {\bibfnamefont {S.}~\bibnamefont {Naimi}}, \bibinfo {author} {\bibfnamefont
  {T.}~\bibnamefont {Niwase}}, \ and\ \bibinfo {author} {\bibfnamefont
  {H.}~\bibnamefont {Wollnik}},\ }\href {\doibase
  https://doi.org/10.48550/arXiv.2110.11507} {\bibfield  {journal} {\bibinfo
  {journal} {Nucl. Instrum. Methods. Phys. Res. A}\ ,\ \bibinfo {pages} {under
  review, see arxiv: https://doi.org/10.48550/arXiv.2110.11507}} (\bibinfo
  {year} {2022})}\BibitemShut {NoStop}%
\bibitem [{\citenamefont {Wada}\ \emph {et~al.}(2011)\citenamefont {Wada},
  \citenamefont {Takamine}, \citenamefont {Sonoda}, \citenamefont {Okada},
  \citenamefont {Schury},\ and\ \citenamefont {for~the
  SLOWRI~Collaboration}}]{Wada2011}%
  \BibitemOpen
  \bibfield  {author} {\bibinfo {author} {\bibfnamefont {M.}~\bibnamefont
  {Wada}}, \bibinfo {author} {\bibfnamefont {A.}~\bibnamefont {Takamine}},
  \bibinfo {author} {\bibfnamefont {T.}~\bibnamefont {Sonoda}}, \bibinfo
  {author} {\bibfnamefont {K.}~\bibnamefont {Okada}}, \bibinfo {author}
  {\bibfnamefont {P.}~\bibnamefont {Schury}}, \ and\ \bibinfo {author}
  {\bibnamefont {for~the SLOWRI~Collaboration}},\ }\href {\doibase
  10.1007/s10751-011-0322-8} {\bibfield  {journal} {\bibinfo  {journal}
  {Hyperfine Interact.}\ }\textbf {\bibinfo {volume} {199}},\ \bibinfo {pages}
  {269} (\bibinfo {year} {2011})}\BibitemShut {NoStop}%
\bibitem [{\citenamefont {Wimmer}\ \emph {et~al.}(2021)\citenamefont {Wimmer},
  \citenamefont {Doornenbal}, \citenamefont {Aoi}, \citenamefont {Baba},
  \citenamefont {Browne}, \citenamefont {Campell}, \citenamefont {Crawford},
  \citenamefont {De~Witte}, \citenamefont {Fransen}, \citenamefont {Hess},
  \citenamefont {Iwazaki}, \citenamefont {Kim}, \citenamefont {Kohda},
  \citenamefont {Koiwai}, \citenamefont {Mauss}, \citenamefont {Moon},
  \citenamefont {Parry}, \citenamefont {Reiter}, \citenamefont {Suzuki},
  \citenamefont {Taniuchi}, \citenamefont {Thiel},\ and\ \citenamefont
  {Yamamoto}}]{Wimmer2021}%
  \BibitemOpen
  \bibfield  {author} {\bibinfo {author} {\bibfnamefont {K.}~\bibnamefont
  {Wimmer}}, \bibinfo {author} {\bibfnamefont {P.}~\bibnamefont {Doornenbal}},
  \bibinfo {author} {\bibfnamefont {N.}~\bibnamefont {Aoi}}, \bibinfo {author}
  {\bibfnamefont {H.}~\bibnamefont {Baba}}, \bibinfo {author} {\bibfnamefont
  {F.}~\bibnamefont {Browne}}, \bibinfo {author} {\bibfnamefont
  {P.}~\bibnamefont {Campell}}, \bibinfo {author} {\bibfnamefont
  {H.}~\bibnamefont {Crawford}}, \bibinfo {author} {\bibfnamefont
  {H.}~\bibnamefont {De~Witte}}, \bibinfo {author} {\bibfnamefont
  {C.}~\bibnamefont {Fransen}}, \bibinfo {author} {\bibfnamefont
  {H.}~\bibnamefont {Hess}}, \bibinfo {author} {\bibfnamefont {S.}~\bibnamefont
  {Iwazaki}}, \bibinfo {author} {\bibfnamefont {J.}~\bibnamefont {Kim}},
  \bibinfo {author} {\bibfnamefont {A.}~\bibnamefont {Kohda}}, \bibinfo
  {author} {\bibfnamefont {T.}~\bibnamefont {Koiwai}}, \bibinfo {author}
  {\bibfnamefont {B.}~\bibnamefont {Mauss}}, \bibinfo {author} {\bibfnamefont
  {B.}~\bibnamefont {Moon}}, \bibinfo {author} {\bibfnamefont {T.}~\bibnamefont
  {Parry}}, \bibinfo {author} {\bibfnamefont {P.}~\bibnamefont {Reiter}},
  \bibinfo {author} {\bibfnamefont {D.}~\bibnamefont {Suzuki}}, \bibinfo
  {author} {\bibfnamefont {R.}~\bibnamefont {Taniuchi}}, \bibinfo {author}
  {\bibfnamefont {S.}~\bibnamefont {Thiel}}, \ and\ \bibinfo {author}
  {\bibfnamefont {Y.}~\bibnamefont {Yamamoto}},\ }\href@noop {} {\bibfield
  {journal} {\bibinfo  {journal} {RIKEN Accel. Prog. Rep.}\ }\textbf {\bibinfo
  {volume} {54}} (\bibinfo {year} {2021})}\BibitemShut {NoStop}%
\bibitem [{\citenamefont {Sakurai}(2010)}]{Sakurai2010}%
  \BibitemOpen
  \bibfield  {author} {\bibinfo {author} {\bibfnamefont {H.}~\bibnamefont
  {Sakurai}},\ }\href {\doibase 10.1063/1.3485213} {\bibfield  {journal}
  {\bibinfo  {journal} {AIP Conf. Proc.}\ }\textbf {\bibinfo {volume} {1269}},\
  \bibinfo {pages} {84} (\bibinfo {year} {2010})}\BibitemShut {NoStop}%
\bibitem [{\citenamefont {Okuno}\ \emph {et~al.}(2012)\citenamefont {Okuno},
  \citenamefont {Fukunishi},\ and\ \citenamefont {Kamigaito}}]{Okuno2012}%
  \BibitemOpen
  \bibfield  {author} {\bibinfo {author} {\bibfnamefont {H.}~\bibnamefont
  {Okuno}}, \bibinfo {author} {\bibfnamefont {N.}~\bibnamefont {Fukunishi}}, \
  and\ \bibinfo {author} {\bibfnamefont {O.}~\bibnamefont {Kamigaito}},\ }\href
  {\doibase 10.1093/ptep/pts046} {\bibfield  {journal} {\bibinfo  {journal}
  {Prog. Theor. Exp. Phys.}\ }\textbf {\bibinfo {volume} {2012}},\ \bibinfo
  {pages} {03C002} (\bibinfo {year} {2012})}\BibitemShut {NoStop}%
\bibitem [{\citenamefont {Chen}(2021)}]{Chen2021}%
  \BibitemOpen
  \bibfield  {author} {\bibinfo {author} {\bibfnamefont {S.}~\bibnamefont
  {Chen}},\ }\href
  {https://www.nishina.riken.jp/researcher/APR/APR054/pdf/97.pdf} {\bibfield
  {journal} {\bibinfo  {journal} {RIKEN Accel. Prog. Rep.}\ }\textbf {\bibinfo
  {volume} {54}},\ \bibinfo {pages} {97} (\bibinfo {year} {2021})}\BibitemShut
  {NoStop}%
\bibitem [{\citenamefont {Wada}\ \emph {et~al.}(2003)\citenamefont {Wada},
  \citenamefont {Ishida}, \citenamefont {Nakamura}, \citenamefont {Yamazaki},
  \citenamefont {Kambara}, \citenamefont {Ohyama}, \citenamefont {Kanai},
  \citenamefont {Kojima}, \citenamefont {Nakai}, \citenamefont {Ohshima},
  \citenamefont {Yoshida}, \citenamefont {Kubo}, \citenamefont {Matsuo},
  \citenamefont {Fukuyama}, \citenamefont {Okada}, \citenamefont {Sonoda},
  \citenamefont {Ohtani}, \citenamefont {Noda}, \citenamefont {Kawakami},\ and\
  \citenamefont {Katayama}}]{WADA2003570}%
  \BibitemOpen
  \bibfield  {author} {\bibinfo {author} {\bibfnamefont {M.}~\bibnamefont
  {Wada}}, \bibinfo {author} {\bibfnamefont {Y.}~\bibnamefont {Ishida}},
  \bibinfo {author} {\bibfnamefont {T.}~\bibnamefont {Nakamura}}, \bibinfo
  {author} {\bibfnamefont {Y.}~\bibnamefont {Yamazaki}}, \bibinfo {author}
  {\bibfnamefont {T.}~\bibnamefont {Kambara}}, \bibinfo {author} {\bibfnamefont
  {H.}~\bibnamefont {Ohyama}}, \bibinfo {author} {\bibfnamefont
  {Y.}~\bibnamefont {Kanai}}, \bibinfo {author} {\bibfnamefont {T.~M.}\
  \bibnamefont {Kojima}}, \bibinfo {author} {\bibfnamefont {Y.}~\bibnamefont
  {Nakai}}, \bibinfo {author} {\bibfnamefont {N.}~\bibnamefont {Ohshima}},
  \bibinfo {author} {\bibfnamefont {A.}~\bibnamefont {Yoshida}}, \bibinfo
  {author} {\bibfnamefont {T.}~\bibnamefont {Kubo}}, \bibinfo {author}
  {\bibfnamefont {Y.}~\bibnamefont {Matsuo}}, \bibinfo {author} {\bibfnamefont
  {Y.}~\bibnamefont {Fukuyama}}, \bibinfo {author} {\bibfnamefont
  {K.}~\bibnamefont {Okada}}, \bibinfo {author} {\bibfnamefont
  {T.}~\bibnamefont {Sonoda}}, \bibinfo {author} {\bibfnamefont
  {S.}~\bibnamefont {Ohtani}}, \bibinfo {author} {\bibfnamefont
  {K.}~\bibnamefont {Noda}}, \bibinfo {author} {\bibfnamefont {H.}~\bibnamefont
  {Kawakami}}, \ and\ \bibinfo {author} {\bibfnamefont {I.}~\bibnamefont
  {Katayama}},\ }\href {\doibase https://doi.org/10.1016/S0168-583X(02)02151-1}
  {\bibfield  {journal} {\bibinfo  {journal} {Nucl. Instrum. Methods. Phys.
  Res. B}\ }\textbf {\bibinfo {volume} {204}},\ \bibinfo {pages} {570 }
  (\bibinfo {year} {2003})}\BibitemShut {NoStop}%
\bibitem [{\citenamefont {Takamine}\ \emph {et~al.}(2005)\citenamefont
  {Takamine}, \citenamefont {Wada}, \citenamefont {Ishida}, \citenamefont
  {Nakamura}, \citenamefont {Okada}, \citenamefont {Yamazaki}, \citenamefont
  {Kambara}, \citenamefont {Kanai}, \citenamefont {Kojima}, \citenamefont
  {Nakai}, \citenamefont {Oshima}, \citenamefont {Yoshida}, \citenamefont
  {Kubo}, \citenamefont {Ohtani}, \citenamefont {Noda}, \citenamefont
  {Katayama}, \citenamefont {Hostain}, \citenamefont {Varentsov},\ and\
  \citenamefont {Wollnik}}]{Takamine2005}%
  \BibitemOpen
  \bibfield  {author} {\bibinfo {author} {\bibfnamefont {A.}~\bibnamefont
  {Takamine}}, \bibinfo {author} {\bibfnamefont {M.}~\bibnamefont {Wada}},
  \bibinfo {author} {\bibfnamefont {Y.}~\bibnamefont {Ishida}}, \bibinfo
  {author} {\bibfnamefont {T.}~\bibnamefont {Nakamura}}, \bibinfo {author}
  {\bibfnamefont {K.}~\bibnamefont {Okada}}, \bibinfo {author} {\bibfnamefont
  {Y.}~\bibnamefont {Yamazaki}}, \bibinfo {author} {\bibfnamefont
  {T.}~\bibnamefont {Kambara}}, \bibinfo {author} {\bibfnamefont
  {Y.}~\bibnamefont {Kanai}}, \bibinfo {author} {\bibfnamefont {T.~M.}\
  \bibnamefont {Kojima}}, \bibinfo {author} {\bibfnamefont {Y.}~\bibnamefont
  {Nakai}}, \bibinfo {author} {\bibfnamefont {N.}~\bibnamefont {Oshima}},
  \bibinfo {author} {\bibfnamefont {A.}~\bibnamefont {Yoshida}}, \bibinfo
  {author} {\bibfnamefont {T.}~\bibnamefont {Kubo}}, \bibinfo {author}
  {\bibfnamefont {S.}~\bibnamefont {Ohtani}}, \bibinfo {author} {\bibfnamefont
  {K.}~\bibnamefont {Noda}}, \bibinfo {author} {\bibfnamefont {I.}~\bibnamefont
  {Katayama}}, \bibinfo {author} {\bibfnamefont {P.}~\bibnamefont {Hostain}},
  \bibinfo {author} {\bibfnamefont {V.}~\bibnamefont {Varentsov}}, \ and\
  \bibinfo {author} {\bibfnamefont {H.}~\bibnamefont {Wollnik}},\ }\href
  {\doibase 10.1063/1.2090290} {\bibfield  {journal} {\bibinfo  {journal} {Rev.
  Sci. Instrum.}\ }\textbf {\bibinfo {volume} {76}},\ \bibinfo {pages} {103503}
  (\bibinfo {year} {2005})}\BibitemShut {NoStop}%
\bibitem [{\citenamefont {Arai}\ \emph {et~al.}(2014)\citenamefont {Arai},
  \citenamefont {Ito}, \citenamefont {Wada}, \citenamefont {Schury},
  \citenamefont {Sonoda},\ and\ \citenamefont {Mita}}]{ARAI201456}%
  \BibitemOpen
  \bibfield  {author} {\bibinfo {author} {\bibfnamefont {F.}~\bibnamefont
  {Arai}}, \bibinfo {author} {\bibfnamefont {Y.}~\bibnamefont {Ito}}, \bibinfo
  {author} {\bibfnamefont {M.}~\bibnamefont {Wada}}, \bibinfo {author}
  {\bibfnamefont {P.}~\bibnamefont {Schury}}, \bibinfo {author} {\bibfnamefont
  {T.}~\bibnamefont {Sonoda}}, \ and\ \bibinfo {author} {\bibfnamefont
  {H.}~\bibnamefont {Mita}},\ }\href {\doibase
  https://doi.org/10.1016/j.ijms.2014.01.005} {\bibfield  {journal} {\bibinfo
  {journal} {Int. J. Mass Spectrom.}\ }\textbf {\bibinfo {volume} {362}},\
  \bibinfo {pages} {56 } (\bibinfo {year} {2014})}\BibitemShut {NoStop}%
\bibitem [{\citenamefont {Bollen}(2011)}]{BOLLEN2011131}%
  \BibitemOpen
  \bibfield  {author} {\bibinfo {author} {\bibfnamefont {G.}~\bibnamefont
  {Bollen}},\ }\href {\doibase https://doi.org/10.1016/j.ijms.2010.09.032}
  {\bibfield  {journal} {\bibinfo  {journal} {Int. J. Mass Spectrom.}\ }\textbf
  {\bibinfo {volume} {299}},\ \bibinfo {pages} {131 } (\bibinfo {year}
  {2011})}\BibitemShut {NoStop}%
\bibitem [{\citenamefont {Schury}\ \emph {et~al.}(2014)\citenamefont {Schury},
  \citenamefont {Ito}, \citenamefont {Wada},\ and\ \citenamefont
  {Wollnik}}]{SCHURY201419}%
  \BibitemOpen
  \bibfield  {author} {\bibinfo {author} {\bibfnamefont {P.}~\bibnamefont
  {Schury}}, \bibinfo {author} {\bibfnamefont {Y.}~\bibnamefont {Ito}},
  \bibinfo {author} {\bibfnamefont {M.}~\bibnamefont {Wada}}, \ and\ \bibinfo
  {author} {\bibfnamefont {H.}~\bibnamefont {Wollnik}},\ }\href {\doibase
  https://doi.org/10.1016/j.ijms.2013.11.005} {\bibfield  {journal} {\bibinfo
  {journal} {Int. J. Mass Spectrom.}\ }\textbf {\bibinfo {volume} {359}},\
  \bibinfo {pages} {19 } (\bibinfo {year} {2014})}\BibitemShut {NoStop}%
\bibitem [{\citenamefont {Ito}\ \emph {et~al.}(2018)\citenamefont {Ito},
  \citenamefont {Schury}, \citenamefont {Wada}, \citenamefont {Arai},
  \citenamefont {Haba}, \citenamefont {Hirayama}, \citenamefont {Ishizawa},
  \citenamefont {Kaji}, \citenamefont {Kimura}, \citenamefont {Koura},
  \citenamefont {MacCormick}, \citenamefont {Miyatake}, \citenamefont {Moon},
  \citenamefont {Morimoto}, \citenamefont {Morita}, \citenamefont {Mukai},
  \citenamefont {Murray}, \citenamefont {Niwase}, \citenamefont {Okada},
  \citenamefont {Ozawa}, \citenamefont {Rosenbusch}, \citenamefont {Takamine},
  \citenamefont {Tanaka}, \citenamefont {Watanabe}, \citenamefont {Wollnik},\
  and\ \citenamefont {Yamaki}}]{Ito2018}%
  \BibitemOpen
  \bibfield  {author} {\bibinfo {author} {\bibfnamefont {Y.}~\bibnamefont
  {Ito}}, \bibinfo {author} {\bibfnamefont {P.}~\bibnamefont {Schury}},
  \bibinfo {author} {\bibfnamefont {M.}~\bibnamefont {Wada}}, \bibinfo {author}
  {\bibfnamefont {F.}~\bibnamefont {Arai}}, \bibinfo {author} {\bibfnamefont
  {H.}~\bibnamefont {Haba}}, \bibinfo {author} {\bibfnamefont {Y.}~\bibnamefont
  {Hirayama}}, \bibinfo {author} {\bibfnamefont {S.}~\bibnamefont {Ishizawa}},
  \bibinfo {author} {\bibfnamefont {D.}~\bibnamefont {Kaji}}, \bibinfo {author}
  {\bibfnamefont {S.}~\bibnamefont {Kimura}}, \bibinfo {author} {\bibfnamefont
  {H.}~\bibnamefont {Koura}}, \bibinfo {author} {\bibfnamefont
  {M.}~\bibnamefont {MacCormick}}, \bibinfo {author} {\bibfnamefont
  {H.}~\bibnamefont {Miyatake}}, \bibinfo {author} {\bibfnamefont {J.~Y.}\
  \bibnamefont {Moon}}, \bibinfo {author} {\bibfnamefont {K.}~\bibnamefont
  {Morimoto}}, \bibinfo {author} {\bibfnamefont {K.}~\bibnamefont {Morita}},
  \bibinfo {author} {\bibfnamefont {M.}~\bibnamefont {Mukai}}, \bibinfo
  {author} {\bibfnamefont {I.}~\bibnamefont {Murray}}, \bibinfo {author}
  {\bibfnamefont {T.}~\bibnamefont {Niwase}}, \bibinfo {author} {\bibfnamefont
  {K.}~\bibnamefont {Okada}}, \bibinfo {author} {\bibfnamefont
  {A.}~\bibnamefont {Ozawa}}, \bibinfo {author} {\bibfnamefont
  {M.}~\bibnamefont {Rosenbusch}}, \bibinfo {author} {\bibfnamefont
  {A.}~\bibnamefont {Takamine}}, \bibinfo {author} {\bibfnamefont
  {T.}~\bibnamefont {Tanaka}}, \bibinfo {author} {\bibfnamefont {Y.~X.}\
  \bibnamefont {Watanabe}}, \bibinfo {author} {\bibfnamefont {H.}~\bibnamefont
  {Wollnik}}, \ and\ \bibinfo {author} {\bibfnamefont {S.}~\bibnamefont
  {Yamaki}},\ }\href {\doibase 10.1103/PhysRevLett.120.152501} {\bibfield
  {journal} {\bibinfo  {journal} {Phys. Rev. Lett.}\ }\textbf {\bibinfo
  {volume} {120}},\ \bibinfo {pages} {152501} (\bibinfo {year}
  {2018})}\BibitemShut {NoStop}%
\bibitem [{\citenamefont {Ito}\ \emph {et~al.}(2013)\citenamefont {Ito},
  \citenamefont {Schury}, \citenamefont {Wada}, \citenamefont {Naimi},
  \citenamefont {Sonoda}, \citenamefont {Mita}, \citenamefont {Arai},
  \citenamefont {Takamine}, \citenamefont {Okada}, \citenamefont {Ozawa},\ and\
  \citenamefont {Wollnik}}]{Ito2013}%
  \BibitemOpen
  \bibfield  {author} {\bibinfo {author} {\bibfnamefont {Y.}~\bibnamefont
  {Ito}}, \bibinfo {author} {\bibfnamefont {P.}~\bibnamefont {Schury}},
  \bibinfo {author} {\bibfnamefont {M.}~\bibnamefont {Wada}}, \bibinfo {author}
  {\bibfnamefont {S.}~\bibnamefont {Naimi}}, \bibinfo {author} {\bibfnamefont
  {T.}~\bibnamefont {Sonoda}}, \bibinfo {author} {\bibfnamefont
  {H.}~\bibnamefont {Mita}}, \bibinfo {author} {\bibfnamefont {F.}~\bibnamefont
  {Arai}}, \bibinfo {author} {\bibfnamefont {A.}~\bibnamefont {Takamine}},
  \bibinfo {author} {\bibfnamefont {K.}~\bibnamefont {Okada}}, \bibinfo
  {author} {\bibfnamefont {A.}~\bibnamefont {Ozawa}}, \ and\ \bibinfo {author}
  {\bibfnamefont {H.}~\bibnamefont {Wollnik}},\ }\href {\doibase
  10.1103/PhysRevC.88.011306} {\bibfield  {journal} {\bibinfo  {journal} {Phys.
  Rev. C}\ }\textbf {\bibinfo {volume} {88}},\ \bibinfo {pages} {011306(R)}
  (\bibinfo {year} {2013})}\BibitemShut {NoStop}%
\bibitem [{\citenamefont {Schury}\ \emph {et~al.}(2017)\citenamefont {Schury},
  \citenamefont {Wada}, \citenamefont {Ito}, \citenamefont {Kaji},
  \citenamefont {Haba}, \citenamefont {Hirayama}, \citenamefont {Kimura},
  \citenamefont {Koura}, \citenamefont {MacCormick}, \citenamefont {Miyatake},
  \citenamefont {Moon}, \citenamefont {Morimoto}, \citenamefont {Morita},
  \citenamefont {Murray}, \citenamefont {Ozawa}, \citenamefont {Rosenbusch},
  \citenamefont {Reponen}, \citenamefont {Takamine}, \citenamefont {Tanaka},
  \citenamefont {Watanabe},\ and\ \citenamefont {Wollnik}}]{Schury2017}%
  \BibitemOpen
  \bibfield  {author} {\bibinfo {author} {\bibfnamefont {P.}~\bibnamefont
  {Schury}}, \bibinfo {author} {\bibfnamefont {M.}~\bibnamefont {Wada}},
  \bibinfo {author} {\bibfnamefont {Y.}~\bibnamefont {Ito}}, \bibinfo {author}
  {\bibfnamefont {D.}~\bibnamefont {Kaji}}, \bibinfo {author} {\bibfnamefont
  {H.}~\bibnamefont {Haba}}, \bibinfo {author} {\bibfnamefont {Y.}~\bibnamefont
  {Hirayama}}, \bibinfo {author} {\bibfnamefont {S.}~\bibnamefont {Kimura}},
  \bibinfo {author} {\bibfnamefont {H.}~\bibnamefont {Koura}}, \bibinfo
  {author} {\bibfnamefont {M.}~\bibnamefont {MacCormick}}, \bibinfo {author}
  {\bibfnamefont {H.}~\bibnamefont {Miyatake}}, \bibinfo {author}
  {\bibfnamefont {J.~Y.}\ \bibnamefont {Moon}}, \bibinfo {author}
  {\bibfnamefont {K.}~\bibnamefont {Morimoto}}, \bibinfo {author}
  {\bibfnamefont {K.}~\bibnamefont {Morita}}, \bibinfo {author} {\bibfnamefont
  {I.}~\bibnamefont {Murray}}, \bibinfo {author} {\bibfnamefont
  {A.}~\bibnamefont {Ozawa}}, \bibinfo {author} {\bibfnamefont
  {M.}~\bibnamefont {Rosenbusch}}, \bibinfo {author} {\bibfnamefont
  {M.}~\bibnamefont {Reponen}}, \bibinfo {author} {\bibfnamefont
  {A.}~\bibnamefont {Takamine}}, \bibinfo {author} {\bibfnamefont
  {T.}~\bibnamefont {Tanaka}}, \bibinfo {author} {\bibfnamefont {Y.~X.}\
  \bibnamefont {Watanabe}}, \ and\ \bibinfo {author} {\bibfnamefont
  {H.}~\bibnamefont {Wollnik}},\ }\href {\doibase
  https://doi.org/10.1016/j.nimb.2017.06.014} {\bibfield  {journal} {\bibinfo
  {journal} {Nucl. Instrum. Methods. Phys. Res. B}\ }\textbf {\bibinfo {volume}
  {407}},\ \bibinfo {pages} {160} (\bibinfo {year} {2017})}\BibitemShut
  {NoStop}%
\bibitem [{\citenamefont {Johnson}(1949)}]{Johnson1949a}%
  \BibitemOpen
  \bibfield  {author} {\bibinfo {author} {\bibfnamefont {N.~L.}\ \bibnamefont
  {Johnson}},\ }\href {http://www.jstor.org/stable/2332539} {\bibfield
  {journal} {\bibinfo  {journal} {Biometrika}\ }\textbf {\bibinfo {volume}
  {36}},\ \bibinfo {pages} {149} (\bibinfo {year} {1949})}\BibitemShut
  {NoStop}%
\bibitem [{\citenamefont {Verkerke}\ and\ \citenamefont
  {Kirkby}(2006)}]{Verkerke2006}%
  \BibitemOpen
  \bibfield  {author} {\bibinfo {author} {\bibfnamefont {W.}~\bibnamefont
  {Verkerke}}\ and\ \bibinfo {author} {\bibfnamefont {D.}~\bibnamefont
  {Kirkby}},\ }\href {\doibase 10.1142/9781860948985_0039} {\bibfield
  {journal} {\bibinfo  {journal} {Stat. Prob. Part. Phys., Astro. Cosmo. -
  Proceedings of PHYSTAT 2005}\ ,\ \bibinfo {pages} {186}} (\bibinfo {year}
  {2006})}\BibitemShut {NoStop}%
\bibitem [{\citenamefont {Brun}\ and\ \citenamefont
  {Rademakers}(1997)}]{BRUN199781}%
  \BibitemOpen
  \bibfield  {author} {\bibinfo {author} {\bibfnamefont {R.}~\bibnamefont
  {Brun}}\ and\ \bibinfo {author} {\bibfnamefont {F.}~\bibnamefont
  {Rademakers}},\ }\href {\doibase
  https://doi.org/10.1016/S0168-9002(97)00048-X} {\bibfield  {journal}
  {\bibinfo  {journal} {Nucl. Instrum. Methods. Phys. Res. B}\ }\textbf
  {\bibinfo {volume} {389}},\ \bibinfo {pages} {81} (\bibinfo {year}
  {1997})}\BibitemShut {NoStop}%
\bibitem [{\citenamefont {Wang}\ \emph {et~al.}(2021)\citenamefont {Wang},
  \citenamefont {Huang}, \citenamefont {Kondev}, \citenamefont {Audi},\ and\
  \citenamefont {Naimi}}]{Wang_2021}%
  \BibitemOpen
  \bibfield  {author} {\bibinfo {author} {\bibfnamefont {M.}~\bibnamefont
  {Wang}}, \bibinfo {author} {\bibfnamefont {W.}~\bibnamefont {Huang}},
  \bibinfo {author} {\bibfnamefont {F.}~\bibnamefont {Kondev}}, \bibinfo
  {author} {\bibfnamefont {G.}~\bibnamefont {Audi}}, \ and\ \bibinfo {author}
  {\bibfnamefont {S.}~\bibnamefont {Naimi}},\ }\href {\doibase
  10.1088/1674-1137/abddaf} {\bibfield  {journal} {\bibinfo  {journal} {Chinese
  Phys. C}\ }\textbf {\bibinfo {volume} {45}},\ \bibinfo {pages} {030003}
  (\bibinfo {year} {2021})}\BibitemShut {NoStop}%
\bibitem [{\citenamefont {Rosenbusch}\ \emph {et~al.}(2015)\citenamefont
  {Rosenbusch}, \citenamefont {Ascher}, \citenamefont {Atanasov}, \citenamefont
  {Barbieri}, \citenamefont {Beck}, \citenamefont {Blaum}, \citenamefont
  {Borgmann}, \citenamefont {Breitenfeldt}, \citenamefont {Cakirli},
  \citenamefont {Cipollone}, \citenamefont {George}, \citenamefont {Herfurth},
  \citenamefont {Kowalska}, \citenamefont {Kreim}, \citenamefont {Lunney},
  \citenamefont {Manea}, \citenamefont {Navr\'atil}, \citenamefont {Neidherr},
  \citenamefont {Schweikhard}, \citenamefont {Som\`a}, \citenamefont {Stanja},
  \citenamefont {Wienholtz}, \citenamefont {Wolf},\ and\ \citenamefont
  {Zuber}}]{Rosenbusch2015}%
  \BibitemOpen
  \bibfield  {author} {\bibinfo {author} {\bibfnamefont {M.}~\bibnamefont
  {Rosenbusch}}, \bibinfo {author} {\bibfnamefont {P.}~\bibnamefont {Ascher}},
  \bibinfo {author} {\bibfnamefont {D.}~\bibnamefont {Atanasov}}, \bibinfo
  {author} {\bibfnamefont {C.}~\bibnamefont {Barbieri}}, \bibinfo {author}
  {\bibfnamefont {D.}~\bibnamefont {Beck}}, \bibinfo {author} {\bibfnamefont
  {K.}~\bibnamefont {Blaum}}, \bibinfo {author} {\bibfnamefont
  {C.}~\bibnamefont {Borgmann}}, \bibinfo {author} {\bibfnamefont
  {M.}~\bibnamefont {Breitenfeldt}}, \bibinfo {author} {\bibfnamefont {R.~B.}\
  \bibnamefont {Cakirli}}, \bibinfo {author} {\bibfnamefont {A.}~\bibnamefont
  {Cipollone}}, \bibinfo {author} {\bibfnamefont {S.}~\bibnamefont {George}},
  \bibinfo {author} {\bibfnamefont {F.}~\bibnamefont {Herfurth}}, \bibinfo
  {author} {\bibfnamefont {M.}~\bibnamefont {Kowalska}}, \bibinfo {author}
  {\bibfnamefont {S.}~\bibnamefont {Kreim}}, \bibinfo {author} {\bibfnamefont
  {D.}~\bibnamefont {Lunney}}, \bibinfo {author} {\bibfnamefont
  {V.}~\bibnamefont {Manea}}, \bibinfo {author} {\bibfnamefont
  {P.}~\bibnamefont {Navr\'atil}}, \bibinfo {author} {\bibfnamefont
  {D.}~\bibnamefont {Neidherr}}, \bibinfo {author} {\bibfnamefont
  {L.}~\bibnamefont {Schweikhard}}, \bibinfo {author} {\bibfnamefont
  {V.}~\bibnamefont {Som\`a}}, \bibinfo {author} {\bibfnamefont
  {J.}~\bibnamefont {Stanja}}, \bibinfo {author} {\bibfnamefont
  {F.}~\bibnamefont {Wienholtz}}, \bibinfo {author} {\bibfnamefont {R.~N.}\
  \bibnamefont {Wolf}}, \ and\ \bibinfo {author} {\bibfnamefont
  {K.}~\bibnamefont {Zuber}},\ }\href {\doibase 10.1103/PhysRevLett.114.202501}
  {\bibfield  {journal} {\bibinfo  {journal} {Phys. Rev. Lett.}\ }\textbf
  {\bibinfo {volume} {114}},\ \bibinfo {pages} {202501} (\bibinfo {year}
  {2015})}\BibitemShut {NoStop}%
\bibitem [{\citenamefont {Otsuka}\ \emph {et~al.}(2010)\citenamefont {Otsuka},
  \citenamefont {Suzuki}, \citenamefont {Honma}, \citenamefont {Utsuno},
  \citenamefont {Tsunoda}, \citenamefont {Tsukiyama},\ and\ \citenamefont
  {Hjorth-Jensen}}]{Otsuka2010}%
  \BibitemOpen
  \bibfield  {author} {\bibinfo {author} {\bibfnamefont {T.}~\bibnamefont
  {Otsuka}}, \bibinfo {author} {\bibfnamefont {T.}~\bibnamefont {Suzuki}},
  \bibinfo {author} {\bibfnamefont {M.}~\bibnamefont {Honma}}, \bibinfo
  {author} {\bibfnamefont {Y.}~\bibnamefont {Utsuno}}, \bibinfo {author}
  {\bibfnamefont {N.}~\bibnamefont {Tsunoda}}, \bibinfo {author} {\bibfnamefont
  {K.}~\bibnamefont {Tsukiyama}}, \ and\ \bibinfo {author} {\bibfnamefont
  {M.}~\bibnamefont {Hjorth-Jensen}},\ }\href {\doibase
  10.1103/PhysRevLett.104.012501} {\bibfield  {journal} {\bibinfo  {journal}
  {Phys. Rev. Lett.}\ }\textbf {\bibinfo {volume} {104}},\ \bibinfo {pages}
  {012501} (\bibinfo {year} {2010})}\BibitemShut {NoStop}%
\bibitem [{\citenamefont {Otsuka}\ and\ \citenamefont
  {Tsunoda}(2016)}]{Otsuka2016}%
  \BibitemOpen
  \bibfield  {author} {\bibinfo {author} {\bibfnamefont {T.}~\bibnamefont
  {Otsuka}}\ and\ \bibinfo {author} {\bibfnamefont {Y.}~\bibnamefont
  {Tsunoda}},\ }\href {\doibase 10.1088/0954-3899/43/2/024009} {\bibfield
  {journal} {\bibinfo  {journal} {J. Phys. G: Nucl. Part. Phys.}\ }\textbf
  {\bibinfo {volume} {43}},\ \bibinfo {pages} {024009} (\bibinfo {year}
  {2016})}\BibitemShut {NoStop}%
\bibitem [{\citenamefont {Shimizu}\ \emph {et~al.}(2012)\citenamefont
  {Shimizu}, \citenamefont {Abe}, \citenamefont {Tsunoda}, \citenamefont
  {Utsuno}, \citenamefont {Yoshida}, \citenamefont {Mizusaki}, \citenamefont
  {Honma},\ and\ \citenamefont {Otsuka}}]{Shimizu2012}%
  \BibitemOpen
  \bibfield  {author} {\bibinfo {author} {\bibfnamefont {N.}~\bibnamefont
  {Shimizu}}, \bibinfo {author} {\bibfnamefont {T.}~\bibnamefont {Abe}},
  \bibinfo {author} {\bibfnamefont {Y.}~\bibnamefont {Tsunoda}}, \bibinfo
  {author} {\bibfnamefont {Y.}~\bibnamefont {Utsuno}}, \bibinfo {author}
  {\bibfnamefont {T.}~\bibnamefont {Yoshida}}, \bibinfo {author} {\bibfnamefont
  {T.}~\bibnamefont {Mizusaki}}, \bibinfo {author} {\bibfnamefont
  {M.}~\bibnamefont {Honma}}, \ and\ \bibinfo {author} {\bibfnamefont
  {T.}~\bibnamefont {Otsuka}},\ }\href {https://doi.org/10.1093/ptep/pts012}
  {\bibfield  {journal} {\bibinfo  {journal} {Prog. Theor. Exp. Phys.}\
  }\textbf {\bibinfo {volume} {2012}},\ \bibinfo {pages} {01A205} (\bibinfo
  {year} {2012})}\BibitemShut {NoStop}%
\bibitem [{\citenamefont {Hjorth-Jensen}\ \emph {et~al.}(1995)\citenamefont
  {Hjorth-Jensen}, \citenamefont {Kuo},\ and\ \citenamefont
  {Osnes}}]{Hjorth95}%
  \BibitemOpen
  \bibfield  {author} {\bibinfo {author} {\bibfnamefont {M.}~\bibnamefont
  {Hjorth-Jensen}}, \bibinfo {author} {\bibfnamefont {T.~T.}\ \bibnamefont
  {Kuo}}, \ and\ \bibinfo {author} {\bibfnamefont {E.}~\bibnamefont {Osnes}},\
  }\href {\doibase https://doi.org/10.1016/0370-1573(95)00012-6} {\bibfield
  {journal} {\bibinfo  {journal} {Physics Reports}\ }\textbf {\bibinfo {volume}
  {261}},\ \bibinfo {pages} {125} (\bibinfo {year} {1995})}\BibitemShut
  {NoStop}%
\bibitem [{\citenamefont {Honma}\ \emph {et~al.}(2005)\citenamefont {Honma},
  \citenamefont {Otsuka}, \citenamefont {Brown},\ and\ \citenamefont
  {Mizusaki}}]{Honma05}%
  \BibitemOpen
  \bibfield  {author} {\bibinfo {author} {\bibfnamefont {M.}~\bibnamefont
  {Honma}}, \bibinfo {author} {\bibfnamefont {T.}~\bibnamefont {Otsuka}},
  \bibinfo {author} {\bibfnamefont {B.~A.}\ \bibnamefont {Brown}}, \ and\
  \bibinfo {author} {\bibfnamefont {T.}~\bibnamefont {Mizusaki}},\ }\href
  {\doibase 10.1140/epjad/i2005-06-032-2} {\bibfield  {journal} {\bibinfo
  {journal} {Eur. Phys. J. A}\ }\textbf {\bibinfo {volume} {25}},\ \bibinfo
  {pages} {499} (\bibinfo {year} {2005})}\BibitemShut {NoStop}%
\bibitem [{\citenamefont {Honma}\ \emph {et~al.}(2009)\citenamefont {Honma},
  \citenamefont {Otsuka}, \citenamefont {Mizusaki},\ and\ \citenamefont
  {Hjorth-Jensen}}]{Honma09}%
  \BibitemOpen
  \bibfield  {author} {\bibinfo {author} {\bibfnamefont {M.}~\bibnamefont
  {Honma}}, \bibinfo {author} {\bibfnamefont {T.}~\bibnamefont {Otsuka}},
  \bibinfo {author} {\bibfnamefont {T.}~\bibnamefont {Mizusaki}}, \ and\
  \bibinfo {author} {\bibfnamefont {M.}~\bibnamefont {Hjorth-Jensen}},\ }\href
  {\doibase 10.1103/PhysRevC.80.064323} {\bibfield  {journal} {\bibinfo
  {journal} {Phys. Rev. C}\ }\textbf {\bibinfo {volume} {80}},\ \bibinfo
  {pages} {064323} (\bibinfo {year} {2009})}\BibitemShut {NoStop}%
\bibitem [{\citenamefont {Tsunoda}\ \emph {et~al.}(2014)\citenamefont
  {Tsunoda}, \citenamefont {Otsuka}, \citenamefont {Shimizu}, \citenamefont
  {Honma},\ and\ \citenamefont {Utsuno}}]{Tsunoda2014}%
  \BibitemOpen
  \bibfield  {author} {\bibinfo {author} {\bibfnamefont {Y.}~\bibnamefont
  {Tsunoda}}, \bibinfo {author} {\bibfnamefont {T.}~\bibnamefont {Otsuka}},
  \bibinfo {author} {\bibfnamefont {N.}~\bibnamefont {Shimizu}}, \bibinfo
  {author} {\bibfnamefont {M.}~\bibnamefont {Honma}}, \ and\ \bibinfo {author}
  {\bibfnamefont {Y.}~\bibnamefont {Utsuno}},\ }\href {\doibase
  10.1103/PhysRevC.89.031301} {\bibfield  {journal} {\bibinfo  {journal} {Phys.
  Rev. C}\ }\textbf {\bibinfo {volume} {89}},\ \bibinfo {pages} {031301(R)}
  (\bibinfo {year} {2014})}\BibitemShut {NoStop}%
\bibitem [{\citenamefont {Shimizu}\ \emph {et~al.}(2019)\citenamefont
  {Shimizu}, \citenamefont {Mizusaki}, \citenamefont {Utsuno},\ and\
  \citenamefont {Tsunoda}}]{Shimizu2019}%
  \BibitemOpen
  \bibfield  {author} {\bibinfo {author} {\bibfnamefont {N.}~\bibnamefont
  {Shimizu}}, \bibinfo {author} {\bibfnamefont {T.}~\bibnamefont {Mizusaki}},
  \bibinfo {author} {\bibfnamefont {Y.}~\bibnamefont {Utsuno}}, \ and\ \bibinfo
  {author} {\bibfnamefont {Y.}~\bibnamefont {Tsunoda}},\ }\href {\doibase
  https://doi.org/10.1016/j.cpc.2019.06.011} {\bibfield  {journal} {\bibinfo
  {journal} {Comput. Phys. Commun.}\ }\textbf {\bibinfo {volume} {244}},\
  \bibinfo {pages} {372} (\bibinfo {year} {2019})}\BibitemShut {NoStop}%
\bibitem [{\citenamefont {M{\"{o}}ller}\ \emph {et~al.}(2016)\citenamefont
  {M{\"{o}}ller}, \citenamefont {Sierk}, \citenamefont {Ichikawa},\ and\
  \citenamefont {Sagawa}}]{Moeller2016}%
  \BibitemOpen
  \bibfield  {author} {\bibinfo {author} {\bibfnamefont {P.}~\bibnamefont
  {M{\"{o}}ller}}, \bibinfo {author} {\bibfnamefont {A.}~\bibnamefont {Sierk}},
  \bibinfo {author} {\bibfnamefont {T.}~\bibnamefont {Ichikawa}}, \ and\
  \bibinfo {author} {\bibfnamefont {H.}~\bibnamefont {Sagawa}},\ }\href
  {\doibase https://doi.org/10.1016/j.adt.2015.10.002} {\bibfield  {journal}
  {\bibinfo  {journal} {At. Data Nucl. Data Tables}\ }\textbf {\bibinfo
  {volume} {109-110}},\ \bibinfo {pages} {1} (\bibinfo {year}
  {2016})}\BibitemShut {NoStop}%
\bibitem [{\citenamefont {Goriely}\ \emph {et~al.}(2016)\citenamefont
  {Goriely}, \citenamefont {Chamel},\ and\ \citenamefont
  {Pearson}}]{Goriely2016}%
  \BibitemOpen
  \bibfield  {author} {\bibinfo {author} {\bibfnamefont {S.}~\bibnamefont
  {Goriely}}, \bibinfo {author} {\bibfnamefont {N.}~\bibnamefont {Chamel}}, \
  and\ \bibinfo {author} {\bibfnamefont {J.~M.}\ \bibnamefont {Pearson}},\
  }\href {\doibase 10.1088/1742-6596/665/1/012038} {\bibfield  {journal}
  {\bibinfo  {journal} {J. Phys. Conf. Ser.}\ }\textbf {\bibinfo {volume}
  {665}},\ \bibinfo {pages} {012038} (\bibinfo {year} {2016})}\BibitemShut
  {NoStop}%
\bibitem [{\citenamefont {Koura}\ \emph {et~al.}(2005)\citenamefont {Koura},
  \citenamefont {Tachibana}, \citenamefont {Uno},\ and\ \citenamefont
  {Yamada}}]{Koura2005}%
  \BibitemOpen
  \bibfield  {author} {\bibinfo {author} {\bibfnamefont {H.}~\bibnamefont
  {Koura}}, \bibinfo {author} {\bibfnamefont {T.}~\bibnamefont {Tachibana}},
  \bibinfo {author} {\bibfnamefont {M.}~\bibnamefont {Uno}}, \ and\ \bibinfo
  {author} {\bibfnamefont {M.}~\bibnamefont {Yamada}},\ }\href {\doibase
  10.1143/PTP.113.305} {\bibfield  {journal} {\bibinfo  {journal} {Prog. Theor.
  Phys.}\ }\textbf {\bibinfo {volume} {113}},\ \bibinfo {pages} {305} (\bibinfo
  {year} {2005})}\BibitemShut {NoStop}%
\bibitem [{\citenamefont {Stroberg}\ \emph {et~al.}(2021)\citenamefont
  {Stroberg}, \citenamefont {Holt}, \citenamefont {Schwenk},\ and\
  \citenamefont {Simonis}}]{Stroberg2021}%
  \BibitemOpen
  \bibfield  {author} {\bibinfo {author} {\bibfnamefont {S.~R.}\ \bibnamefont
  {Stroberg}}, \bibinfo {author} {\bibfnamefont {J.~D.}\ \bibnamefont {Holt}},
  \bibinfo {author} {\bibfnamefont {A.}~\bibnamefont {Schwenk}}, \ and\
  \bibinfo {author} {\bibfnamefont {J.}~\bibnamefont {Simonis}},\ }\href
  {\doibase 10.1103/PhysRevLett.126.022501} {\bibfield  {journal} {\bibinfo
  {journal} {Phys. Rev. Lett.}\ }\textbf {\bibinfo {volume} {126}},\ \bibinfo
  {pages} {022501} (\bibinfo {year} {2021})}\BibitemShut {NoStop}%
\end{thebibliography}

\end{document}